\documentclass{emulateapj}

\def\deg{{$^{\circ}$}}

\def\kmsec{\mbox{km~s$^{\rm -1}$}}
\def\logg{\mbox{log~{\it g}}}
\def\teff{\mbox{$T_{\rm eff}$}}

\def\rpro{\mbox{$r$-process}}
\def\spro{\mbox{$s$-process}}
\def\ncap{\mbox{$n$-capture}}
\def\msun{$M_{\odot}$}
\def\lsun{$L_{\odot}$}

\shorttitle{Heavy Element Dispersion in M92}
\shortauthors{Roederer \& Sneden}
\submitted{Accepted for publication in the Astronomical Journal}

\begin{document}

\title{Heavy Element Dispersion in the Metal-Poor Globular Cluster 
M92\footnote{
The WIYN Observatory is a joint facility of the 
University of Wisconsin-Madison, Indiana University, Yale University, 
and the National Optical Astronomy Observatory.}}

\author{
Ian U.\ Roederer\altaffilmark{1,3} and
Christopher Sneden\altaffilmark{2}
}

\altaffiltext{1}{Carnegie Observatories,
813 Santa Barbara Street, Pasadena, CA 91101 USA;
iur@obs.carnegiescience.edu}

\altaffiltext{2}{Department of Astronomy, University of Texas at Austin,
1 University Station, C1400, Austin, TX 78712 USA}

\altaffiltext{3}{Visiting Astronomer, Kitt Peak National Observatory, 
National Optical Astronomy Observatories, which is operated by the 
Association of Universities for Research in Astronomy, Inc.\ (AURA) 
under cooperative agreement with the National Science Foundation.}

\begin{abstract}
Dispersion among the light elements 
is common in globular clusters (GCs), while
dispersion among heavier elements is less common.
We present detection of \rpro\ dispersion relative to Fe 
in 19 red giants of the metal-poor GC M92.
Using spectra obtained with 
the Hydra multi-object spectrograph on the WIYN Telescope 
at Kitt Peak National Observatory,
we derive differential abundances for 21~species of 19~elements.
The Fe-group elements, plus Y and Zr, are
homogeneous at a level of 0.07--0.16~dex.
The heavy elements La, Eu, and Ho exhibit clear
star-to-star dispersion spanning 0.5--0.8~dex.
The abundances of these elements are correlated with one another,
and we demonstrate that they were produced by \rpro\ nucleosynthesis.
This \rpro\ dispersion is not correlated with the dispersion in
C, N, or Na in M92, indicating that \rpro\ inhomogeneities
were present in the gas throughout star formation.
The \rpro\ dispersion is similar to that previously 
observed in the metal-poor GC M15, but 
its origin in M15 or M92 is unknown at present.
\end{abstract}


\section{Introduction}
\label{intro}

Dispersion observed among the light elements 
(Li, C, N, O, Na, Mg, Al, and Si)
in Galactic globular clusters (GCs) has motivated
numerous attempts to characterize it,
both in terms of the internal star-to-star dispersion
and the range from one GC to another.
An order of magnitude increase in the amount of observational data
of these elements
in the last 5~years has led to an explosion of attempts
to model the light element dispersion
and understand its implications for GC formation.
The uniformity of heavier $\alpha$, Fe-group,
and neutron ($n$) capture elements in GCs has 
provided important constraints for these models, 
but characterizing this homogeneity has usually been 
of secondary importance when designing observational studies.

Instruments for multi-object observations 
($\sim$~20--100~stars per GC) dictate that
a choice of wavelength range must be made.
Wavelength ranges appropriate
for the O--Al absorption lines
have allowed simultaneous study of heavier elements only
when their absorption lines fortuitously fall in the same 
wavelength range.
Other studies have examined many elements per star
by obtaining complete wavelength coverage and 
high spectral resolution at the cost
of studying a limited number of stars ($\sim$~5--20~stars per GC).
Together, these approaches have allowed observers to
identify infrequent but genuine dispersion among the heavy elements.

For example, several massive GCs exhibit significant sub-populations
of stars whose Ca or Fe-group abundances are different from one another 
(M22, e.g., \citealt{marino11} and references therein;
M54, e.g., \citealt{carretta10a} and references therein;
\mbox{NGC~1851}, e.g., \citealt{carretta10b} and references therein;
\mbox{NGC~2419}, \citealt{cohen10};
$\omega$~Centauri, e.g., \citealt{johnson10} and references therein).
Two studies have reported individual stars in M92 whose
Fe-group abundances are higher by 0.15--0.20~dex than other 
members \citep{king98,langer98}.
While the heaviest elements in most metal-poor GCs have 
been produced primarily by rapid ($r$) \ncap\ nucleosynthesis, 
some GCs have been enriched by a significant amount of 
material produced in the slow ($s$) \ncap\ process
(M4, \citealt{ivans99}, \citealt{yong08a,yong08b};
M22, \citealt{marino09,marino11};
\mbox{NGC~1851}, \citealt{yong08c}, \citealt{yong09},
\citealt{carretta10b}, \citealt{villanova10}; 
$\omega$~Centauri, e.g., \citealt{smith00}, \citealt{johnson10}).
Finally, the \ncap\ elements in M15, produced by \rpro\ nucleosynthesis,
exhibit significant 
star-to-star dispersion (nearly $\sim$~1~dex;
\citealt{sneden97,sneden00,otsuki06,sobeck11}).

How anomalous is M15?
In this paper we revisit the heavy \ncap\ element
abundances in M92, a metal-poor GC similar 
(metallicity, age, luminosity, orbital kinematics) to M15.
Table~\ref{basicdata} summarizes the basic properties of M92.
The \ncap\ elements in M92 are relatively understudied considering
that it is one of the brightest and most metal-poor GCs
accessible to northern hemisphere telescopes.

\begin{deluxetable}{lcc}
\tablecaption{M92 Basic Parameters
\label{basicdata}}
\tablewidth{0pt}
%\scriptsize
\tablehead{
\colhead{Quantity} &
\colhead{Value} &
\colhead{References} }
\startdata
R.A.\ (J2000)   & 17:17:07            & 1 \\
Dec.\ (J2000)   & $+$43:08:11         & 1 \\
$\ell$          & 68.3\deg            & 1 \\
\textit{b}      & 34.9\deg            & 1 \\
$M_{V}$         & $-$8.20             & 1, 2, 3, 4 \\
($m-M$)$_{V}$   & 14.67~$\pm$~0.08    & 5 \\
E($B-V$)        & 0.02                & 1, 6, 7, 8, 9 \\
R$_{\odot}$     & 8.2~kpc             & 1 \\
R$_{\rm G.C.}$  & 9.6~kpc             & 1 \\
R$_{\rm peri}$  & 1.4~$\pm$~0.2~kpc   & 10 \\
R$_{\rm apo}$   & 9.9~$\pm$~0.4~kpc   & 10 \\
Z$_{\rm max}$   & 3.8~$\pm$~0.5~kpc   & 10 \\
$T_{\rm orbit}$ & 0.20~$\pm$~0.01~Gyr & 10 \\
\enddata
\tablerefs{
(1)~\citealt{harris96};
(2)~\citealt{webbink85};
(3)~\citealt{peterson87};
(4)~\citealt{vandenbergh91};
(5)~\citealt{pont98};
(6)~\citealt{sandage69};
(7)~\citealt{zinn80};
(8)~\citealt{reed88};
(9)~\citealt{schlegel98};
(10)~\citealt{dinescu99}
}
\end{deluxetable}

\citet{cohen79} performed the first study of \ncap\ elements in M92,
deriving abundances of Y~\textsc{ii}, Zr~\textsc{ii}, Ba~\textsc{ii}, 
La~\textsc{ii}, and Nd~\textsc{ii} in 4 red giant branch (RGB) stars.
She found a general decrease in these abundances 
relative to the more metal-rich GC M13, 
but the overall pattern was unchanged.
\citet{peterson90} derived abundances of Y~\textsc{ii} and Ba~\textsc{ii} 
in 2 M92 RGB stars. 
\citet{armosky94} examined Ba~\textsc{ii} and Nd~\textsc{ii} 
in 9 and 4 RGB stars, respectively.
That study found no dispersion in either element and, accordingly,
no correlation with the light element dispersion in M92.
At this point, the observations
were still inadequate to discern the nucleosynthetic 
origin of the heavy elements in M92.

\citet{shetrone96} and \citet{shetrone98} derived abundances of Eu~\textsc{ii}
in 3 RGB stars and Ba~\textsc{ii} in 5 RGB stars, respectively.
\citet{sneden97} used the Ba from \citet{armosky94} and the
Eu from \citet{shetrone96} to infer that 
\rpro\ nucleosynthesis dominated the production of
the heavy elements in M92 (and M13)
more than in the solar system (S.S.).
\citet{sneden00} derived Ba~\textsc{ii} abundances for 32 stars in M92;
the dispersion in [Ba/Fe],\footnote{
For elements X and Y, 
[X/Y]~$\equiv \log_{10} (N_{\rm X}/N_{\rm Y})_{\star} -
\log_{10} (N_{\rm X}/N_{\rm Y})_{\odot}$ and
log~$\epsilon$(X)~$\equiv$ 
log$_{10}$(N$_{\rm X}$/N$_{\rm H}$)~$+$~12.0.}
0.16~dex, was only slightly less than that of 31~stars in M15, 0.21~dex.

Over the last 10~years, a few other investigators 
\citep{shetrone01,johnson02,sadakane04}
have made detailed abundance analyses of small numbers of
M92 giants,
but no study has examined enough stars
to show conclusively whether a dispersion exists among the 
heaviest elements.
This is our motivation for the present study.
Sections~\ref{observations} and \ref{analysis} describe the characteristics
of the new M92 spectra obtained for this study
and the details of our abundance analysis.
Section~\ref{results} presents evidence that genuine dispersion
exists among the \ncap\ elements.
Section~\ref{discussion} demonstrates that the heavy elements in M92 were
produced by \rpro\ nucleosynthesis and compares the
M92 dispersion with that in M15.
We present our conclusions in Section~\ref{conclusions}.

\section{Observations}
\label{observations}

\begin{figure}
\includegraphics[angle=0,width=3.4in]{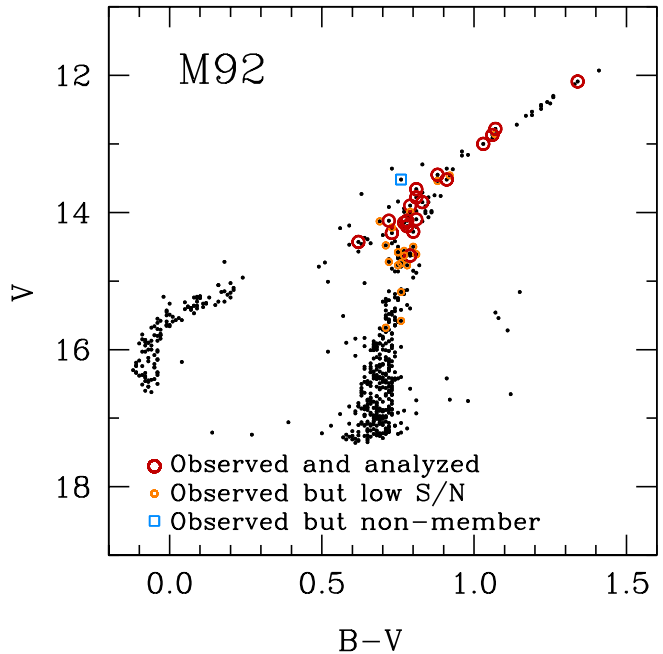}
\caption{
\label{cmdplot}
CMD of the RGB, HB, and AGB in M92 with
photometry from \citet{buonanno83}.
Stars with large red circles indicate stars observed by us
with sufficient S/N to perform an abundance analysis,
stars with small orange circles indicate stars observed by us
that lack the S/N necessary to perform an abundance analysis, and
the blue square indicates the non-member star \mbox{VI-7}.
}
\end{figure}

Previous studies of GCs have generally 
focused on red regions of the spectrum 
(5600--7000\AA)
that are nearer to the peak of the stellar flux distribution
and allow the study of the light element dispersion.
Since many useful transitions of \ncap\ elements are
in the blue around 4000\AA, 
we study this spectral region instead.
%we selected a telescope and instrument for this study that 
%provide multi-object capability with good throughput at these wavelengths.
%
All observations were taken 28--31~May 2010 
using the Hydra multi-object fiber positioner
and bench spectrograph \citep{barden95,bershady08}
on the WIYN 3.5~m Telescope at Kitt Peak National Observatory.
We used the blue fiber cable (3.1'' fibers on sky) and
316$@$63.4 echelle grating to obtain
a resolution of R~$\equiv \lambda/\Delta\lambda \sim$~14,000
as measured from isolated ThAr lines.
The X7.5 filter provides wavelength coverage from 
3850~$< \lambda <$~4050\AA\ with steep drop-off beyond these limits.

\citet{rees92} measured proper motions for 365 stars with 
$V \lesssim$~16 in the M92 field, and this is our primary source
for assessing GC membership probability.
We used two fiber position settings to observe a total of
39 stars classified as proper motion members ($P >$~90\%).
These stars are marked on the color-magnitude diagram (CMD)
shown in Figure~\ref{cmdplot}.
Remaining unused fibers were placed on blank sky to 
assist in sky subtraction.
Most stars (28) were included in both fiber settings, 
but to increase the total number of stars observed
a few additional stars were substituted for the brightest ones
after they had achieved adequate signal-to-noise (S/N).
Exposure times are listed in Table~\ref{phottab}.
S/N per pixel in the continuum near 4000\AA, 
derived assuming Poisson statistics,
are also listed in Table~\ref{phottab}.  
The S/N ranges from 110/1 for the brightest star ($V =$~12.1)
to 30/1 for the faintest star ($V =$~15.7).
A few stars, such as \mbox{V-45}, have lower S/N than would be
expected based on their magnitudes;
our fiber position astrometry may have been slightly in
error for these stars.

\begin{center}

\begin{deluxetable*}{llccccccccc}
\tablecaption{Photometry, Exposure Times, and S/N Estimates
\label{phottab}}
\tablewidth{0pt}
\tabletypesize{\scriptsize}
%\rotate
\tablehead{
\colhead{Star Name} &
\colhead{Alt.\ Name} &
\colhead{$V$} &
\colhead{$B-V$} &
\colhead{$V-J$} &
\colhead{$V-H$} &
\colhead{$V-K$} &
\colhead{Refs.} &
\colhead{No.} &
\colhead{Exp.\ Time} &
\colhead{S/N} \\
%\colhead{$V_{\rm rad}$} &
%\colhead{$\sigma$} &
%\colhead{T$_{\rm eff}$} &
%\colhead{log $g$} &
%\colhead{$v_{t}$} &
%\colhead{[M/H]} &
%\colhead{Notes} \\
\colhead{(Ref.\ 1)} &
\colhead{(Ref.\ 2)} &
\colhead{} &
\colhead{} &
\colhead{} &
\colhead{} &
\colhead{} &
\colhead{} &
\colhead{Exp.} &
\colhead{(sec)} &
\colhead{(4000\AA)} }
%\colhead{(\kmsec)} &
%\colhead{(\kmsec)} &
%\colhead{(K)} &
%\colhead{} &
%\colhead{(\kmsec)} &
%\colhead{} &
%\colhead{} }
\startdata
VII-18   & Bu488 & 12.09 & 1.34 & 2.20  & 2.86  & 2.97  & 2, 4 & 4 & 52700  & 110/1 \\
XII-8    & Bu510 & 12.78 & 1.07 & 2.05  & 2.65  & 2.74  & 2, 4 & 4 & 52700  & 100/1 \\
V-45     & Bu92  & 12.86 & 1.07 & 2.08  & 2.64  & 2.76  & 2, 4 & 4 & 52700  & 35/1  \\
XI-19    & Bu550 & 12.87 & 1.06 & 2.01  & 2.56  & 2.69  & 2, 4 & 4 & 52700  & 95/1  \\
XI-80    & Bu454 & 13.00 & 1.03 & 2.03  & 2.62  & 2.71  & 2, 4 & 4 & 52700  & 90/1  \\
XII-34   & Bu330 & 13.45 & 0.88 & 1.85  & 2.38  & 2.46  & 2, 4 & 4 & 52700  & 80/1  \\
IV-10    & Bu21  & 13.46 & 0.92 & 1.93  & 2.49  & 2.60  & 2, 4 & 4 & 52700  & 20/1  \\
IV-79    & Bu109 & 13.52 & 0.91 & 1.95  & 2.51  & 2.60  & 2, 4 & 4 & 52700  & 50/1  \\
VI-7     &\nodata& 13.52 & 0.76 & 1.40  & 1.85  & 1.94  & 3, 4 & 8 & 106300 & 105/1 \\
IV-2     & Bu12  & 13.54 & 0.88 & 1.85  & 2.43  & 2.52  & 2, 4 & 8 & 106300 & 55/1  \\
VII-10   &\nodata& 13.66 & 0.81 &\nodata&\nodata&\nodata& 3    & 8 & 106300 & 80/1  \\
VI-18    & Bu271 & 13.78 & 0.81 & 1.80  & 2.34  & 2.38  & 2, 4 & 8 & 106300 & 85/1  \\
IX-49    & Bu539 & 13.85 & 0.83 & 1.80  & 2.30  & 2.43  & 2, 4 & 8 & 106300 & 85/1  \\
IV-40    & Bu43  & 13.90 & 0.79 & 1.78  & 2.29  & 2.34  & 2, 4 & 8 & 106300 & 70/1  \\
XII-31   & Bu322 & 13.99 & 0.79 & 1.81  & 2.31  & 2.43  & 2, 4 & 4 & 53600  & 45/1  \\
VIII-44  & Bu545 & 14.10 & 0.81 & 1.76  & 2.30  & 2.41  & 2, 4 & 8 & 106300 & 80/1  \\
XI-10    & Bu395 & 14.12 & 0.78 & 1.74  & 2.29  & 2.35  & 2, 4 & 8 & 106300 & 75/1  \\
VIII-24  & Bu507 & 14.12 & 0.72 & 1.67  & 2.15  & 2.21  & 2, 4 & 8 & 106300 & 70/1  \\
III-4    & Bu14  & 14.13 & 0.69 & 1.59  & 2.06  & 2.14  & 2, 4 & 8 & 106300 & 50/1  \\
\nodata  & Bu166 & 14.15 & 0.77 & 1.80  & 2.32  & 2.39  & 2, 4 & 8 & 106300 & 65/1  \\
IX-89    & Bu497 & 14.20 & 0.78 & 1.74  & 2.27  & 2.37  & 2, 4 & 8 & 106300 & 65/1  \\
VII-79   & Bu429 & 14.22 & 0.73 & 1.78  & 2.27  & 2.38  & 2, 4 & 8 & 106300 & 65/1  \\
II-39    & Bu71  & 14.28 & 0.80 & 1.66  & 2.16  & 2.28  & 2, 4 & 8 & 106300 & 60/1  \\
VII-68   & Bu305 & 14.30 & 0.73 & 1.79  & 2.37  & 2.38  & 2, 4 & 8 & 106300 & 70/1  \\
XII-18   & Bu386 & 14.43 & 0.62 & 1.49  & 1.94  & 2.04  & 2, 4 & 8 & 106300 & 70/1  \\
V-78     & Bu104 & 14.48 & 0.71 & 1.54  & 1.98  & 2.07  & 2, 4 & 8 & 106300 & 55/1  \\
II-24    & Bu37  & 14.50 & 0.80 & 1.63  & 2.16  & 2.23  & 2, 4 & 8 & 106300 & 30/1  \\
X-28     & Bu544 & 14.56 & 0.77 & 1.58  & 2.08  & 2.20  & 2, 4 & 4 & 53600  & 35/1  \\
II-12    & Bu31  & 14.58 & 0.75 & 1.62  & 2.09  & 2.18  & 2, 4 & 8 & 106300 & 50/1  \\
XI-38    & Bu512 & 14.58 & 0.75 & 1.60  & 2.07  & 2.16  & 2, 4 & 4 & 53600  & 35/1  \\
IX-6     &\nodata& 14.61 & 0.81 &\nodata&\nodata&\nodata& 3    & 8 & 106300 & 60/1  \\
X-3      &\nodata& 14.63 & 0.79 &\nodata&\nodata&\nodata& 3    & 8 & 106300 & 65/1  \\
IX-10    &\nodata& 14.63 & 0.77 &\nodata&\nodata&\nodata& 3    & 8 & 106300 & 60/1  \\
IX-2     &\nodata& 14.72 & 0.72 &\nodata&\nodata&\nodata& 3    & 8 & 106300 & 60/1  \\
\nodata  & Bu486 & 14.74 & 0.76 & 1.65  & 2.17  & 2.24  & 2, 4 & 8 & 106300 & 60/1  \\
I-40     & Bu254 & 14.77 & 0.78 & 1.51  & 1.99  & 2.13  & 2, 4 & 8 & 106300 & 35/1  \\
XII-5    & Bu396 & 14.77 & 0.75 & 1.54  & 2.09  & 2.11  & 2, 4 & 4 & 53600  & 30/1  \\
III-11   & Bu27  & 15.16 & 0.76 & 1.52  & 1.97  & 2.02  & 2, 4 & 8 & 106300 & 35/1  \\
%III-49   & Bu42  & 15.58 & 0.76 & 1.47  & 2.00  & 1.99  & 2, 4 & 8 & 106300 & 20/1  \\
V-84     & Bu183 & 15.68 & 0.71 & 1.49  & 1.94  & 2.01  & 2, 4 & 8 & 106300 & 30/1  \\
\enddata
\tablerefs{
(1)~\citealt{sandage66};
(2)~\citealt{buonanno83};
(3)~\citealt{rees92};
(4)~2MASS
}
\end{deluxetable*}

\end{center}

\citet{buonanno83} is our primary source for broadband $BV$ photometry.
They provide a nearly-complete census of stars along the
RGB, horizontal branch (HB), and 
asymptotic giant branch (AGB) in M92 in a 14~$\times$~14 arcmin
field centered on M92 (excluding the crowded central region).
A few stars not covered by Buonanno et al.\ were also observed.
\citet{rees92} provided $BV$ photometry for these six stars, and the
conversion from the Rees to Buonanno et al.\ scale
is $\Delta\,V \approx$~0.00~$\pm$~0.05 for the stars in this 
magnitude range (see Figure~1 of Rees),
so we apply no correction.
In Section~\ref{atmosphere} we assess the impact of mixing
photometric scales on the derived model atmosphere parameters.
Finally, $JHK$ photometry from the 
Two-Micron All Sky Survey (2MASS; \citealt{skrutskie06}) 
is available for nearly all (34) stars in our study.
Photometry for our targets is listed 
in Table~\ref{phottab}, which also gives the
cross-identification between \citet{sandage66} and \citet{buonanno83}.

We use the IRAF environment\footnote{
IRAF is distributed by the National Optical Astronomy Observatories,
which are operated by the Association of Universities for Research
in Astronomy, Inc., under cooperative agreement with the National
Science Foundation.}
to perform standard data reduction, including bias subtraction,
flat fielding, image coaddition, 
order extraction, wavelength calibration, sky subtraction,
radial velocity (RV) cross-correlation,
and continuum normalization of the spectra.
Scattered light was not removed because we found that
this produced negative counts in the cores of the Ca~\textsc{ii} H and K lines.
Each exposure consists of 3--4 sub-exposures (of length
3600--3900~s) coadded to facilitate removal of cosmic rays.

\section{Analysis}
\label{analysis}

\subsection{Radial Velocities}
\label{rv}

We measure the RV of each exposure of each
star by cross-correlating against a template.
We produce this template by measuring wavelengths of individual spectral
lines in the best exposure of the brightest star (\mbox{VII-18})
and shifting this spectrum to zero RV.
The template zeropoint has a precision of about 0.6~\kmsec.
The mean RV and standard deviation for each star are listed in 
Table~\ref{atmtab}.
Heliocentric corrections are computed with the IRAF task 
\textit{rvcorrect}.
The mean (heliocentric) RV is $-$119.7~$\pm$~0.8
($\sigma =$~4.8) \kmsec.

\begin{center}

\begin{deluxetable*}{lccccccc}
\tablecaption{Radial Velocities and Stellar Parameters
\label{atmtab}}
\tablewidth{0pt}
\tabletypesize{\scriptsize}
%\rotate
\tablehead{
\colhead{Star Name} &
%\colhead{Alt.\ Name} &
%\colhead{$V$} &
%\colhead{$B-V$} &
%\colhead{$V-J$} &
%\colhead{$V-H$} &
%\colhead{$V-K$} &
%\colhead{Refs.} &
%\colhead{No.} &
%\colhead{Exp.\ Time} &
%\colhead{S/N} &
\colhead{$V_{\rm rad}$} &
\colhead{$\sigma$} &
\colhead{T$_{\rm eff}$} &
\colhead{log $g$} &
\colhead{$v_{t}$} &
\colhead{[M/H]} &
\colhead{Notes} \\
%\colhead{(Ref.\ 1)} &
%\colhead{(Ref.\ 2)} &
\colhead{} &
%\colhead{} &
%\colhead{} &
%\colhead{} &
%\colhead{} &
%\colhead{} &
%\colhead{Exp.} &
%\colhead{(sec)} &
%\colhead{(4000\AA)} &
\colhead{(\kmsec)} &
\colhead{(\kmsec)} &
\colhead{(K)} &
\colhead{} &
\colhead{(\kmsec)} &
\colhead{} &
\colhead{} }
\startdata
VII-18   & $-$116.6 & 0.3  & 4300 & 0.60 & 2.60 & $-$2.4 &            \\
XII-8    & $-$117.0 & 0.8  & 4450 & 1.00 & 2.40 & $-$2.4 &            \\
V-45     & $-$118.2 & 0.5  & 4440 & 1.00 & 2.35 & $-$2.4 &            \\
XI-19    & $-$115.1 & 0.6  & 4500 & 1.05 & 2.40 & $-$2.4 &            \\
XI-80    & $-$122.7 & 0.5  & 4470 & 1.10 & 2.35 & $-$2.4 &            \\
XII-34   & $-$113.6 & 0.5  & 4670 & 1.40 & 2.30 & $-$2.4 &            \\
IV-10    & $-$116.4 & 1.3  & 4570 & 1.35 & 2.25 & $-$2.4 &            \\
IV-79    & $-$118.3 & 1.0  & 4560 & 1.35 & 2.20 & $-$2.4 &            \\
VI-7     & $-$137.8 & 0.4  &\nodata&\nodata&\nodata&\nodata& Non-member \\
IV-2     & $-$122.9 & 0.3  & 4640 & 1.40 & 2.25 & $-$2.4 &            \\
VII-10   & $-$118.1 & 0.9  & 4680 & 1.45 & 2.25 & $-$2.4 & RHB/AGB    \\
VI-18    & $-$123.3 & 0.8  & 4730 & 1.55 & 2.25 & $-$2.4 & RHB/AGB    \\
IX-49    & $-$126.7 & 0.9  & 4720 & 1.60 & 2.20 & $-$2.4 &            \\
IV-40    & $-$117.3 & 1.0  & 4760 & 1.60 & 2.20 & $-$2.4 & RHB/AGB    \\
XII-31   & $-$113.6 & 1.3  & 4720 & 1.65 & 2.15 & $-$2.4 & RHB/AGB    \\
VIII-44  & $-$117.1 & 0.6  & 4750 & 1.70 & 2.15 & $-$2.4 &            \\
XI-10    & $-$127.0 & 0.6  & 4780 & 1.70 & 2.15 & $-$2.4 &            \\
VIII-24  & $-$116.2 & 0.4  & 4900 & 1.80 & 2.25 & $-$2.4 & RHB/AGB    \\
III-4    & $-$118.4 & 0.9  & 5000 & 1.85 & 2.30 & $-$2.4 & RHB/AGB    \\
Bu166    & $-$124.2 & 0.9  & 4730 & 1.70 & 2.10 & $-$2.4 &            \\
IX-89    & $-$115.0 & 0.5  & 4780 & 1.75 & 2.10 & $-$2.4 &            \\
VII-79   & $-$114.1 & 0.5  & 4760 & 1.75 & 2.10 & $-$2.4 & RHB/AGB    \\
II-39    & $-$118.8 & 0.7  & 4880 & 1.85 & 2.15 & $-$2.4 &            \\
VII-68   & $-$115.9 & 1.3  & 4740 & 1.75 & 2.05 & $-$2.4 & RHB/AGB    \\
XII-18   & $-$123.0 & 0.8  & 5140 & 2.00 & 2.30 & $-$2.4 & RHB/AGB    \\
V-78     & $-$123.3 & 0.9  & 5080 & 2.00 & 2.25 & $-$2.4 & RHB/AGB    \\
II-24    & $-$118.3 & 1.3  & 4910 & 1.95 & 2.10 & $-$2.4 &            \\
X-28     & $-$116.3 & 0.7  & 4970 & 2.00 & 2.15 & $-$2.4 &            \\
II-12    & $-$117.6 & 1.1  & 4960 & 1.95 & 2.10 & $-$2.4 &            \\
XI-38    & $-$112.8 & 1.5  & 4980 & 2.00 & 2.15 & $-$2.4 &            \\
IX-6     & $-$123.3 & 1.4  & 4930 & 2.00 & 2.05 & $-$2.4 &            \\
X-3      & $-$123.1 & 0.5  & 4940 & 2.00 & 2.10 & $-$2.4 &            \\
IX-10    & $-$121.6 & 1.2  & 4940 & 2.00 & 2.10 & $-$2.4 &            \\
IX-2     & $-$123.3 & 0.9  & 4960 & 2.05 & 2.05 & $-$2.4 &            \\
Bu486    & $-$121.4 & 0.9  & 4890 & 2.00 & 2.00 & $-$2.4 &            \\
I-40     & $-$122.0 & 1.1  & 5070 & 2.10 & 2.15 & $-$2.4 &            \\
XII-5    & $-$116.9 & 0.5  & 5030 & 2.10 & 2.10 & $-$2.4 &            \\
III-11   & $-$117.6 & 1.5  & 5120 & 2.30 & 2.05 & $-$2.4 &            \\
%III-49   &  \nodata &\nodata& 5150 & 2.50 & 1.90 & $-$2.4 &            \\
V-84     & $-$122.1 & 1.5  & 5150 & 2.50 & 1.90 & $-$2.4 &            \\
\enddata
\end{deluxetable*}

\end{center}

No telluric lines are covered in any of these spectra, so we
cannot assess the absolute zeropoint of the RV measurements, 
but the mean RV is in good agreement with previous studies.
\citet{drukier07} observed all of the M92 stars in our sample,
and we find a mean offset of 2.1~$\pm$~0.3~\kmsec\ (our study minus theirs).
\citet{meszaros09} observed 30 M92 stars in common with us,
and we find a mean offset of 2.1~$\pm$~0.4~\kmsec.
\citet{soderberg99} and \citet{pilachowski00} 
measured the RV of 19 and 23 stars in common with 
our sample, and we find an offset of 1.4~$\pm$~0.3~\kmsec\ 
(note that Pilachowski et al.\
normalized their RV measurements to Soderberg et al.).
\citet{shetrone01} report the RV of one star in common,
\mbox{VII-18}, which is different from our measurement by 1~\kmsec.
The observed velocity dispersion that we derive
from 39 stars in M92, $\sigma =$~4.8~$\pm$~0.8, also compares
well with previous estimates
(5.0~$\pm$~0.5~\kmsec, 49~stars, \citealt{rees92} and \citealt{pryor93}
reanalyzing the unpublished data of \citealt{lupton85};
3.3~$\pm$~0.5~\kmsec, 35~stars, \citealt{soderberg99};
4.4~$\pm$~0.6~\kmsec, 61~stars, \citealt{pilachowski00};
4.8~$\pm$~0.4~\kmsec, 64~stars, \citealt{meszaros09};
5.1~$\pm$~2.4~\kmsec, 5~stars, \citealt{cohen97}).

\citet{meszaros09} reported one RV variable star among the 
stars in our sample, \mbox{XI-38}, which is confirmed by our
measurements.
No other stars exhibit any significant ($\gtrsim$~3$\sigma$)
RV drift over the 14~years that span the 4~sets of observations.\footnote{
There has been considerable uncertainty surrounding the membership
of star \mbox{VI-7} ($=$~ZDA1 and ZNG4).
This star appears to be RV variable, with velocities ranging
from $-$90 to $-$158~\kmsec\ 
(\citealt{strom71,zinn73,norris77,pilachowski00}; this study).
Proper motion studies have assigned various probabilities to its
membership: 15\% \citep{cudworth76}, 99\% \citep{rees92}, and 
68\% \citep{tucholke96}.
\citet{carbon82} also considered \mbox{VI-7} a non-member based 
on the available RV and proper motion data and
on account of its stronger Ca~\textsc{ii} H and K lines.
Our abundance analysis finds [Fe/H]~$= -$2.0, 
thus quantifying the Carbon et al.\ assertion and further strengthening
this conclusion.
}

In summary, our mean RV, individual stellar RVs, and 
observed velocity dispersion
are all in reasonably good agreement with previous measurements.

\subsection{Model Atmosphere Parameters}
\label{atmosphere}

Our spectra cover a very narrow wavelength range, and this
naturally restricts the number of methods available
to determine model atmosphere parameters.
Effective temperatures (\teff)
calculated from broadband color-\teff\ relations
provide a satisfactory option.
The sensitivity of the $B$ band to individual stellar
CN and CH band strengths makes the $B-V$ color-\teff\
relation an undesirable option if alternatives exist.
We use $JHK$ broadband photometry from 2MASS, 
available for most of our sample,
to calculate temperatures from the 
$V-J$, $V-H$, and $V-K$ color-\teff\ relations.
We average the temperatures predicted from these three colors
as given by the metallicity-calibrated relations for giants
presented in \citet{ramirez05b}.
These temperatures are listed in Table~\ref{atmtab}.
An uncertainty of $\Delta V =$~0.05~mag translates to 
changes in the \teff\ predicted from $V-K$ of 50~K.
Standard deviations of the residuals after applying a linear
least-squares fit to each relation are each 50--60~K.
We then interpolate temperatures for stars lacking 2MASS photometry 
from these relations.

Since the distance to M92 is well known, we calculate
physical surface gravities,
%according to the relation
%\begin{equation}
%\log g = 0.4(M_{K,\star} + BC_{K} - M_{\rm bol,\odot}) 
%+ \log g_{\odot} + 4\log(T_{\rm eff,\star}/T_{\rm eff,\odot})
%+ \log(m_{\star}/m_{\odot}),
%\end{equation}
where a star's apparent magnitude is related to its absolute
magnitude through the distance modulus and bolometric correction (BC).
We transform $K$~magnitudes
from the 2MASS system to the TCS system
according to Equation~5c of \citet{ramirez05a}
and interpolate (in $V-K$ and [Fe/H]) 
the grid of BCs presented by \citet{alonso99}.
We adopt the M92 distance modulus and reddening listed in 
Table~\ref{basicdata}, 
extinction coefficients given by \citet{mccall04}, 
0.8~\msun\ as the mass of stars on the RGB,
and the solar values 
$M_{\rm bol,\odot} =$~4.74,
$\log g_{\odot} =$~4.44, and
$T_{\rm eff,\odot} =$~5780~K.
Again, since not all stars have 2MASS photometry, we 
interpolate surface gravity from the relationship
between $V$ and $\log g$ established by those stars that do.
The scatter in this relationship 
is only 0.04~dex.
Final $\log g$ values are listed in Table~\ref{atmtab}.
They are relatively insensitive to 
uncertainties in the input parameters.
For $\Delta m_{K}$, $\Delta m-K$, or $\Delta BC = \pm$~0.1, 
          $\Delta \log g = \mp$~0.04;
for $\Delta T_{\rm eff} = \pm$~60~K, 
          $\Delta \log g = \mp$~0.02; and 
for $\Delta m_{\star} = \pm$~0.15~$M_{\odot}$, 
          $\Delta \log g = ^{+0.09}_{-0.08}$.
Several tenths of a solar mass may be lost between the time a star
leaves the main sequence and arrives on the red horizontal branch
(RHB; e.g., \citealt{preston06}),
but this effect
has little impact on the relative surface gravities calculated here.

We calculate microturbulent velocities ($v_{t}$) 
from the empirical relationship 
between \teff, $\log g$, and $v_{t}$ for metal-poor 
field giants derived by \citet{gratton96}.
These results are in good agreement with previous spectroscopically-derived
estimates of $v_{t}$ for red giants 
in M92 \citep{shetrone96,shetrone01,johnson02},
including earlier studies that found 2~\kmsec\ was
an adequate estimate for all stars
\citep{sneden91,sneden00}. 
The mean difference between the empirical relationship and
the latter approach is only
$\Delta v_{t} = $0.18~$\pm$~0.02 ($\sigma =$~0.15)~\kmsec\
(in the sense of Gratton et al.\ relation$-$2.0).
These uncertainties are well within the
precision regularly achieved for analyses of metal-poor giants.

\citet{sneden00} obtained spectra covering 250\AA\ near 5900\AA\ for
34 stars in M92 using Hydra.
That study computed \teff\ and $\log g$ by comparing 
dereddened $B-V$ and $M_{V}$ with the predicted colors
and magnitudes derived from model atmospheres \citep{carbon82}.
For the 18~stars in common with Sneden et al., we find mean differences of 
$\Delta T_{\rm eff} = -$48~$\pm$~16 ($\sigma =$~67) K and
$\Delta$log~$g = -$0.16~$\pm$~0.03 ($\sigma =$~0.15).
There are no significant trends with either \teff\ or log~$g$.

Finally, we uniformly adopt a metallicity of [M/H]~$= -$2.4
for all model atmospheres.
We generate model atmospheres from the MARCS grid of 1-dimensional,
spherical, standard composition (i.e., $\alpha$-enhanced at this
metallicity) models computed assuming 
local thermodynamic equilibrium (LTE) \citep{gustafsson08}
using interpolation software kindly provided by A.\ McWilliam
(2009, private communication).  
We emphasize that our primary goal is to
examine the 
dispersion in the abundance ratios, so the absolute 
temperature and metallicity scales are only of secondary importance.

\subsection{Derivation of Abundances}
\label{abund}

The resolution of our spectra is considerably lower than
that commonly used for detailed abundance analyses,
and the S/N is generally a decreasing function
of luminosity along the RGB.
As such, we perform a differential abundance analysis
to search for star-to-star chemical dispersion among 
M92 red giants.
Only 19~stars in our sample have S/N sufficient to derive
reliable abundances.
The differential abundances are then placed on an absolute scale
by computing abundances in one star by the usual techniques.
Here we describe these methods in more detail.

\begin{deluxetable*}{cccccc}
\tablecaption{Atomic Data
\label{atomictab}}
\tablewidth{0pt}
\tabletypesize{\scriptsize}
\tablehead{
\colhead{Species} &
\colhead{$Z$} &
\colhead{Wavelength (\AA)} &
\colhead{E.P.\ (eV)} &
\colhead{log($gf$)} &
\colhead{Reference} }
\startdata
C ($^{12}$CH)  &  6 & 4000.98$+$4001.07 &  0.64   & $-$1.12, $-$1.10 &  1 \\
C ($^{12}$CH)  &  6 & 4020.02$+$4020.18 &  0.46   & $-$1.38, $-$1.35 &  1 \\
N (CN)         &  7 & 3879.0--3883.5    & \nodata & \nodata          &  2 \\
Si~\textsc{i}  & 14 &    3905.52        &  1.91   & $-$1.04          &  3 \\
Sc~\textsc{i}  & 21 &    3911.82        &  0.02   & $+$0.40          &  2 \\
Sc~\textsc{ii} & 21 &    3989.13\tablenotemark{a} & 0.32 & $-$2.72   &  2, 4 \\
Ti~\textsc{i}  & 22 &    3904.78        &  0.90   & $+$0.03          &  2 \\
Ti~\textsc{i}  & 22 &    3989.76        &  0.02   & $-$0.13          &  5, 6 \\
Ti~\textsc{i}  & 22 &    3998.64        &  0.05   & $+$0.01          &  5, 6 \\
Ti~\textsc{i}  & 22 &    4008.93        &  0.02   & $-$1.02          &  5, 6 \\
Ti~\textsc{ii} & 22 &    3987.61        &  0.61   & $-$2.93          &  7 \\
Ti~\textsc{ii} & 22 &    4025.13        &  0.61   & $-$2.14          &  7 \\
V~\textsc{ii}  & 23 &    3951.96        &  1.48   & $-$0.78          &  8 \\
V~\textsc{ii}  & 23 &    4002.94        &  1.43   & $-$1.45          &  8 \\
V~\textsc{ii}  & 23 &    4005.71        &  1.82   & $-$0.52          &  8 \\
V~\textsc{ii}  & 23 &    4023.38        &  1.80   & $-$0.69          &  8 \\
V~\textsc{ii}  & 23 &    4036.78        &  1.48   & $-$1.59          &  8 \\
Cr~\textsc{i}  & 24 &    3908.76        &  1.00   & $-$1.05          &  9 \\
Mn~\textsc{i}  & 25 &    4018.10        &  2.11   & $-$0.19          & 10 \\
Mn~\textsc{i}  & 25 &    4030.75        &  0.00   & $-$0.47          & 11 \\
Mn~\textsc{i}  & 25 &    4033.06        &  0.00   & $-$0.62          & 11 \\
Mn~\textsc{i}  & 25 &    4034.48        &  0.00   & $-$0.81          & 11 \\
Mn~\textsc{i}  & 25 &    4041.36        &  2.11   & $+$0.28          & 11 \\
Fe~\textsc{i}  & 26 &    3891.93        &  3.41   & $-$0.73          & 12 \\
Fe~\textsc{i}  & 26 &    3899.03        &  2.45   & $-$1.81          & 12 \\
Fe~\textsc{i}  & 26 &    3985.39        &  3.30   & $-$0.99          & 12 \\
Fe~\textsc{i}  & 26 &    4001.66        &  2.17   & $-$1.90          & 12 \\
Fe~\textsc{i}  & 26 &    4007.27        &  2.76   & $-$1.28          & 12 \\
Fe~\textsc{i}  & 26 &    4013.82        &  3.02   & $-$1.70          & 2 \\
Fe~\textsc{i}  & 26 & 4017.08$+$4017.15 & 2.76, 3.05 & $-$1.99, $-$1.06 & 2, 12 \\
Fe~\textsc{i}  & 26 & 4032.45$+$4032.63 & 4.26, 1.48 & $-$0.84, $-$2.38 & 2, 12 \\
Fe~\textsc{i}  & 26 &    4044.61        & 2.83    & $-$1.22          & 12 \\
Co~\textsc{i}  & 27 &    3995.31        & 0.92    & $-$0.14          & 13 \\
Co~\textsc{i}  & 27 &    4020.90        & 0.43    & $-$2.04          & 13 \\
Co~\textsc{i}  & 27 &    4027.02        & 0.17    & $-$2.87          & 2 \\
Ni~\textsc{i}  & 28 &    3912.97        & 0.02    & $-$3.70          & 2 \\
Y~\textsc{ii}  & 39 &    3950.36        & 0.10    & $-$0.49          & 14 \\
Y~\textsc{ii}  & 39 &    3982.60        & 0.13    & $-$0.49          & 14 \\
Zr~\textsc{ii} & 40 &    3991.13        & 0.76    & $-$0.23          & 15 \\
Zr~\textsc{ii} & 40 &    4029.68        & 0.71    & $-$0.74          & 15 \\
Zr~\textsc{ii} & 40 &    4050.33        & 0.71    & $-$1.00          & 15 \\
La~\textsc{ii} & 57 &    3949.10\tablenotemark{a} & 0.40 & $+$0.49   & 16 \\
La~\textsc{ii} & 57 &    3988.51\tablenotemark{a} & 0.40 & $+$0.21   & 16 \\
La~\textsc{ii} & 57 &    3995.74\tablenotemark{a} & 0.17 & $-$0.06   & 16 \\
La~\textsc{ii} & 57 &    4031.69\tablenotemark{a} & 0.32 & $-$0.08   & 16 \\
Ce~\textsc{ii} & 58 &    4042.58        & 0.50    & $+$0.00          & 17 \\
Nd~\textsc{ii} & 60 &    4023.00        & 0.56    & $+$0.04          & 18 \\
Nd~\textsc{ii} & 60 &    4051.14        & 0.38    & $-$0.30          & 18 \\
Eu~\textsc{ii} & 63 &    3907.11\tablenotemark{a} & 0.21 & $+$0.17   & 19 \\
Ho~\textsc{ii} & 67 &    3891.00\tablenotemark{a} & 0.08 & $+$0.46   & 20 \\
Er~\textsc{ii} & 68 &    3896.23        & 0.06    & $-$0.12          & 21 \\
\enddata
\tablenotetext{a}{Includes hyperfine splitting structure}
\tablerefs{
(1)~B.\ Plez, 2007, private communication;
(2)~\citet{kurucz95};
(3)~\citet{obrian91b};
(4)~\citet{lawler89};
(5)~\citet{blackwell82};
(6)~\citet{grevesse89};
(7)~\citet{pickering02};
(8)~\citet{biemont89};
(9)~\citet{sobeck07};
(10)~\citet{blackwellwhitehead07};
(11)~\citet{booth84};
(12)~\citet{obrian91a};
(13)~\citet{nitz99};
(14)~\citet{hannaford82};
(15)~\citet{malcheva06};
(16)~\citet{lawler01a};
(17)~\citet{lawler09};
(18)~\citet{denhartog03};
(19)~\citet{lawler01b};
(20)~\citet{lawler04};
(21)~\citet{lawler08}.
}
\end{deluxetable*}

We adopt \mbox{XII-8} as our abundance reference because
its S/N is among the highest we have attained, 100/1 at 4000\AA.
Nearly all absorption lines are blended at our spectral resolution,
so we derive abundances in \mbox{XII-8} by spectrum synthesis.
We can reliably derive abundances for 21 species of 19 elements
in this star
(C, N, Si~\textsc{i}, Sc~\textsc{i}, Sc~\textsc{ii}, 
Ti~\textsc{i}, Ti~\textsc{ii}, V~\textsc{ii}, Cr~\textsc{i},
Mn~\textsc{i}, Fe~\textsc{i}, Co~\textsc{i}, Ni~\textsc{i}, 
Y~\textsc{ii}, Zr~\textsc{ii}, La~\textsc{ii}, Ce~\textsc{ii},
Nd~\textsc{ii}, Eu~\textsc{ii}, Ho~\textsc{ii}, and Er~\textsc{ii}).
Our linelist is given in Table~\ref{atomictab}.
We use the latest version (2010) of the spectrum analysis code MOOG
\citep{sneden73} to generate synthetic spectra.
Abundances derived for each line in \mbox{XII-8} 
are listed in Table~\ref{refabundtab}.

\begin{deluxetable}{cccc}
\tablecaption{Abundances in Reference Star XII-8
\label{refabundtab}}
\tablewidth{0pt}
\tabletypesize{\scriptsize}
\tablehead{
\colhead{Species} &
\colhead{$Z$} &
\colhead{Wavelength (\AA)} &
\colhead{log $\epsilon$} }
\startdata
C ($^{12}$CH)  &  6 & 4001.03 & 5.05  \\
C ($^{12}$CH)  &  6 & 4020.10 & 5.10  \\
N (CN)         &  7 & 3883.00 & 5.65  \\
Si~\textsc{i}  & 14 & 3905.52 & 5.05  \\
Sc~\textsc{i}  & 21 & 3911.82 & 0.40  \\
Sc~\textsc{ii} & 21 & 3989.13 & 0.80  \\
Ti~\textsc{i}  & 22 & 3904.78 & 2.50  \\
Ti~\textsc{i}  & 22 & 3989.76 & 1.88  \\
Ti~\textsc{i}  & 22 & 3998.64 & 2.13  \\
Ti~\textsc{i}  & 22 & 4008.93 & 2.35  \\
Ti~\textsc{ii} & 22 & 3987.61 & 2.95  \\
Ti~\textsc{ii} & 22 & 4025.13 & 2.91  \\
V~\textsc{ii}  & 23 & 3951.96 & 1.29  \\
V~\textsc{ii}  & 23 & 4002.94 & 1.64  \\
V~\textsc{ii}  & 23 & 4005.71 & 1.91  \\
V~\textsc{ii}  & 23 & 4023.38 & 1.76  \\
V~\textsc{ii}  & 23 & 4036.78 & 1.60  \\
Cr~\textsc{i}  & 24 & 3908.76 & 2.20  \\
Mn~\textsc{i}  & 25 & 4018.10 & 2.08  \\
Mn~\textsc{i}  & 25 & 4030.75 & 2.00  \\
Mn~\textsc{i}  & 25 & 4033.06 & 1.90  \\
Mn~\textsc{i}  & 25 & 4034.48 & 1.80  \\
Mn~\textsc{i}  & 25 & 4041.36 & 2.05  \\
Fe~\textsc{i}  & 26 & 3891.93 & 4.71  \\
Fe~\textsc{i}  & 26 & 3899.03 & 4.44  \\
Fe~\textsc{i}  & 26 & 3985.39 & 4.29  \\
Fe~\textsc{i}  & 26 & 4001.66 & 4.72  \\
Fe~\textsc{i}  & 26 & 4007.27 & 4.53  \\
Fe~\textsc{i}  & 26 & 4013.82 & 4.90  \\
Fe~\textsc{i}  & 26 & 4017.15 & 4.64  \\
Fe~\textsc{i}  & 26 & 4032.63 & 4.64  \\
Fe~\textsc{i}  & 26 & 4044.61 & 4.64  \\
Co~\textsc{i}  & 27 & 3995.31 & 1.77  \\
Co~\textsc{i}  & 27 & 4020.90 & 2.37  \\
Co~\textsc{i}  & 27 & 4027.02 & 2.50  \\
Ni~\textsc{i}  & 28 & 3912.97 & 3.25  \\
Y~\textsc{ii}  & 39 & 3950.36 & $-$0.60 \\
Y~\textsc{ii}  & 39 & 3982.60 & $-$1.00 \\
Zr~\textsc{ii} & 40 & 3991.13 & 0.03  \\
Zr~\textsc{ii} & 40 & 4029.68 & 0.59  \\
Zr~\textsc{ii} & 40 & 4050.33 & 0.20  \\
La~\textsc{ii} & 57 & 3949.10 & $-$1.30 \\
La~\textsc{ii} & 57 & 3988.51 & $-$1.40 \\
La~\textsc{ii} & 57 & 3995.74 & $-$1.45 \\
La~\textsc{ii} & 57 & 4031.69 & $-$1.35 \\
Ce~\textsc{ii} & 58 & 4042.58 & $-$0.85 \\
Nd~\textsc{ii} & 60 & 4023.00 & $-$0.85 \\
Nd~\textsc{ii} & 60 & 4051.14 & $-$1.05 \\
Eu~\textsc{ii} & 63 & 3907.11 & $-$1.75 \\
Ho~\textsc{ii} & 67 & 3891.00 & $-$1.85 \\
Er~\textsc{ii} & 68 & 3896.23 & $-$1.30 \\
\enddata                              
\end{deluxetable}

Next, for each line in each star we calculate differential abundances relative
to the corresponding line in \mbox{XII-8}.
The spectrum generated by MOOG for each line 
is divided by
a synthetic spectrum of that line in \mbox{XII-8}.
This quotient is compared to the quotient of the 
observed spectra of the two stars.
Uncertainties are computed according to 
$\chi^{2}$ statistics regarding the goodness of fit
between the observed and synthetic spectra ratios.

\begin{deluxetable}{ccccc}
\tablecaption{Line-by-Line Abundances
\label{deltatab}}
\tablewidth{0pt}
\tabletypesize{\scriptsize}
\tablehead{
\colhead{Species} &
\colhead{$Z$} &
\colhead{Wavelength (\AA)} &
\colhead{$\Delta$ (dex)} &
\colhead{$\sigma$ (dex)}}
\startdata
%\hline
\multicolumn{5}{c}{VII-18} \\
\hline
C ($^{12}$CH)  &  6 & 4001.03 & $-$0.18 & 0.12 \\
C ($^{12}$CH)  &  6 & 4020.10 & 0.00    & 0.12 \\
N (CN)         &  7 & 3883.00 & 1.03    & 0.10 \\
Si~\textsc{i}  & 14 & 3905.52 & 0.07    & 0.16 \\
Sc~\textsc{i}  & 21 & 3911.82 & 0.12    & 0.17 \\
Sc~\textsc{ii} & 21 & 3989.13 & 0.28    & 0.12 \\
Ti~\textsc{i}  & 22 & 3904.78 & $-$0.02 & 0.16 \\
Ti~\textsc{i}  & 22 & 3989.76 & $-$0.11 & 0.10 \\
Ti~\textsc{i}  & 22 & 3998.64 & 0.04    & 0.21 \\
Ti~\textsc{i}  & 22 & 4008.93 & $-$0.09 & 0.08 \\
Ti~\textsc{ii} & 22 & 3987.61 & $-$0.09 & 0.16 \\
Ti~\textsc{ii} & 22 & 4025.13 & $-$0.08 & 0.13 \\
V~\textsc{ii}  & 23 & 3951.96 & 0.09    & 0.14 \\
V~\textsc{ii}  & 23 & 4002.94 & 0.11    & 0.11 \\
V~\textsc{ii}  & 23 & 4005.71 & $-$0.07 & 0.11 \\
V~\textsc{ii}  & 23 & 4023.38 & $-$0.04 & 0.11 \\
V~\textsc{ii}  & 23 & 4036.78 & 0.08    & 0.16 \\
Cr~\textsc{i}  & 24 & 3908.76 & 0.41    & 0.18 \\
Mn~\textsc{i}  & 25 & 4018.10 & 0.22    & 0.12 \\
Mn~\textsc{i}  & 25 & 4030.75 & $-$0.04 & 0.17 \\
Mn~\textsc{i}  & 25 & 4033.06 & 0.07    & 0.22 \\
Mn~\textsc{i}  & 25 & 4034.48 & 0.19    & 0.10 \\
Mn~\textsc{i}  & 25 & 4041.36 & 0.39    & 0.09 \\
Fe~\textsc{i}  & 26 & 3891.93 & $-$0.02 & 0.13 \\
Fe~\textsc{i}  & 26 & 3899.03 & 0.08    & 0.11 \\
Fe~\textsc{i}  & 26 & 3985.39 & 0.21    & 0.17 \\
Fe~\textsc{i}  & 26 & 4001.66 & 0.07    & 0.09 \\
Fe~\textsc{i}  & 26 & 4007.27 & 0.03    & 0.09 \\
Fe~\textsc{i}  & 26 & 4013.82 & 0.02    & 0.08 \\
Fe~\textsc{i}  & 26 & 4017.15 & 0.10    & 0.09 \\
Fe~\textsc{i}  & 26 & 4032.63 & 0.20    & 0.09 \\
Fe~\textsc{i}  & 26 & 4044.61 & 0.17    & 0.11 \\
Co~\textsc{i}  & 27 & 3995.31 & 0.19    & 0.10 \\
Co~\textsc{i}  & 27 & 4020.90 & 0.09    & 0.14 \\
Co~\textsc{i}  & 27 & 4027.02 & 0.31    & 0.11 \\
Ni~\textsc{i}  & 28 & 3912.97 & 0.09    & 0.16 \\
Y~\textsc{ii}  & 39 & 3950.36 & 0.02    & 0.14 \\
Y~\textsc{ii}  & 39 & 3982.60 & $-$0.14 & 0.13 \\
Zr~\textsc{ii} & 40 & 3991.13 & $-$0.09 & 0.14 \\
Zr~\textsc{ii} & 40 & 4029.68 & $-$0.02 & 0.14 \\
Zr~\textsc{ii} & 40 & 4050.33 & $-$0.08 & 0.18 \\
La~\textsc{ii} & 57 & 3949.10 & $-$0.23 & 0.15 \\
La~\textsc{ii} & 57 & 3988.51 & $-$0.40 & 0.11 \\
La~\textsc{ii} & 57 & 3995.74 & 0.00    & 0.11 \\
La~\textsc{ii} & 57 & 4031.69 & $-$0.02 & 0.19 \\
Ce~\textsc{ii} & 58 & 4042.58 & 0.04    & 0.19 \\
Nd~\textsc{ii} & 60 & 4023.00 & $-$0.20 & 0.15 \\
Nd~\textsc{ii} & 60 & 4051.14 & 0.07    & 0.21 \\
Eu~\textsc{ii} & 63 & 3907.11 & $-$0.08 & 0.11 \\
Ho~\textsc{ii} & 67 & 3891.00 & $-$0.23 & 0.13 \\
Er~\textsc{ii} & 68 & 3896.23 & 0.08    & 0.16 \\
\enddata
\tablecomments{
The complete version of Table~\ref{deltatab} is 
available online.
The data for one example star are shown here.
}
\end{deluxetable}

This technique
yields a differential abundance and relative uncertainty
for each line in each star.
These values are reported in Table~\ref{deltatab}.
We find that 
S/N~$\gtrsim$~65 is generally necessary to derive reliable 
differential abundances of the \ncap\ elements, 
though the minimum S/N varies slightly.
Final abundances are computed by performing a weighted average
of the differentials
for a given species and adding this mean differential
to the mean log~$\epsilon$
value for that species derived in \mbox{XII-8}.
S.S.\ abundances used to compute the [X/Fe] ratios are 
taken from \citet{asplund09}.
Tables~\ref{abundtab1}--\ref{abundtab7} list, for each of the 19~stars, 
log~$\epsilon$ abundances (column~1),
[X/Fe] ratios (column~2), 
standard error ($\sigma/\sqrt{N}$, column~3),
standard deviation ($\sigma$, column~4), and
number of lines used ($N$, column~5).

We stress that the uncertainties (1$\sigma$) are computed with respect to the
differentials, thus 
\textit{they reflect internal uncertainties only}.
In the results and discussion that follows, we reference these 
uncertainties unless otherwise noted.
Absolute uncertainties may be computed by combining 
the standard error of the
species from the star in question with that of \mbox{XII-8}
and the uncertainty associated with the atmospheric parameters.
The magnitude of this final source of uncertainty is assessed
by rederiving the abundance ratios of several key elements
in \mbox{XII-8} after making reasonable variations 
to the model atmosphere parameters.
These values are listed in Table~\ref{deltaatmtab}.

\begin{figure*}
\epsscale{0.80}
\plotone{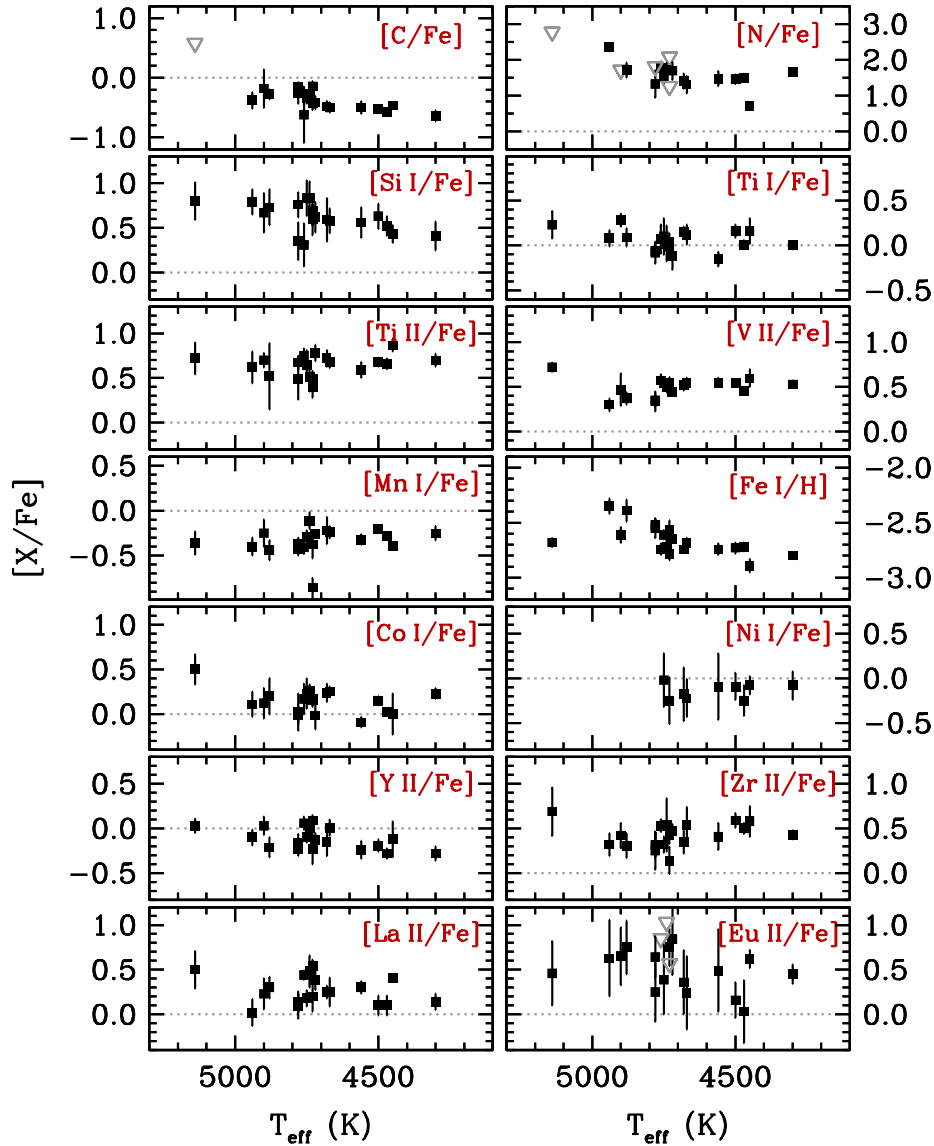}
\caption{
\label{teffabundplot}
Abundance ratios as a function of $T_{\rm eff}$.
Detections are indicated by filled squares, 
and upper limits are indicated by downward-facing open triangles.
Dotted lines represent the S.S.\ ratios.
Species with limited numbers of detections are omitted.
Only the internal (i.e., star-to-star) uncertainties are shown.
}
\end{figure*}

\begin{deluxetable*}{ccccccccccccccccccc}
\tablecaption{Stellar Abundances I
\label{abundtab1}}
\tablewidth{0pt}
%\rotate
\tabletypesize{\scriptsize}
\tablehead{
\colhead{} &
\colhead{} &
\multicolumn{5}{c}{II-39} &
\colhead{} &
\multicolumn{5}{c}{IV-40} &
\colhead{} &
\multicolumn{5}{c}{IV-79} \\
\cline{3-7} \cline{9-13} \cline{15-19}
\colhead{Species} &
\colhead{$Z$} &
\colhead{log~$\epsilon$} &
\colhead{[X/Fe]} &
\colhead{$\sigma_{\mu}$} &
\colhead{$\sigma$} &
\colhead{$N$} &
\colhead{} &
\colhead{log~$\epsilon$} &
\colhead{[X/Fe]} &
\colhead{$\sigma_{\mu}$} &
\colhead{$\sigma$} &
\colhead{$N$} &
\colhead{} &
\colhead{log~$\epsilon$} &
\colhead{[X/Fe]} &
\colhead{$\sigma_{\mu}$} &
\colhead{$\sigma$} &
\colhead{$N$} }
\startdata
C ($^{12}$CH)  &  6 & 5.76     & $-$0.28    & 0.08     & 0.12     & 2        &  & 5.07     & $-$0.62    & 0.47     & 0.47     & 1       & & 5.19     & $-$0.50    & 0.11     & 0.16     & 2       \\
N (CN)         &  7 & 7.15     & 1.71     & 0.21     & 0.21     & 1        &  & 6.85     & 1.76     & 0.20     & 0.20     & 1       & & 6.57     & 1.48     & 0.21     & 0.21     & 1       \\
Si~\textsc{i}  & 14 & 5.85     & 0.73     & 0.20     & 0.20     & 1        &  & 5.08     & 0.31     & 0.24     & 0.24     & 1       & & 5.33     & 0.56     & 0.17     & 0.17     & 1       \\
Sc~\textsc{i}  & 21 & \nodata  & \nodata  & \nodata  & \nodata  & \nodata  &  & \nodata  & \nodata  & \nodata  & \nodata  & \nodata & & \nodata  & \nodata  & \nodata  & \nodata  & \nodata \\
Sc~\textsc{ii} & 21 & \nodata  & \nodata  & \nodata  & \nodata  & \nodata  &  & \nodata  & \nodata  & \nodata  & \nodata  & \nodata & & 1.11     & 0.70     & 0.28     & 0.28     & 1       \\
Ti~\textsc{i}  & 22 & 2.65     & 0.09     & 0.10     & 0.20     & 4        &  & 2.28     & 0.07     & 0.16     & 0.28     & 3       & & 2.07     & $-$0.14    & 0.09     & 0.18     & 4       \\
Ti~\textsc{ii} & 22 & 3.08     & 0.52     & 0.37     & 0.52     & 2        &  & 2.96     & 0.74     & 0.08     & 0.12     & 2       & & 2.80     & 0.59     & 0.09     & 0.13     & 2       \\
V~\textsc{ii}  & 23 & 1.91     & 0.38     & 0.07     & 0.15     & 5        &  & 1.76     & 0.57     & 0.07     & 0.15     & 4       & & 1.74     & 0.55     & 0.05     & 0.11     & 5       \\
Cr~\textsc{i}  & 24 & \nodata  & \nodata  & \nodata  & \nodata  & \nodata  &  & \nodata  & \nodata  & \nodata  & \nodata  & \nodata & & \nodata  & \nodata  & \nodata  & \nodata  & \nodata \\
Mn~\textsc{i}  & 25 & 2.60     & $-$0.43    & 0.11     & 0.22     & 4        &  & 2.28     & $-$0.41    & 0.07     & 0.14     & 4       & & 2.37     & $-$0.33    & 0.06     & 0.12     & 4       \\
Fe~\textsc{i}  & 26 & 5.11     & $-$2.39     & 0.10     & 0.28     & 8        &  & 4.76     & $-$2.74     & 0.05     & 0.13     & 7       & & 4.76     & $-$2.74     & 0.05     & 0.15     & 9       \\
Co~\textsc{i}  & 27 & 2.80     & 0.20     & 0.20     & 0.20     & 1        &  & 2.42     & 0.17     & 0.17     & 0.29     & 3       & & 2.16     & $-$0.09    & 0.08     & 0.14     & 3       \\
Ni~\textsc{i}  & 28 & \nodata  & \nodata  & \nodata  & \nodata  & \nodata  &  & \nodata  & \nodata  & \nodata  & \nodata  & \nodata & & 3.39     & $-$0.09    & 0.37     & 0.37     & 1       \\
Y~\textsc{ii}  & 39 & $-$0.39    & $-$0.20    & 0.11     & 0.16     & 2        &  & $-$0.47    & 0.05     & 0.07     & 0.10     & 2       & & $-$0.77    & $-$0.25    & 0.10     & 0.14     & 2       \\
Zr~\textsc{ii} & 40 & 0.50     & 0.31     & 0.14     & 0.20     & 2        &  & 0.37     & 0.52     & 0.07     & 0.12     & 3       & & 0.25     & 0.41     & 0.15     & 0.26     & 3       \\
La~\textsc{ii} & 57 & $-$0.99    & 0.31     & 0.12     & 0.21     & 3        &  & $-$1.20    & 0.44     & 0.05     & 0.09     & 3       & & $-$1.34    & 0.30     & 0.07     & 0.12     & 3       \\
Ce~\textsc{ii} & 58 & \nodata  & \nodata  & \nodata  & \nodata  & \nodata  &  & \nodata  & \nodata  & \nodata  & \nodata  & \nodata & & \nodata  & \nodata  & \nodata  & \nodata  & \nodata \\
Nd~\textsc{ii} & 60 & \nodata  & \nodata  & \nodata  & \nodata  & \nodata  &  & \nodata  & \nodata  & \nodata  & \nodata  & \nodata & & \nodata  & \nodata  & \nodata  & \nodata  & \nodata \\
Eu~\textsc{ii} & 63 & $-$1.12    & 0.75     & 0.30     & 0.30     & 1        &  & $<-$1.4    & $<$0.85   & \nodata  & \nodata  & \nodata & & $-$1.73    & 0.49     & 0.46     & 0.46     & 1       \\
Ho~\textsc{ii} & 67 & \nodata  & \nodata  & \nodata  & \nodata  & \nodata  &  & \nodata  & \nodata  & \nodata  & \nodata  & \nodata & & \nodata  & \nodata  & \nodata  & \nodata  & \nodata \\
Er~\textsc{ii} & 68 & \nodata  & \nodata  & \nodata  & \nodata  & \nodata  &  & \nodata  & \nodata  & \nodata  & \nodata  & \nodata & & \nodata  & \nodata  & \nodata  & \nodata  & \nodata \\
\enddata
\tablecomments{[Fe/H], not [X/Fe], is indicated for Fe.
The standard deviation is indicated by $\sigma$,
and the standard error is indicated by $\sigma_{\mu}$.
These are internal uncertainties only;
see Section~\ref{abund} for a discussion of absolute uncertainties.
}
\end{deluxetable*}

\begin{deluxetable*}{ccccccccccccccccccc}
\tablecaption{Stellar Abundances II
\label{abundtab2}}
\tablewidth{0pt}
%\rotate
\tabletypesize{\scriptsize}
\tablehead{
\colhead{} &
\colhead{} &
\multicolumn{5}{c}{VI-18} &
\colhead{} &
\multicolumn{5}{c}{VII-10} &
\colhead{} &
\multicolumn{5}{c}{VII-18} \\
\cline{3-7} \cline{9-13} \cline{15-19}
\colhead{Species} &
\colhead{$Z$} &
\colhead{log~$\epsilon$} &
\colhead{[X/Fe]} &
\colhead{$\sigma_{\mu}$} &
\colhead{$\sigma$} &
\colhead{$N$} &
\colhead{} &
\colhead{log~$\epsilon$} &
\colhead{[X/Fe]} &
\colhead{$\sigma_{\mu}$} &
\colhead{$\sigma$} &
\colhead{$N$} &
\colhead{} &
\colhead{log~$\epsilon$} &
\colhead{[X/Fe]} &
\colhead{$\sigma_{\mu}$} &
\colhead{$\sigma$} &
\colhead{$N$} }
\startdata
C ($^{12}$CH)  &  6 & 5.22      &  $-$0.43    & 0.11     & 0.15     & 2        &  & 5.21     & $-$0.48    & 0.09     & 0.13     & 2         & & 4.99     & $-$0.64    & 0.09     & 0.13     & 2     \\
N (CN)         &  7 & $<$7.1      &  $<$2.08    & \nodata  & \nodata  & \nodata  &  & 6.49     & 1.40     & 0.22     & 0.22     & 1         & & 6.68     & 1.65     & 0.10     & 0.10     & 1     \\
Si~\textsc{i}  & 14 & 5.32      &  0.59     & 0.17     & 0.17     & 1        &  & 5.36     & 0.59     & 0.24     & 0.24     & 1         & & 5.12     & 0.41     & 0.16     & 0.16     & 1     \\
Sc~\textsc{i}  & 21 & \nodata   &  \nodata  & \nodata  & \nodata  & \nodata  &  & \nodata  & \nodata  & \nodata  & \nodata  & \nodata   & & 0.52     & 0.17     & 0.17     & 0.17     & 1     \\
Sc~\textsc{ii} & 21 & 0.87      &  0.50     & 0.37     & 0.37     & 1        &  & \nodata  & \nodata  & \nodata  & \nodata  & \nodata   & & 1.03     & 0.68     & 0.12     & 0.12     & 1     \\
Ti~\textsc{i}  & 22 & 2.21      &  0.04     & 0.09     & 0.19     & 4        &  & 2.36     & 0.16     & 0.06     & 0.11     & 3         & & 2.14     & 0.00     & 0.04     & 0.08     & 4     \\
Ti~\textsc{ii} & 22 & 2.57      &  0.41     & 0.12     & 0.17     & 2        &  & 2.93     & 0.72     & 0.09     & 0.12     & 2         & & 2.85     & 0.70     & 0.07     & 0.10     & 2     \\
V~\textsc{ii}  & 23 & 1.66      &  0.51     & 0.04     & 0.08     & 4        &  & 1.71     & 0.53     & 0.06     & 0.13     & 5         & & 1.66     & 0.54     & 0.04     & 0.08     & 5     \\
Cr~\textsc{i}  & 24 & \nodata   &  \nodata  & \nodata  & \nodata  & \nodata  &  & \nodata  & \nodata  & \nodata  & \nodata  & \nodata   & & 2.61     & $-$0.23    & 0.18     & 0.18     & 1     \\
Mn~\textsc{i}  & 25 & 1.79      &  $-$0.86    & 0.11     & 0.25     & 5        &  & 2.47     & $-$0.22    & 0.15     & 0.33     & 5         & & 2.39     & $-$0.24    & 0.08     & 0.18     & 5     \\
Fe~\textsc{i}  & 26 & 4.72      &  $-$2.78    & 0.06     & 0.18     & 8        &  & 4.76     & $-$2.74    & 0.02     & 0.05     & 8         & & 4.70     & $-$2.80    & 0.03     & 0.08     & 9     \\
Co~\textsc{i}  & 27 & 2.37      &  0.16     & 0.06     & 0.10     & 3        &  & 2.49     & 0.25     & 0.10     & 0.17     & 3         & & 2.42     & 0.23     & 0.06     & 0.11     & 3     \\
Ni~\textsc{i}  & 28 & 3.19      &  $-$0.25    & 0.26     & 0.26     & 1        &  & 3.30     & $-$0.18    & 0.30     & 0.30     & 1         & & 3.34     & $-$0.08    & 0.16     & 0.16     & 1     \\
Y~\textsc{ii}  & 39 & $-$0.48     &  0.09     & 0.07     & 0.09     & 2        &  & $-$0.68    & $-$0.15    & 0.16     & 0.22     & 2         & & $-$0.87    & $-$0.27    & 0.08     & 0.11     & 2     \\
Zr~\textsc{ii} & 40 & 0.22      &  0.42     & 0.09     & 0.12     & 2        &  & 0.19     & 0.36     & 0.13     & 0.22     & 3         & & 0.21     & 0.43     & 0.05     & 0.09     & 3     \\
La~\textsc{ii} & 57 & $-$1.14     &  0.55     & 0.07     & 0.15     & 4        &  & $-$1.39    & 0.26     & 0.07     & 0.14     & 4         & & $-$1.57    & 0.14     & 0.10     & 0.19     & 4     \\
Ce~\textsc{ii} & 58 & \nodata   &  \nodata  & \nodata  & \nodata  & \nodata  &  & \nodata  & \nodata  & \nodata  & \nodata  & \nodata   & & $-$0.81    & 0.41     & 0.19     & 0.19     & 1     \\
Nd~\textsc{ii} & 60 & $-$0.82     &  0.54     & 0.31     & 0.31     & 1        &  & \nodata  & \nodata  & \nodata  & \nodata  & \nodata   & & $-$1.06    & 0.32     & 0.14     & 0.20     & 2     \\
Eu~\textsc{ii} & 63 & $-$1.51     &  0.75     & 0.21     & 0.21     & 1        &  & $-$1.86    & 0.36     & 0.36     & 0.36     & 1         & & $-$1.83    & 0.45     & 0.11     & 0.11     & 1     \\
Ho~\textsc{ii} & 67 & $-$1.31     &  0.99     & 0.14     & 0.14     & 1        &  & \nodata  & \nodata  & \nodata  & \nodata  & \nodata   & & $-$2.08    & 0.24     & 0.13     & 0.13     & 1     \\
Er~\textsc{ii} & 68 & $-$0.80     &  1.06     & 0.19     & 0.19     & 1        &  & \nodata  & \nodata  & \nodata  & \nodata  & \nodata   & & $-$1.22    & 0.66     & 0.16     & 0.16     & 1     \\
\enddata
\tablecomments{[Fe/H], not [X/Fe], is indicated for Fe.
The standard deviation is indicated by $\sigma$,
and the standard error is indicated by $\sigma_{\mu}$.
These are internal uncertainties only;
see Section~\ref{abund} for a discussion of absolute uncertainties.
}
\end{deluxetable*}

\begin{deluxetable*}{ccccccccccccccccccc}
\tablecaption{Stellar Abundances III
\label{abundtab3}}
\tablewidth{0pt}
%\rotate
\tabletypesize{\scriptsize}
\tablehead{
\colhead{} &
\colhead{} &
\multicolumn{5}{c}{VII-68} &
\colhead{} &
\multicolumn{5}{c}{VIII-24} &
\colhead{} &
\multicolumn{5}{c}{VIII-44} \\
\cline{3-7} \cline{9-13} \cline{15-19}
\colhead{Species} &
\colhead{$Z$} &
\colhead{log~$\epsilon$} &
\colhead{[X/Fe]} &
\colhead{$\sigma_{\mu}$} &
\colhead{$\sigma$} &
\colhead{$N$} &
\colhead{} &
\colhead{log~$\epsilon$} &
\colhead{[X/Fe]} &
\colhead{$\sigma_{\mu}$} &
\colhead{$\sigma$} &
\colhead{$N$} &
\colhead{} &
\colhead{log~$\epsilon$} &
\colhead{[X/Fe]} &
\colhead{$\sigma_{\mu}$} &
\colhead{$\sigma$} &
\colhead{$N$} }
\startdata
C ($^{12}$CH)  &  6 & 5.44     & $-$0.27    & 0.09     & 0.12     & 2        &  & 5.64     & $-$0.18    & 0.32     & 0.32     & 1        &  & 5.48     & $-$0.35    & 0.10     & 0.13     & 2       \\
N (CN)         &  7 & 6.83     & 1.72     & 0.20     & 0.20     & 1        &  & $<$6.9     & $<$1.71    & \nodata  & \nodata  & \nodata  &  & 6.75     & 1.53     & 0.22     & 0.22     & 1       \\
Si~\textsc{i}  & 14 & 5.62     & 0.83     & 0.19     & 0.19     & 1        &  & 5.57     & 0.67     & 0.22     & 0.22     & 1        &  & 5.73     & 0.83     & 0.20     & 0.20     & 1       \\
Sc~\textsc{i}  & 21 & \nodata  & \nodata  & \nodata  & \nodata  & \nodata  &  & \nodata  & \nodata  & \nodata  & \nodata  & \nodata  &  & \nodata  & \nodata  & \nodata  & \nodata  & \nodata \\
Sc~\textsc{ii} & 21 & \nodata  & \nodata  & \nodata  & \nodata  & \nodata  &  & \nodata  & \nodata  & \nodata  & \nodata  & \nodata  &  & \nodata  & \nodata  & \nodata  & \nodata  & \nodata \\
Ti~\textsc{i}  & 22 & 2.25     & 0.02     & 0.20     & 0.34     & 3        &  & 2.62     & 0.28     & 0.07     & 0.12     & 3        &  & 2.44     & 0.10     & 0.20     & 0.29     & 2       \\
Ti~\textsc{ii} & 22 & 2.74     & 0.51     & 0.16     & 0.23     & 2        &  & 3.04     & 0.70     & 0.10     & 0.14     & 2        &  & 2.98     & 0.64     & 0.17     & 0.24     & 2       \\
V~\textsc{ii}  & 23 & 1.71     & 0.50     & 0.06     & 0.12     & 4        &  & 1.79     & 0.46     & 0.18     & 0.25     & 2        &  & 1.86     & 0.53     & 0.05     & 0.12     & 5       \\
Cr~\textsc{i}  & 24 & \nodata  & \nodata  & \nodata  & \nodata  & \nodata  &  & \nodata  & \nodata  & \nodata  & \nodata  & \nodata  &  & \nodata  & \nodata  & \nodata  & \nodata  & \nodata \\
Mn~\textsc{i}  & 25 & 2.60     & $-$0.10    & 0.09     & 0.15     & 3        &  & 2.57     & $-$0.25    & 0.15     & 0.30     & 4        &  & 2.53     & $-$0.29    & 0.07     & 0.14     & 4       \\
Fe~\textsc{i}  & 26 & 4.78     & $-$2.72    & 0.07     & 0.17     & 7        &  & 4.89     & $-$2.61    & 0.06     & 0.17     & 8        &  & 4.89     & $-$2.61    & 0.04     & 0.12     & 9       \\
Co~\textsc{i}  & 27 & 2.53     & 0.26     & 0.07     & 0.13     & 3        &  & 2.50     & 0.12     & 0.17     & 0.23     & 2        &  & 2.61     & 0.23     & 0.17     & 0.30     & 3       \\
Ni~\textsc{i}  & 28 & \nodata  & \nodata  & \nodata  & \nodata  & \nodata  &  & \nodata  & \nodata  & \nodata  & \nodata  & \nodata  &  & 3.59     & $-$0.02    & 0.30     & 0.30     & 1       \\
Y~\textsc{ii}  & 39 & $-$0.51    & 0.00     & 0.08     & 0.12     & 2        &  & $-$0.37    & 0.03     & 0.10     & 0.14     & 2        &  & $-$0.49    & $-$0.09    & 0.07     & 0.09     & 2       \\
Zr~\textsc{ii} & 40 & 0.40     & 0.55     & 0.30     & 0.42     & 2        &  & 0.39     & 0.42     & 0.14     & 0.19     & 2        &  & 0.29     & 0.32     & 0.09     & 0.13     & 2       \\
La~\textsc{ii} & 57 & $-$1.14    & 0.48     & 0.18     & 0.18     & 1        &  & $-$1.28    & 0.23     & 0.17     & 0.25     & 2        &  & $-$1.33    & 0.18     & 0.09     & 0.16     & 3       \\
Ce~\textsc{ii} & 58 & \nodata  & \nodata  & \nodata  & \nodata  & \nodata  &  & \nodata  & \nodata  & \nodata  & \nodata  & \nodata  &  & \nodata  & \nodata  & \nodata  & \nodata  & \nodata \\
Nd~\textsc{ii} & 60 & \nodata  & \nodata  & \nodata  & \nodata  & \nodata  &  & \nodata  & \nodata  & \nodata  & \nodata  & \nodata  &  & \nodata  & \nodata  & \nodata  & \nodata  & \nodata \\
Eu~\textsc{ii} & 63 & $<-$1.2    & $<$1.03    & \nodata  & \nodata  & \nodata  &  & $-$1.44    & 0.65     & 0.32     & 0.32     & 1        &  & $-$1.70    & 0.39     & 0.39     & 0.39     & 1       \\
Ho~\textsc{ii} & 67 & $-$1.55    & 0.69     & 0.21     & 0.21     & 1        &  & \nodata  & \nodata  & \nodata  & \nodata  & \nodata  &  & \nodata  & \nodata  & \nodata  & \nodata  & \nodata \\ 
Er~\textsc{ii} & 68 & \nodata  & \nodata  & \nodata  & \nodata  & \nodata  &  & \nodata  & \nodata  & \nodata  & \nodata  & \nodata  &  & \nodata  & \nodata  & \nodata  & \nodata  & \nodata \\
\enddata
\tablecomments{[Fe/H], not [X/Fe], is indicated for Fe.
The standard deviation is indicated by $\sigma$,
and the standard error is indicated by $\sigma_{\mu}$.
These are internal uncertainties only;
see Section~\ref{abund} for a discussion of absolute uncertainties.
}
\end{deluxetable*}

\begin{deluxetable*}{ccccccccccccccccccc}
\tablecaption{Stellar Abundances IV
\label{abundtab4}}
\tablewidth{0pt}
%\rotate
\tabletypesize{\scriptsize}
\tablehead{
\colhead{} &
\colhead{} &
\multicolumn{5}{c}{IX-49} &
\colhead{} &
\multicolumn{5}{c}{IX-89} &
\colhead{} &
\multicolumn{5}{c}{X-3} \\
\cline{3-7} \cline{9-13} \cline{15-19}
\colhead{Species} &
\colhead{$Z$} &
\colhead{log~$\epsilon$} &
\colhead{[X/Fe]} &
\colhead{$\sigma_{\mu}$} &
\colhead{$\sigma$} &
\colhead{$N$} &
\colhead{} &
\colhead{log~$\epsilon$} &
\colhead{[X/Fe]} &
\colhead{$\sigma_{\mu}$} &
\colhead{$\sigma$} &
\colhead{$N$} &
\colhead{} &
\colhead{log~$\epsilon$} &
\colhead{[X/Fe]} &
\colhead{$\sigma_{\mu}$} &
\colhead{$\sigma$} &
\colhead{$N$} }
\startdata
C ($^{12}$CH)  &  6 & 5.36       & $-$0.42    & 0.11     & 0.15     & 2        &  & 5.62     & $-$0.26    & 0.18     & 0.25     & 2        &  & 5.70     & $-$0.38    & 0.14     & 0.20     & 2        \\
N (CN)         &  7 & 6.87       & 1.69     & 0.24     & 0.24     & 1        &  & 7.09     & 1.81     & 0.15     & 0.15     & 1        &  & 7.83     & 2.35     & 0.11     & 0.11     & 1        \\
Si~\textsc{i}  & 14 & 5.48       & 0.62     & 0.17     & 0.17     & 1        &  & 5.31     & 0.35     & 0.21     & 0.21     & 1        &  & 5.95     & 0.79     & 0.14     & 0.14     & 1        \\
Sc~\textsc{i}  & 21 & \nodata    & \nodata  & \nodata  & \nodata  & \nodata  &  & \nodata  & \nodata  & \nodata  & \nodata  & \nodata  &  & \nodata  & \nodata  & \nodata  & \nodata  & \nodata  \\
Sc~\textsc{ii} & 21 & 1.09       & 0.59     & 0.39     & 0.39     & 1        &  & \nodata  & \nodata  & \nodata  & \nodata  & \nodata  &  & \nodata  & \nodata  & \nodata  & \nodata  & \nodata  \\
Ti~\textsc{i}  & 22 & 2.18       & $-$0.12    & 0.15     & 0.30     & 4        &  & 2.32     & $-$0.08    & 0.12     & 0.21     & 3        &  & 2.68     & 0.08     & 0.09     & 0.16     & 3        \\
Ti~\textsc{ii} & 22 & 3.07       & 0.77     & 0.09     & 0.13     & 2        &  & 2.89     & 0.49     & 0.23     & 0.32     & 2        &  & 3.23     & 0.63     & 0.18     & 0.25     & 2        \\
V~\textsc{ii}  & 23 & 1.72       & 0.44     & 0.04     & 0.09     & 5        &  & 1.72     & 0.34     & 0.11     & 0.21     & 4        &  & 1.88     & 0.30     & 0.07     & 0.12     & 3        \\
Cr~\textsc{i}  & 24 & \nodata    & \nodata  & \nodata  & \nodata  & \nodata  &  & \nodata  & \nodata  & \nodata  & \nodata  & \nodata  &  & \nodata  & \nodata  & \nodata  & \nodata  & \nodata  \\
Mn~\textsc{i}  & 25 & 2.52       & $-$0.27    & 0.04     & 0.10     & 5        &  & 2.45     & $-$0.43    & 0.08     & 0.15     & 4        &  & 2.68     & $-$0.40    & 0.10     & 0.23     & 5        \\
Fe~\textsc{i}  & 26 & 4.85       & $-$2.65    & 0.03     & 0.10     & 8        &  & 4.95     & $-$2.55    & 0.09     & 0.25     & 8        &  & 5.15     & $-$2.35    & 0.07     & 0.19     & 8        \\
Co~\textsc{i}  & 27 & 2.32       & $-$0.02    & 0.15     & 0.26     & 3        &  & 2.46     & 0.03     & 0.21     & 0.36     & 3        &  & 2.75     & 0.11     & 0.14     & 0.19     & 2        \\
Ni~\textsc{i}  & 28 & \nodata    & \nodata  & \nodata  & \nodata  & \nodata  &  & \nodata  & \nodata  & \nodata  & \nodata  & \nodata  &  & \nodata  & \nodata  & \nodata  & \nodata  & \nodata  \\
Y~\textsc{ii}  & 39 & $-$0.57      & $-$0.13    & 0.08     & 0.11     & 2        &  & $-$0.50    & $-$0.16    & 0.11     & 0.15     & 2        &  & $-$0.24    & $-$0.10    & 0.09     & 0.12     & 2        \\
Zr~\textsc{ii} & 40 & 0.40       & 0.47     & 0.08     & 0.14     & 3        &  & 0.35     & 0.33     & 0.14     & 0.19     & 2        &  & 0.55     & 0.32     & 0.13     & 0.18     & 2        \\
La~\textsc{ii} & 57 & $-$1.16      & 0.39     & 0.12     & 0.16     & 2        &  & $-$1.31    & 0.14     & 0.12     & 0.17     & 2        &  & $-$1.23    & 0.02     & 0.15     & 0.21     & 2        \\
Ce~\textsc{ii} & 58 & \nodata    & \nodata  & \nodata  & \nodata  & \nodata  &  & \nodata  & \nodata  & \nodata  & \nodata  & \nodata  &  & \nodata  & \nodata  & \nodata  & \nodata  & \nodata  \\
Nd~\textsc{ii} & 60 & \nodata    & \nodata  & \nodata  & \nodata  & \nodata  &  & \nodata  & \nodata  & \nodata  & \nodata  & \nodata  &  & \nodata  & \nodata  & \nodata  & \nodata  & \nodata  \\
Eu~\textsc{ii} & 63 & $-$1.29      & 0.84     & 0.40     & 0.40     & 1        &  & $-$1.39    & 0.64     & 0.23     & 0.23     & 1        &  & $-$1.20    & 0.63     & 0.43     & 0.43     & 1        \\
Ho~\textsc{ii} & 67 & \nodata    & \nodata  & \nodata  & \nodata  & \nodata  &  & $-$1.41    & 0.66     & 0.38     & 0.38     & 1        &  & \nodata  & \nodata  & \nodata  & \nodata  & \nodata  \\
Er~\textsc{ii} & 68 & \nodata    & \nodata  & \nodata  & \nodata  & \nodata  &  & \nodata  & \nodata  & \nodata  & \nodata  & \nodata  &  & \nodata  & \nodata  & \nodata  & \nodata  & \nodata  \\
\enddata
\tablecomments{[Fe/H], not [X/Fe], is indicated for Fe.
The standard deviation is indicated by $\sigma$,
and the standard error is indicated by $\sigma_{\mu}$.
These are internal uncertainties only;
see Section~\ref{abund} for a discussion of absolute uncertainties.
}
\end{deluxetable*}

\begin{deluxetable*}{ccccccccccccccccccc}
\tablecaption{Stellar Abundances V
\label{abundtab5}}
\tablewidth{0pt}
%\rotate
\tabletypesize{\scriptsize}
\tablehead{
\colhead{} &
\colhead{} &
\multicolumn{5}{c}{XI-10} &
\colhead{} &
\multicolumn{5}{c}{XI-19} &
\colhead{} &
\multicolumn{5}{c}{XI-80} \\
\cline{3-7} \cline{9-13} \cline{15-19}
\colhead{Species} &
\colhead{$Z$} &
\colhead{log~$\epsilon$} &
\colhead{[X/Fe]} &
\colhead{$\sigma_{\mu}$} &
\colhead{$\sigma$} &
\colhead{$N$} &
\colhead{} &
\colhead{log~$\epsilon$} &
\colhead{[X/Fe]} &
\colhead{$\sigma_{\mu}$} &
\colhead{$\sigma$} &
\colhead{$N$} &
\colhead{} &
\colhead{log~$\epsilon$} &
\colhead{[X/Fe]} &
\colhead{$\sigma_{\mu}$} &
\colhead{$\sigma$} &
\colhead{$N$} }
\startdata
C ($^{12}$CH)  &  6 & 5.76     & $-$0.15    & 0.07     & 0.10     & 2        &  & 5.18     & $-$0.51    & 0.06     & 0.09     & 2   &  & 5.13     & $-$0.58    & 0.06     & 0.08     & 2     \\
N (CN)         &  7 & 6.64     & 1.33     & 0.38     & 0.38     & 1        &  & 6.56     & 1.46     & 0.13     & 0.13     & 1   &  & 6.61     & 1.50     & 0.11     & 0.11     & 1     \\
Si~\textsc{i}  & 14 & 5.75     & 0.76     & 0.14     & 0.14     & 1        &  & 5.41     & 0.63     & 0.14     & 0.14     & 1   &  & 5.31     & 0.52     & 0.11     & 0.11     & 1     \\
Sc~\textsc{i}  & 21 & \nodata  & \nodata  & \nodata  & \nodata  & \nodata  &  & 0.46     & 0.04     & 0.18     & 0.18     & 1   &  & 0.63     & 0.20     & 0.16     & 0.16     & 1     \\
Sc~\textsc{ii} & 21 & \nodata  & \nodata  & \nodata  & \nodata  & \nodata  &  & 1.13     & 0.71     & 0.28     & 0.28     & 1   &  & 0.95     & 0.52     & 0.26     & 0.26     & 1     \\
Ti~\textsc{i}  & 22 & 2.37     & $-$0.06    & 0.07     & 0.15     & 4        &  & 2.38     & 0.17     & 0.07     & 0.14     & 4   &  & 2.46     & 0.23     & 0.05     & 0.10     & 4     \\
Ti~\textsc{ii} & 22 & 3.10     & 0.67     & 0.08     & 0.12     & 2        &  & 2.90     & 0.69     & 0.05     & 0.07     & 2   &  & 2.89     & 0.66     & 0.06     & 0.09     & 2     \\
V~\textsc{ii}  & 23 & 1.76     & 0.35     & 0.07     & 0.15     & 4        &  & 1.75     & 0.55     & 0.03     & 0.07     & 5   &  & 1.67     & 0.46     & 0.03     & 0.06     & 5     \\
Cr~\textsc{i}  & 24 & \nodata  & \nodata  & \nodata  & \nodata  & \nodata  &  & 2.69     & $-$0.22    & 0.20     & 0.20     & 1   &  & 2.54     & $-$0.38    & 0.15     & 0.15     & 1     \\
Mn~\textsc{i}  & 25 & 2.54     & $-$0.38    & 0.07     & 0.14     & 4        &  & 2.50     & $-$0.19    & 0.04     & 0.09     & 5   &  & 2.43     & $-$0.28    & 0.04     & 0.09     & 5     \\
Fe~\textsc{i}  & 26 & 4.98     & $-$2.52    & 0.05     & 0.15     & 8        &  & 4.77     & $-$2.73    & 0.05     & 0.14     & 9   &  & 4.78     & $-$2.72    & 0.04     & 0.12     & 9     \\
Co~\textsc{i}  & 27 & 2.46     & $-$0.01    & 0.10     & 0.14     & 2        &  & 2.41     & 0.15     & 0.05     & 0.08     & 3   &  & 2.29     & 0.02     & 0.04     & 0.08     & 3     \\
Ni~\textsc{i}  & 28 & \nodata  & \nodata  & \nodata  & \nodata  & \nodata  &  & 3.40     & $-$0.09    & 0.15     & 0.15     & 1   &  & 3.25     & $-$0.25    & 0.17     & 0.17     & 1     \\
Y~\textsc{ii}  & 39 & $-$0.54    & $-$0.23    & 0.08     & 0.12     & 2        &  & $-$0.71    & $-$0.19    & 0.07     & 0.10     & 2   &  & $-$0.79    & $-$0.28    & 0.06     & 0.08     & 2     \\
Zr~\textsc{ii} & 40 & 0.31     & 0.24     & 0.21     & 0.29     & 2        &  & 0.44     & 0.60     & 0.08     & 0.13     & 3   &  & 0.36     & 0.50     & 0.06     & 0.11     & 3     \\
La~\textsc{ii} & 57 & $-$1.37    & 0.04     & 0.14     & 0.20     & 2        &  & $-$1.53    & 0.10     & 0.11     & 0.21     & 4   &  & $-$1.51    & 0.11     & 0.11     & 0.23     & 4     \\
Ce~\textsc{ii} & 58 & \nodata  & \nodata  & \nodata  & \nodata  & \nodata  &  & $-$0.69    & 0.46     & 0.14     & 0.14     & 1   &  & $-$0.73    & 0.41     & 0.23     & 0.23     & 1     \\
Nd~\textsc{ii} & 60 & \nodata  & \nodata  & \nodata  & \nodata  & \nodata  &  & $-$0.93    & 0.38     & 0.13     & 0.18     & 2   &  & $-$1.21    & 0.09     & 0.32     & 0.32     & 1     \\
Eu~\textsc{ii} & 63 & $-$1.75    & 0.25     & 0.33     & 0.33     & 1        &  & $-$2.05    & 0.16     & 0.20     & 0.20     & 1   &  & $-$2.17    & 0.03     & 0.35     & 0.35     & 1     \\
Ho~\textsc{ii} & 67 & \nodata  & \nodata  & \nodata  & \nodata  & \nodata  &  & $-$2.02    & 0.23     & 0.19     & 0.19     & 1   &  & $-$1.86    & 0.38     & 0.24     & 0.24     & 1     \\
Er~\textsc{ii} & 68 & \nodata  & \nodata  & \nodata  & \nodata  & \nodata  &  & $-$1.05    & 0.76     & 0.15     & 0.15     & 1   &  & $-$0.98    & 0.82     & 0.17     & 0.17     & 1     \\
\enddata
\tablecomments{[Fe/H], not [X/Fe], is indicated for Fe.
The standard deviation is indicated by $\sigma$,
and the standard error is indicated by $\sigma_{\mu}$.
These are internal uncertainties only;
see Section~\ref{abund} for a discussion of absolute uncertainties.
}
\end{deluxetable*}

\begin{deluxetable*}{ccccccccccccccccccc}
\tablecaption{Stellar Abundances VI
\label{abundtab6}}
\tablewidth{0pt}
%\rotate
\tabletypesize{\scriptsize}
\tablehead{
\colhead{} &
\colhead{} &
\multicolumn{5}{c}{XII-8} &
\colhead{} &
\multicolumn{5}{c}{XII-18} &
\colhead{} &
\multicolumn{5}{c}{XII-34} \\
\cline{3-7} \cline{9-13} \cline{15-19}
\colhead{Species} &
\colhead{$Z$} &
\colhead{log~$\epsilon$} &
\colhead{[X/Fe]} &
\colhead{$\sigma_{\mu}$} &
\colhead{$\sigma$} &
\colhead{$N$} &
\colhead{} &
\colhead{log~$\epsilon$} &
\colhead{[X/Fe]} &
\colhead{$\sigma_{\mu}$} &
\colhead{$\sigma$} &
\colhead{$N$} &
\colhead{} &
\colhead{log~$\epsilon$} &
\colhead{[X/Fe]} &
\colhead{$\sigma_{\mu}$} &
\colhead{$\sigma$} &
\colhead{$N$} }
\startdata
C ($^{12}$CH)  &  6 & 5.08     & $-$0.46    & 0.04     & 0.04     & 2   &  & $<$6.3     & $<$0.58    & \nodata  & \nodata  & \nodata  &  & 5.24     & $-$0.50    & 0.09     & 0.13     & 2         \\
N (CN)         &  7 & 5.65     & 0.71     & 0.10     & 0.10     & 1   &  & $<$7.9     & $<$2.78    & \nodata  & \nodata  & \nodata  &  & 6.44     & 1.30     & 0.24     & 0.24     & 1         \\
Si~\textsc{i}  & 14 & 5.05     & 0.43     & 0.10     & 0.10     & 1   &  & 5.63     & 0.80     & 0.21     & 0.21     & 1        &  & 5.40     & 0.58     & 0.14     & 0.14     & 1         \\
Sc~\textsc{i}  & 21 & 0.40     & 0.14     & 0.10     & 0.10     & 1   &  & \nodata  & \nodata  & \nodata  & \nodata  & \nodata  &  & \nodata  & \nodata  & \nodata  & \nodata  & \nodata   \\
Sc~\textsc{ii} & 21 & 0.80     & 0.54     & 0.10     & 0.10     & 1   &  & \nodata  & \nodata  & \nodata  & \nodata  & \nodata  &  & 1.03     & 0.57     & 0.19     & 0.19     & 1         \\
Ti~\textsc{i}  & 22 & 2.22     & 0.16     & 0.14     & 0.27     & 4   &  & 2.50     & 0.23     & 0.15     & 0.25     & 3        &  & 2.38     & 0.12     & 0.11     & 0.23     & 4         \\
Ti~\textsc{ii} & 22 & 2.93     & 0.87     & 0.07     & 0.10     & 2   &  & 2.99     & 0.71     & 0.18     & 0.25     & 2        &  & 2.94     & 0.68     & 0.07     & 0.09     & 2         \\
V~\textsc{ii}  & 23 & 1.64     & 0.60     & 0.10     & 0.23     & 5   &  & 1.97     & 0.71     & 0.05     & 0.11     & 5        &  & 1.78     & 0.54     & 0.06     & 0.12     & 5         \\
Cr~\textsc{i}  & 24 & 2.20     & $-$0.55    & 0.10     & 0.10     & 1   &  & \nodata  & \nodata  & \nodata  & \nodata  & \nodata  &  & \nodata  & \nodata  & \nodata  & \nodata  & \nodata   \\
Mn~\textsc{i}  & 25 & 2.15     & $-$0.39    & 0.04     & 0.10     & 5   &  & 2.39     & $-$0.37    & 0.13     & 0.25     & 4        &  & 2.50     & $-$0.24    & 0.07     & 0.15     & 5         \\
Fe~\textsc{i}  & 26 & 4.61     & $-$2.89    & 0.06     & 0.18     & 9   &  & 4.82     & $-$2.68    & 0.04     & 0.09     & 5        &  & 4.81     & $-$2.69    & 0.05     & 0.15     & 9         \\
Co~\textsc{i}  & 27 & 2.21     & 0.11     & 0.23     & 0.39     & 3   &  & 2.81     & 0.50     & 0.17     & 0.23     & 2        &  & 2.55     & 0.25     & 0.06     & 0.10     & 3         \\
Ni~\textsc{i}  & 28 & 3.25     & $-$0.08    & 0.10     & 0.10     & 1   &  & \nodata  & \nodata  & \nodata  & \nodata  & \nodata  &  & 3.31     & $-$0.22    & 0.21     & 0.21     & 1         \\
Y~\textsc{ii}  & 39 & $-$0.80    & $-$0.12    & 0.20     & 0.28     & 2   &  & $-$0.44    & 0.03     & 0.08     & 0.12     & 2        &  & $-$0.47    & 0.01     & 0.09     & 0.13     & 2         \\
Zr~\textsc{ii} & 40 & 0.27     & 0.58     & 0.17     & 0.29     & 3   &  & 0.59     & 0.69     & 0.27     & 0.27     & 1        &  & 0.43     & 0.54     & 0.11     & 0.20     & 3         \\
La~\textsc{ii} & 57 & $-$1.38    & 0.41     & 0.03     & 0.06     & 4   &  & $-$1.08    & 0.50     & 0.21     & 0.21     & 1        &  & $-$1.34    & 0.25     & 0.17     & 0.33     & 4         \\
Ce~\textsc{ii} & 58 & $-$0.85    & 0.46     & 0.10     & 0.10     & 1   &  & \nodata  & \nodata  & \nodata  & \nodata  & \nodata  &  & \nodata  & \nodata  & \nodata  & \nodata  & \nodata   \\
Nd~\textsc{ii} & 60 & $-$0.95    & 0.52     & 0.10     & 0.14     & 2   &  & \nodata  & \nodata  & \nodata  & \nodata  & \nodata  &  & $-$1.12    & 0.15     & 0.30     & 0.30     & 1         \\
Eu~\textsc{ii} & 63 & $-$1.75    & 0.62     & 0.10     & 0.10     & 1   &  & $-$1.70    & 0.46     & 0.36     & 0.36     & 1        &  & $-$1.93    & 0.24     & 0.41     & 0.41     & 1         \\
Ho~\textsc{ii} & 67 & $-$1.85    & 0.56     & 0.10     & 0.10     & 1   &  & $-$1.37    & 0.83     & 0.37     & 0.37     & 1        &  & $-$1.94    & 0.27     & 0.36     & 0.36     & 1         \\
Er~\textsc{ii} & 68 & $-$1.30    & 0.67     & 0.10     & 0.10     & 1   &  & \nodata  & \nodata  & \nodata  & \nodata  & \nodata  &  & $-$1.16    & 0.61     & 0.36     & 0.36     & 1         \\
\enddata
\tablecomments{[Fe/H], not [X/Fe], is indicated for Fe.
The standard deviation is indicated by $\sigma$,
and the standard error is indicated by $\sigma_{\mu}$.
These are internal uncertainties only;
see Section~\ref{abund} for a discussion of absolute uncertainties.
}
\end{deluxetable*}

\begin{deluxetable}{ccccccc}
\tablecaption{Stellar Abundances VII
\label{abundtab7}}
\tablewidth{0pt}
%\rotate
\tabletypesize{\scriptsize}
\tablehead{
\colhead{} &
\colhead{} &
\multicolumn{5}{c}{Bu166} \\
\cline{3-7} 
\colhead{Species} &
\colhead{$Z$} &
\colhead{log~$\epsilon$} &
\colhead{[X/Fe]} &
\colhead{$\sigma_{\mu}$} &
\colhead{$\sigma$} &
\colhead{$N$} }
\startdata
C ($^{12}$CH)  &  6 & 5.73     & $-$0.14    & 0.08     & 0.11     & 2       \\
N (CN)         &  7 & $<$6.5     & $<$1.26    & \nodata  & \nodata  & \nodata \\
Si~\textsc{i}  & 14 & 5.64     & 0.69     & 0.16     & 0.16     & 1       \\
Sc~\textsc{i}  & 21 & \nodata  & \nodata  & \nodata  & \nodata  & \nodata \\
Sc~\textsc{ii} & 21 & \nodata  & \nodata  & \nodata  & \nodata  & \nodata \\
Ti~\textsc{i}  & 22 & 2.38     & $-$0.01    & 0.09     & 0.18     & 4       \\
Ti~\textsc{ii} & 22 & 2.88     & 0.49     & 0.09     & 0.12     & 2       \\
V~\textsc{ii}  & 23 & 1.92     & 0.55     & 0.06     & 0.12     & 4       \\
Cr~\textsc{i}  & 24 & \nodata  & \nodata  & \nodata  & \nodata  & \nodata \\
Mn~\textsc{i}  & 25 & 2.49     & $-$0.38    & 0.15     & 0.25     & 3       \\
Fe~\textsc{i}  & 26 & 4.94     & $-$2.56    & 0.08     & 0.19     & 6       \\
Co~\textsc{i}  & 27 & 2.60     & 0.17     & 0.09     & 0.16     & 3       \\
Ni~\textsc{i}  & 28 & \nodata  & \nodata  & \nodata  & \nodata  & \nodata \\
Y~\textsc{ii}  & 39 & $-$0.58    & $-$0.23    & 0.17     & 0.24     & 2       \\
Zr~\textsc{ii} & 40 & 0.16     & 0.14     & 0.15     & 0.21     & 2       \\
La~\textsc{ii} & 57 & $-$1.26    & 0.20     & 0.17     & 0.24     & 2       \\
Ce~\textsc{ii} & 58 & \nodata  & \nodata  & \nodata  & \nodata  & \nodata \\
Nd~\textsc{ii} & 60 & \nodata  & \nodata  & \nodata  & \nodata  & \nodata \\
Eu~\textsc{ii} & 63 & $<-$1.5    & $<$0.57    & \nodata  & \nodata  & \nodata \\
Ho~\textsc{ii} & 67 & \nodata  & \nodata  & \nodata  & \nodata  & \nodata \\
Er~\textsc{ii} & 68 & \nodata  & \nodata  & \nodata  & \nodata  & \nodata \\
\enddata
\tablecomments{[Fe/H], not [X/Fe], is indicated for Fe.
The standard deviation is indicated by $\sigma$,
and the standard error is indicated by $\sigma_{\mu}$.
These are internal uncertainties only;
see Section~\ref{abund} for a discussion of absolute uncertainties.
}
\end{deluxetable}

\subsection{Comments on Individual Species}
\label{comments}

A few comments regarding these abundances are warranted.
Figure~\ref{teffabundplot} demonstrates that our analysis
produces an artificial relation between [Fe/H] and \teff.
Our Fe abundances span a range from 
$-$2.89~$<$~[Fe/H]~$< -$2.35
with a mean of [Fe/H]~$= -$2.70~$\pm$~0.03 ($\sigma =$~0.14).
This mean metallicity is lower than has been derived in previous
studies, and we address this point in detail in the Appendix.
We have detected no lines of Fe~\textsc{ii}, the dominant Fe 
species in these stars, in our spectra;
however, many of the elements we are interested in studying
are only detected in the singly-ionized state.
This casts considerable uncertainty on the accuracy of [X/Fe] ratios
when X is a singly-ionized species (e.g., 
[La~\textsc{ii}/Fe~\textsc{i}] or [Eu~\textsc{ii}/Fe~\textsc{i}]),
but ratios among 
species of the same ionization state should be more robust.

The star-to-star dispersion among metal ratios is robust, as 
illustrated in Figure~\ref{teffabundplot}.
Our [Ti~\textsc{i}/Fe~\textsc{i}] ratios are persistently lower than
the [Ti~\textsc{ii}/Fe~\textsc{i}] ratios by $\sim$~0.5~dex, likely indicating
that LTE treatments of the level populations of 
these neutral species are inadequate
(see, e.g., \citealt{bergemann11} and references therein).
This may also account for the [V~\textsc{ii}/Fe~\textsc{i}]
ratios that are higher than typically found in metal-poor field stars.
If so, we should expect that all [X~\textsc{ii}/Fe~\textsc{i}]
ratios may be similarly overestimated in our results.

Figure~\ref{teffabundplot} also shows 
that most [X/Fe] ratios have no \teff\ dependence.
C and N may be expected to exhibit such trends resulting
from internal processing.
The [Si~\textsc{i}/Fe] ratio, derived from the Si~\textsc{i} 
3905\AA\ line, increases by $\sim$~0.3~dex from the coolest stars
to the warmest ones in our sample.
This trend is in the opposite sense of what has been found by
previous investigators 
\citep{cohen04,preston06,lai08,bonifacio09,roederer10a}.
The Si~\textsc{i} 3905\AA\ line is very strong and blended,
particularly with CH.
Our tests indicate that modeling these blends 
with a 1D LTE approach could account for the trend
in Si (i.e., reasonable variations in the CH abundance
can account for the deficiency of Si in the coolest stars),
so we discard the Si abundances from further consideration.
No heavier species exhibits a correlation with \teff\ in [X/Fe], 
except perhaps [Y~\textsc{ii}/Fe].
This apparent correlation rests strongly on the warmest star, \mbox{XII-18}.
Other [X/Fe] ratios in this star also appear to be higher than
their respective means for M92 and the [Fe/H] ratio seems to be 
lower than the mean trend defined by the other 18 stars, so
we dismiss this trend as well.

Following recommendations by \citet{cayrel04} and \citet{roederer10a}, 
we artificially
increase the Mn~\textsc{i} abundances derived from the
4030, 4033, and 4034\AA\ lines by 0.3~dex to bring them
in better agreement with abundances derived from other Mn~\textsc{i} 
indicators.
This offset is reflected in all tables and figures.

\subsection{Comparison of Individual Stars with Previous Studies}
\label{compare}

\mbox{VII-18} has been studied in detail
by \citet{shetrone01} and \citet{johnson02}.
Both of these studies derived [Fe~\textsc{i}/H] higher by $\approx$~0.5~dex.
The neutral to neutral metal ratios 
(i.e., [X~\textsc{i}/Fe~\textsc{i}])
are generally in agreement to better than 0.2~dex.
Correcting for differences in the log($gf$) values, we find
that the singly-ionized to singly-ionized ratios
among the heavy \ncap\ elements are also in agreement within 0.2~dex.
The [X~\textsc{ii}/Fe~\textsc{i}] ratios
are higher by 0.4--0.6~dex in Shetrone et al.\ and higher by 0.2--0.4~dex
in Johnson.
This leads us to suspect that our [X~\textsc{ii}/Fe~\textsc{i}] ratios
may be overestimated by $\sim$~0.4~dex in \mbox{VII-18}.
Since this is the coolest star in our sample, we do not 
apply a universal offset based on these comparisons to all of our 
[X~\textsc{ii}/Fe~\textsc{i}] ratios, but we caution that the
absolute values of these ratios are likely overestimated.

\citet{langer98} presented clear evidence for a
0.18~$\pm$~0.01~dex ($\sigma =$~0.12) overabundance in 
Ca~\textsc{i}, Sc~\textsc{ii}, Ti~\textsc{i}, Ti~\textsc{ii},
Cr~\textsc{i}, Fe~\textsc{i}, Fe~\textsc{ii}, Fe~\textsc{ii}, 
Co~\textsc{i}, and Ni~\textsc{i} 
in \mbox{XI-19} relative to \mbox{XII-8} and \mbox{V-45} in M92.
These three stars have very similar $B-V$ colors and $V$ magnitudes.
\mbox{XI-19} and \mbox{XII-8} are also included in our study.
Examining the line-by-line differential abundances 
for Sc~\textsc{i} to Ni~\textsc{i} (31~lines) in these two stars, 
we find a mean difference
of 0.17~$\pm$~0.03~dex ($\sigma =$~0.16), identical to that found
by Langer et al.
Since there is a clear systematic trend of Fe abundance with \teff\ in 
our sample, our offset should not be taken too literally.
After removing the trend there is still a residual
dispersion of 0.11~dex in Fe at a given \teff.
We should not expect to probe star-to-star dispersion
at a level smaller than this, which renders the 
offset between \mbox{XI-19} and \mbox{XII-8}
only mildly significant.
The offset discovered by Langer et al.\ is still unexplained at present.

\section{Results}
\label{results}

We have derived abundances of up to 21~species of 19~elements
in each of 19~stars in M92 from a differential analysis.
Our goal is to search for star-to-star dispersion
among the \ncap\ abundances.
In this section we analyze the degree of homogeneity
of our abundance measurements.

\subsection{Homogeneity of the Fe-group Elements}
\label{fegroup}

As discussed in Section~\ref{comments}, we find a range of 
[Fe~\textsc{i}/H] ratios in M92;
this is an artifact of our analysis and does not reflect a
genuine spread in Fe.
The other Fe-group elements examined
(Sc--Ni) exhibit similar trends, but the [X/Fe] ratios
are generally constant across all \teff.
The dispersion in each of these ratios,
listed in Table~\ref{meantab}, 
is accordingly very small.
For well-determined means
(i.e., [X/Fe] is measured in $\geq$~5 stars), 
the standard deviation is modest, 
0.07~$\leq \sigma \leq$~0.16~dex, 
and consistent with observational uncertainty.
Our data reveal no star-to-star
dispersion among the Fe-group elements, and we
regard this range of standard deviations as the smallest
level of dispersion that can be probed by our data.

\subsection{Dispersion of the Heavy \ncap\ Elements}
\label{ncaps}

The elements Y and Zr, which in principle must have
been produced at least in part by \ncap\ nucleosynthesis,
show no evidence of dispersion in [Y/Fe] or [Zr/Fe].
Their standard deviation, 0.12~$\leq \sigma \leq$~0.14~dex,
is the same as found for the Fe-group elements.

The heavy elements La--Er paint a more complex picture.
Their [X/Fe] star-to-star dispersion is larger, 
0.17~$\leq \sigma \leq$~0.28~dex, but this
may reflect the difficulty in deriving their abundances.
Fewer lines are typically available for analysis
(only 1 in the case of Ce~\textsc{ii}, Eu~\textsc{ii}, 
Ho~\textsc{ii}, and Er~\textsc{ii}), and these lines are often 
weaker and more blended than the Fe-group element lines
that we have analyzed.

\begin{figure}
\includegraphics[angle=0,width=3.4in]{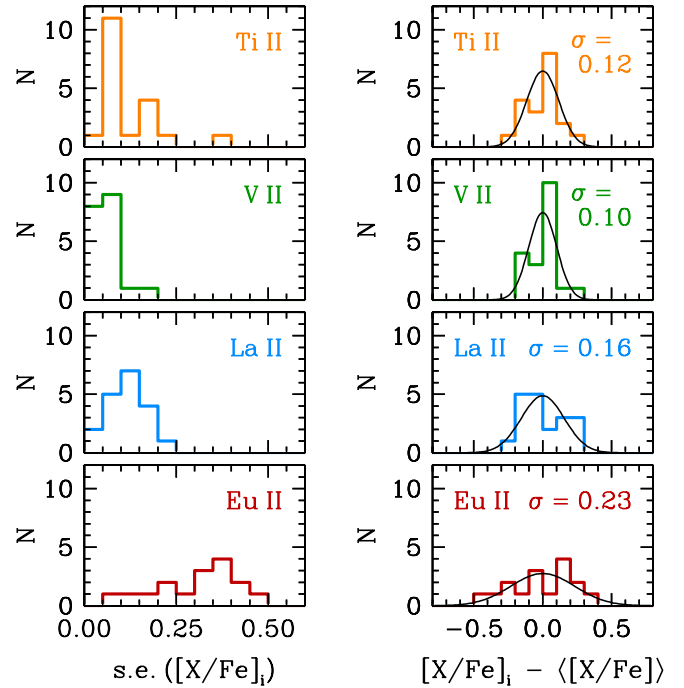}
\caption{
\label{histogram}
Histograms of the standard error (\textit{left}) and 
deviation from the mean [X/Fe] abundance ratio (\textit{right})
for Ti~\textsc{ii}, V~\textsc{ii}, La~\textsc{ii}, and Eu~\textsc{ii}.
The left panels indicate that the median and mode of the (internal)
standard error distributions increase from approximately 0.10~dex
for Ti~\textsc{ii} and V~\textsc{ii} to approximately 0.15~dex
for La~\textsc{ii} and 0.35~dex for Eu~\textsc{ii}.
The (internal) standard deviation, $\sigma$, is
shown in the right set of panels along with a Gaussian fit
to each distribution.
The increase in individual uncertainties
can account for much of the broadening of the
[X/Fe] distributions for La~\textsc{ii} and Eu~\textsc{ii}, 
but it cannot account for 
correlations between the abundance ratios shown
in subsequent figures.
}
\end{figure}

To demonstrate that this may explain at least part of the
larger dispersion, in Figure~\ref{histogram} we
plot histograms of the standard error on [X/Fe]$_{i}$
and the deviation of [X/Fe]$_{i}$ from the mean [X/Fe].
Four ratios are shown: [Ti~\textsc{ii}/Fe~\textsc{i}],
[V~\textsc{ii}/Fe~\textsc{i}],
[La~\textsc{ii}/Fe~\textsc{i}], and
[Eu~\textsc{ii}/Fe~\textsc{i}].
The mean, median, and mode of the standard error distributions for 
[Ti/Fe] and [V/Fe] are all $\leq$~0.10~dex, and 
this is reflected in the dispersion about the mean [Ti/Fe] and [V/Fe]
ratios, $\sigma =$~0.12 and 0.10~dex, respectively.
These measures of the standard error distribution
are larger for La and Eu, $\approx$~0.15 and 0.35~dex, 
respectively, and this is also reflected in the 
dispersion about the mean [La/Fe] and [Eu/Fe] ratios,
$\sigma =$~0.16 and 0.23~dex, respectively.
Thus it would seem that the larger dispersion observed in  
[La/Fe] and [Eu/Fe] can be partially attributed to the
larger measurement uncertainties.

\begin{figure}
\begin{center}
\includegraphics[angle=0,width=2.5in]{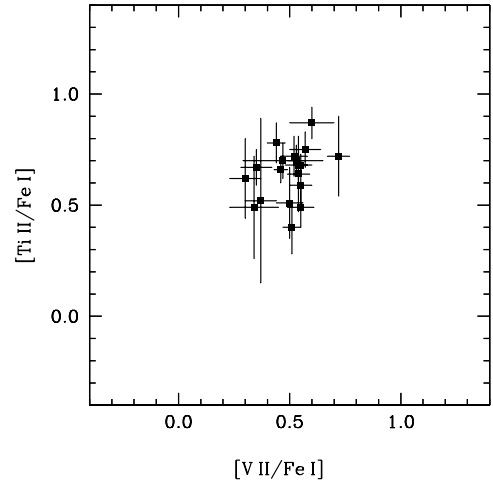}
\end{center}
\caption{
\label{tivplot}
Comparison of the [Ti~\textsc{ii}/Fe~\textsc{i}]
and [V~\textsc{ii}/Fe~\textsc{i}] ratios.
Only the internal (i.e., star-to-star) uncertainties are shown.
There is no significant correlation.
}
\end{figure}

\begin{figure}
\begin{center}
\includegraphics[angle=0,width=2.5in]{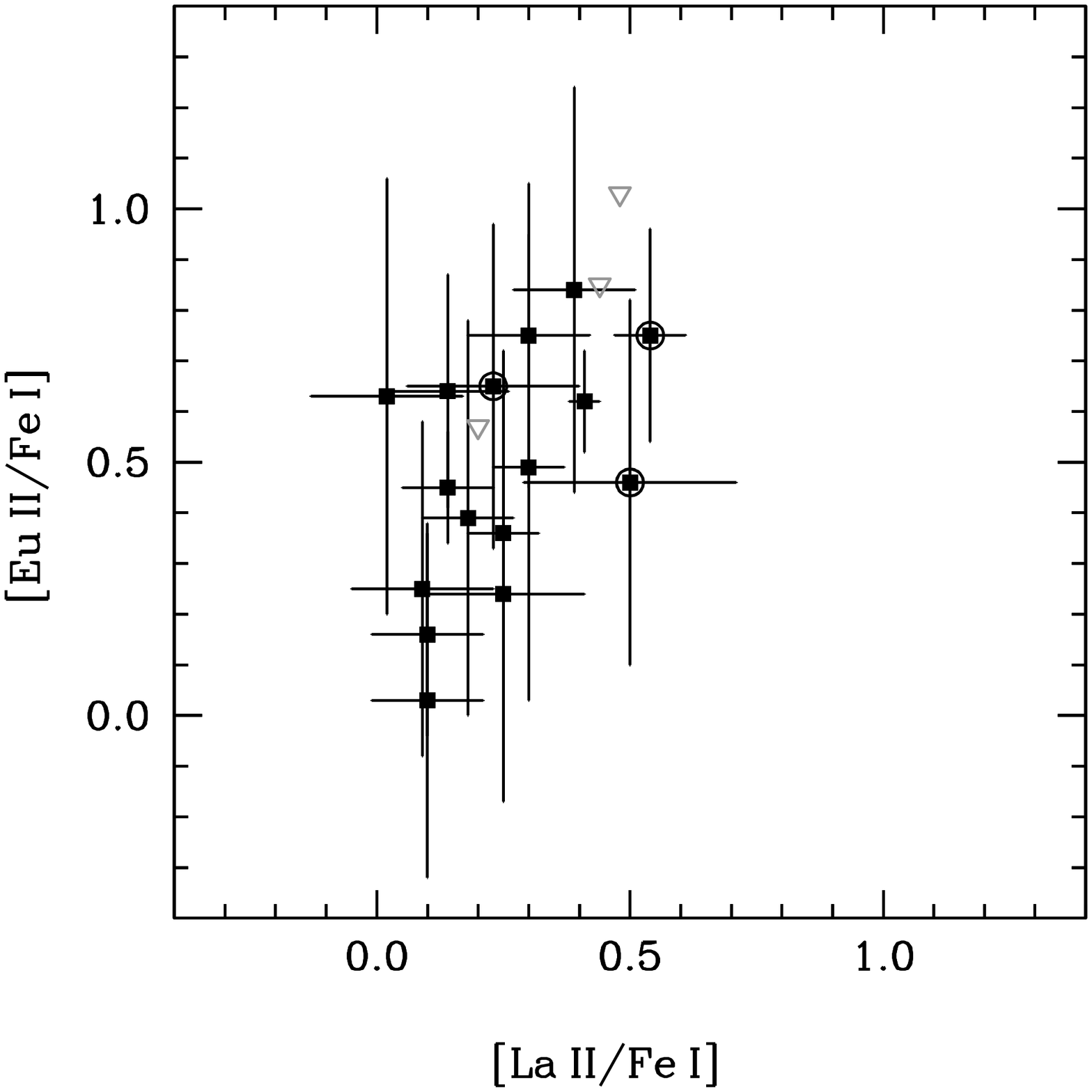} \\
\vspace*{0.2in}
\includegraphics[angle=0,width=2.5in]{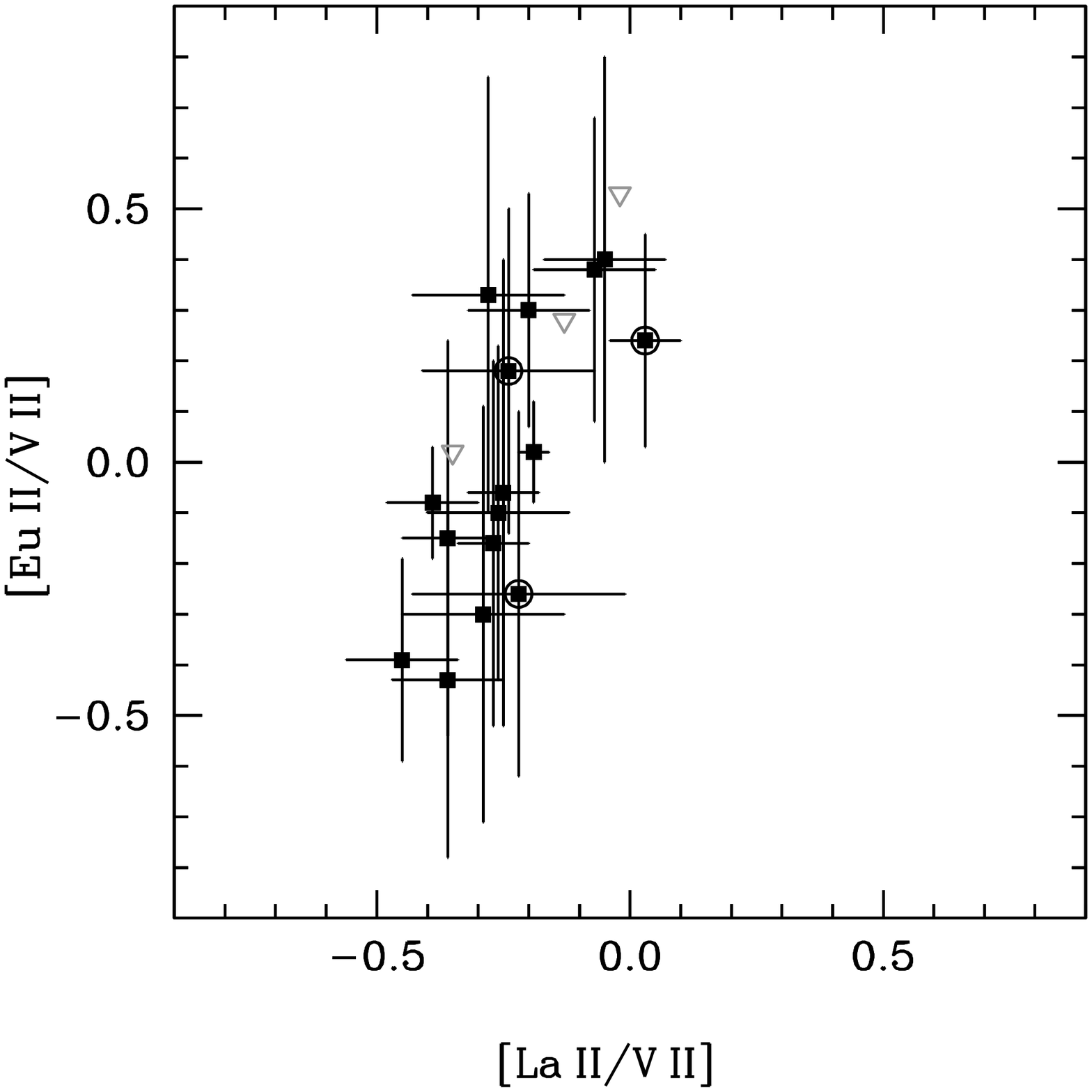}
\end{center}
\clearpage
\caption{
\label{comparev}
\scriptsize
Comparison of the [La~\textsc{ii}/Fe~\textsc{i}] and 
[Eu~\textsc{ii}/Fe~\textsc{i}] ratios (\textit{top}) and the
[La~\textsc{ii}/V~\textsc{ii}] and 
[Eu~\textsc{ii}/V~\textsc{ii}] ratios (\textit{bottom}).
Filled squares represent detections and downward-facing open
triangles represent upper limits.
Open circles indicate probable AGB stars.
The correlation is significant whether using a neutral
Fe-group abundance indicator or a singly-ionized one,
indicating that the correlation is unlikely to be an artifact
of our abundance analysis.
Only the internal (i.e., star-to-star) uncertainties are shown.
}
\end{figure}

\begin{figure*}
\begin{center}
\includegraphics[angle=0,width=2.5in]{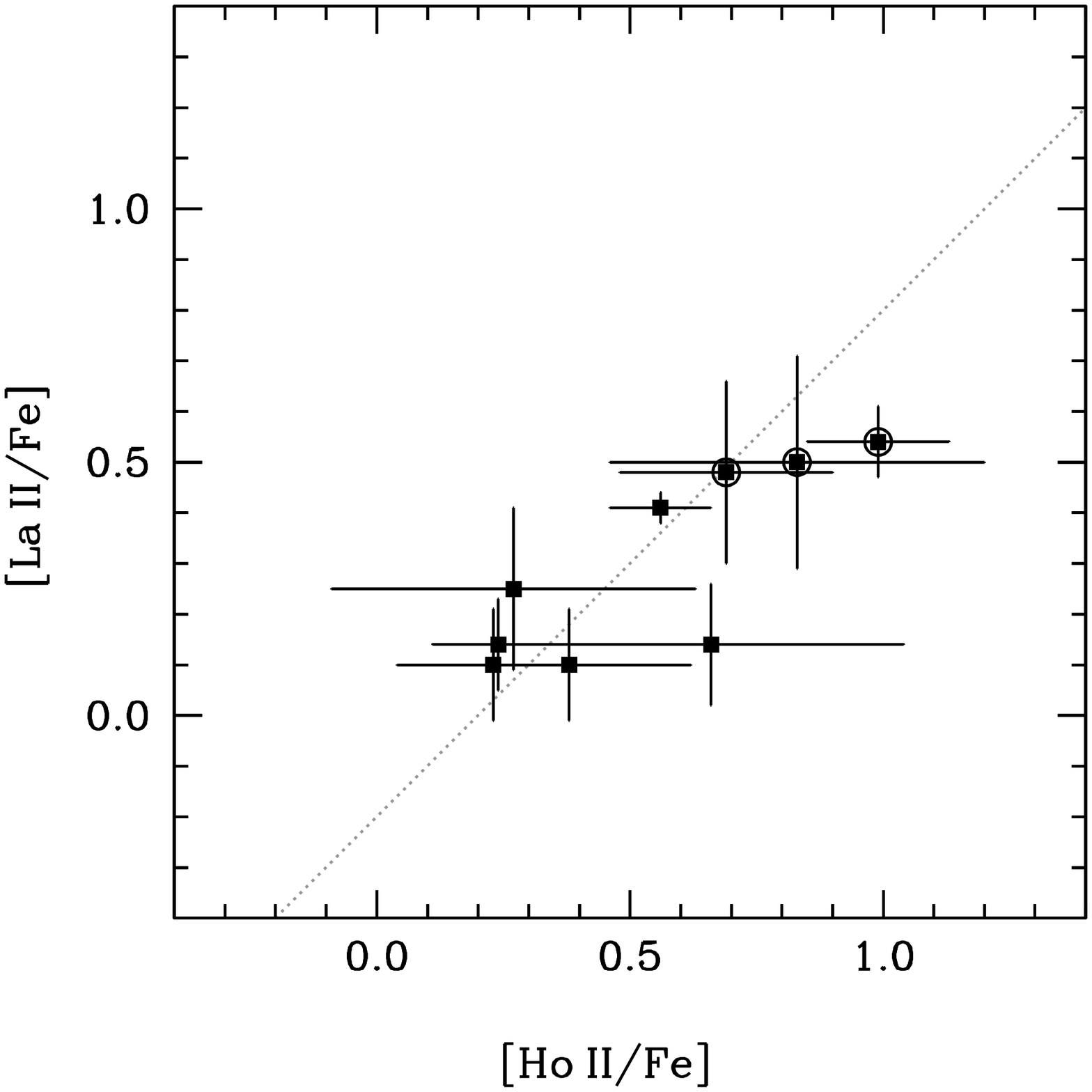} \hspace*{0.2in} 
\includegraphics[angle=0,width=2.5in]{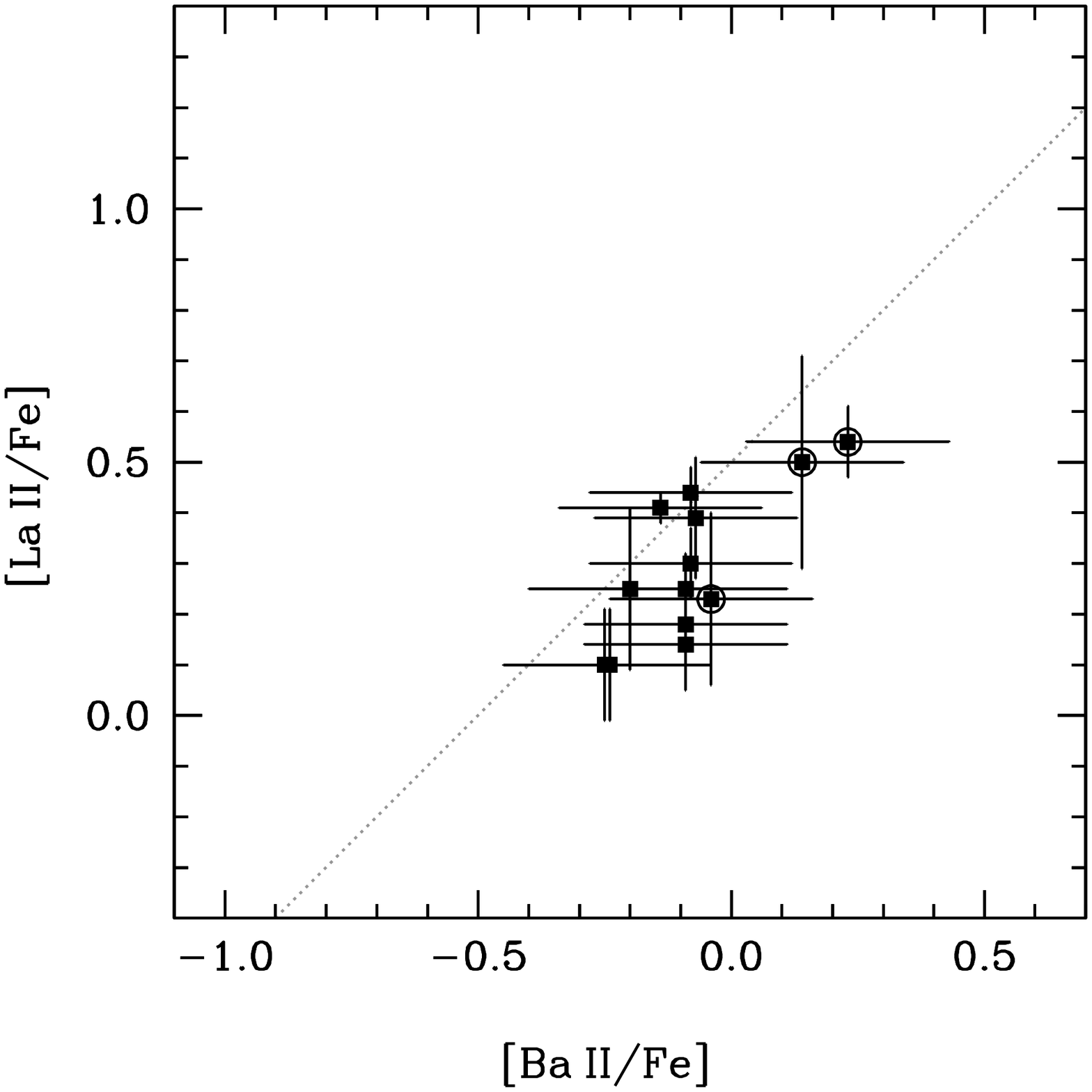} \\
\vspace*{0.2in}
\includegraphics[angle=0,width=2.5in]{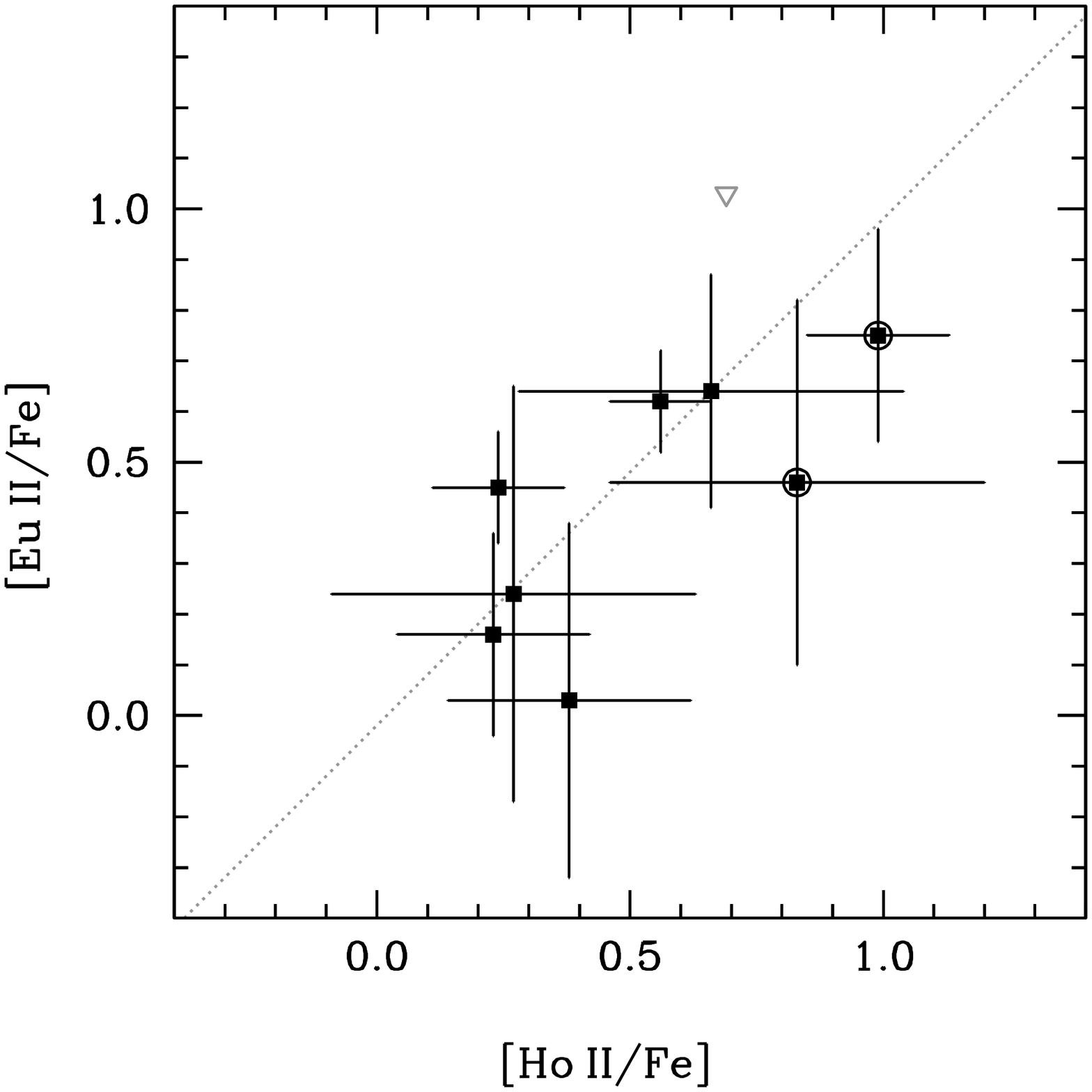} \hspace*{0.2in} 
\includegraphics[angle=0,width=2.5in]{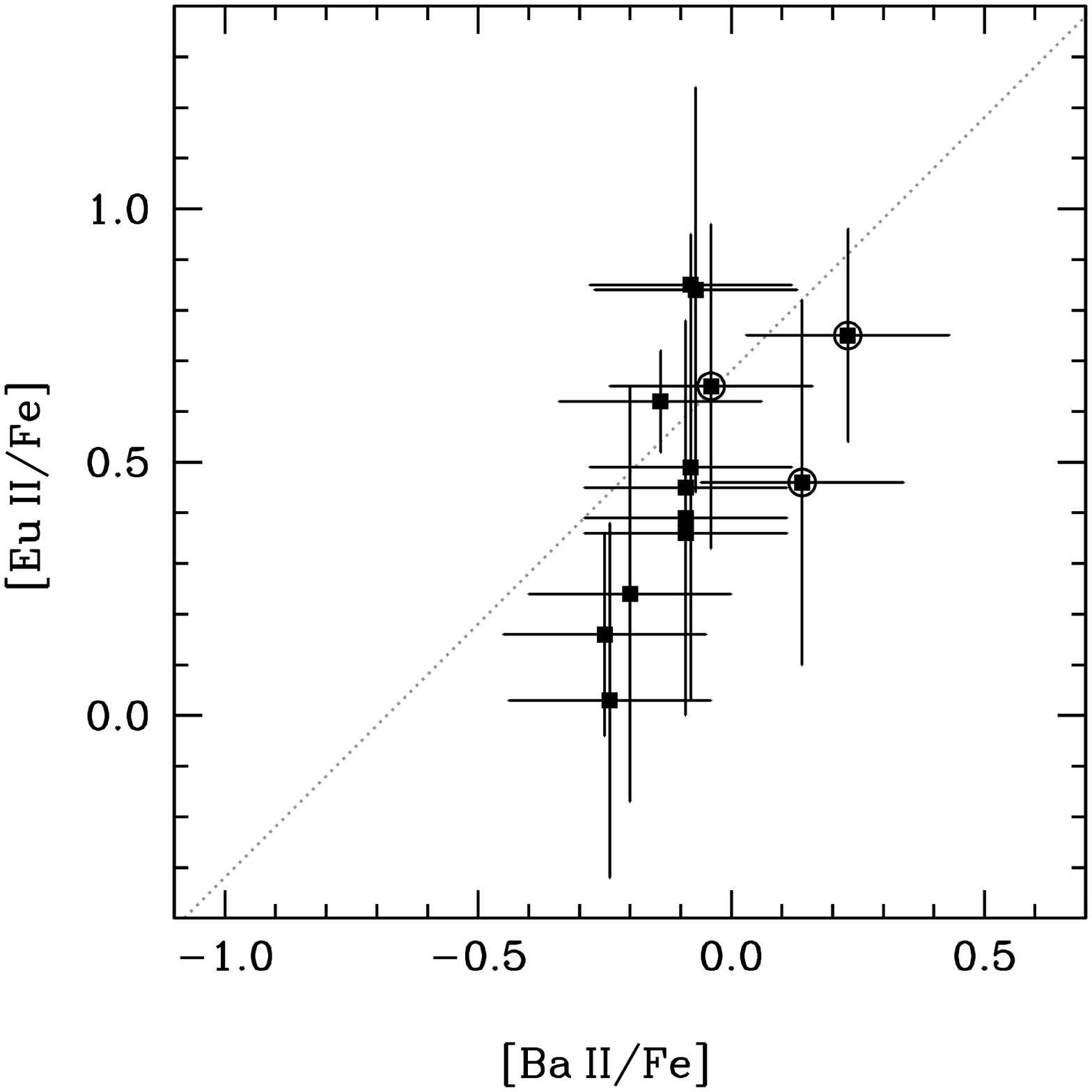} \\
\end{center}
\clearpage
\caption{
\label{compareho}
%\scriptsize
Comparison of the [La/Fe] and [Eu/Fe] ratios against [Ho/Fe] and [Ba/Fe].
Open circles indicate probable AGB stars.
Dotted lines indicate a 1:1 correlation offset by the
mean [Ho/La] and [Ho/Eu] ratios, $+$0.20 and $+$0.02~dex, respectively, 
or the
mean [Ba/La] and [Ba/Eu] ratios, $-$0.50 and $-$0.68~dex, respectively.
All correlations here are significant, and the [Ba/Fe] ratios
were derived from a previous, independent abundance study
of these M92 giants \citep{sneden00}.
Only the internal (i.e., star-to-star) uncertainties are shown.
}
\end{figure*}

\begin{figure*}
\begin{center}
\includegraphics[angle=270,width=4.7in]{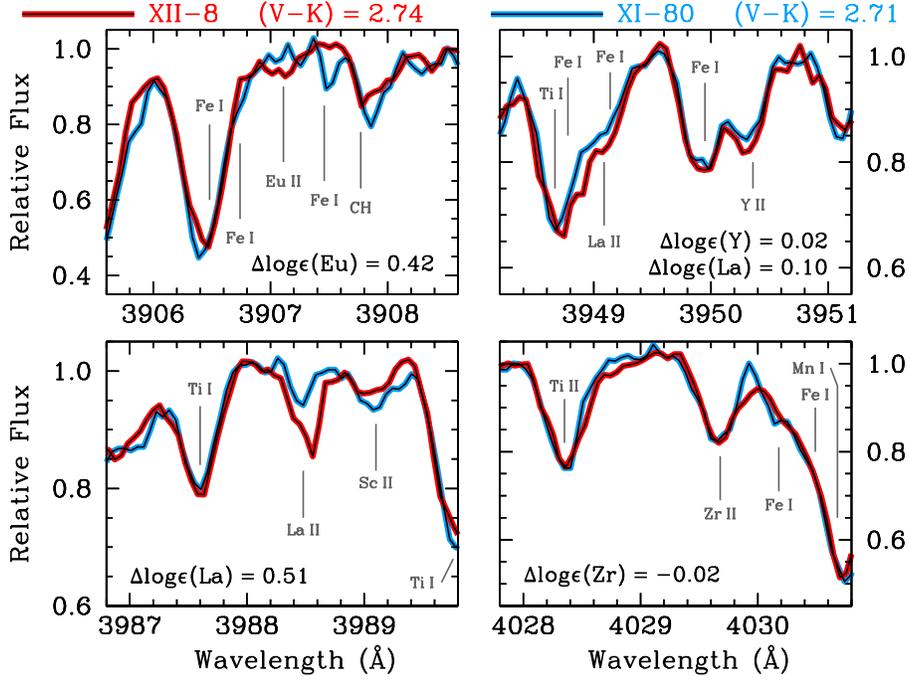} \\
%\vspace*{0.3in}
%\includegraphics[angle=270,width=4.7in]{f07b.eps}
\end{center}
\clearpage
\caption{
\label{overplotspec}
%\scriptsize
Comparison of the spectra of 2 stars with contrasting heavy
element abundances.
The two stars, \mbox{XII-8} and \mbox{XI-80}, 
have very similar colors and temperatures, yet their Eu and La abundance 
differentials are large.
%The two stars in the bottom set of panels, \mbox{XII-34} and \mbox{VI-18},
%also have similar (but not identical) 
%colors and temperatures, yet their Eu and La
%abundance differentials are also clearly distinct.
%In both cases, the Y and Zr differentials are very small.
}
\end{figure*}

\begin{figure*}
\begin{center}
\includegraphics[angle=0,width=2.0in]{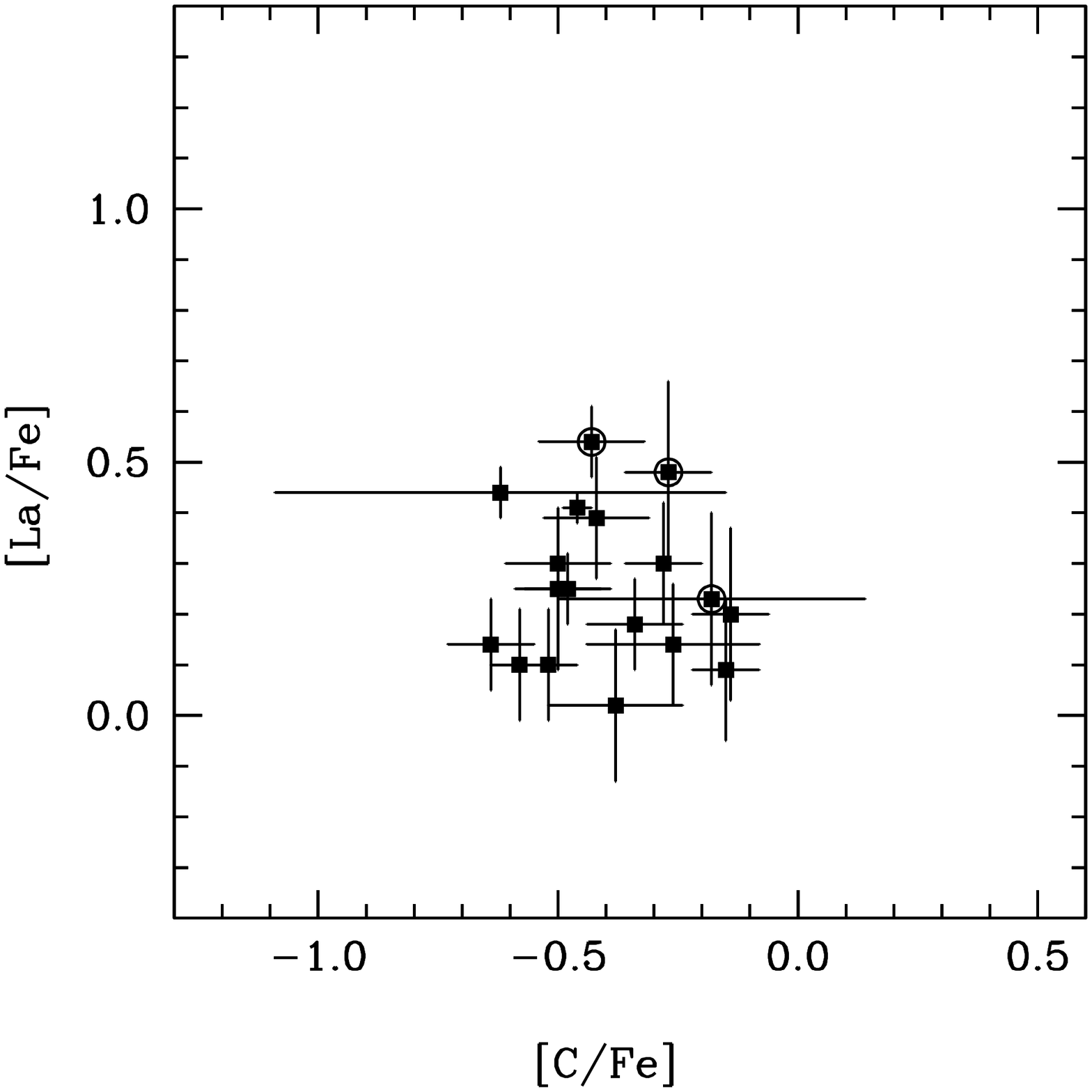} \hspace*{0.1in}
\includegraphics[angle=0,width=2.0in]{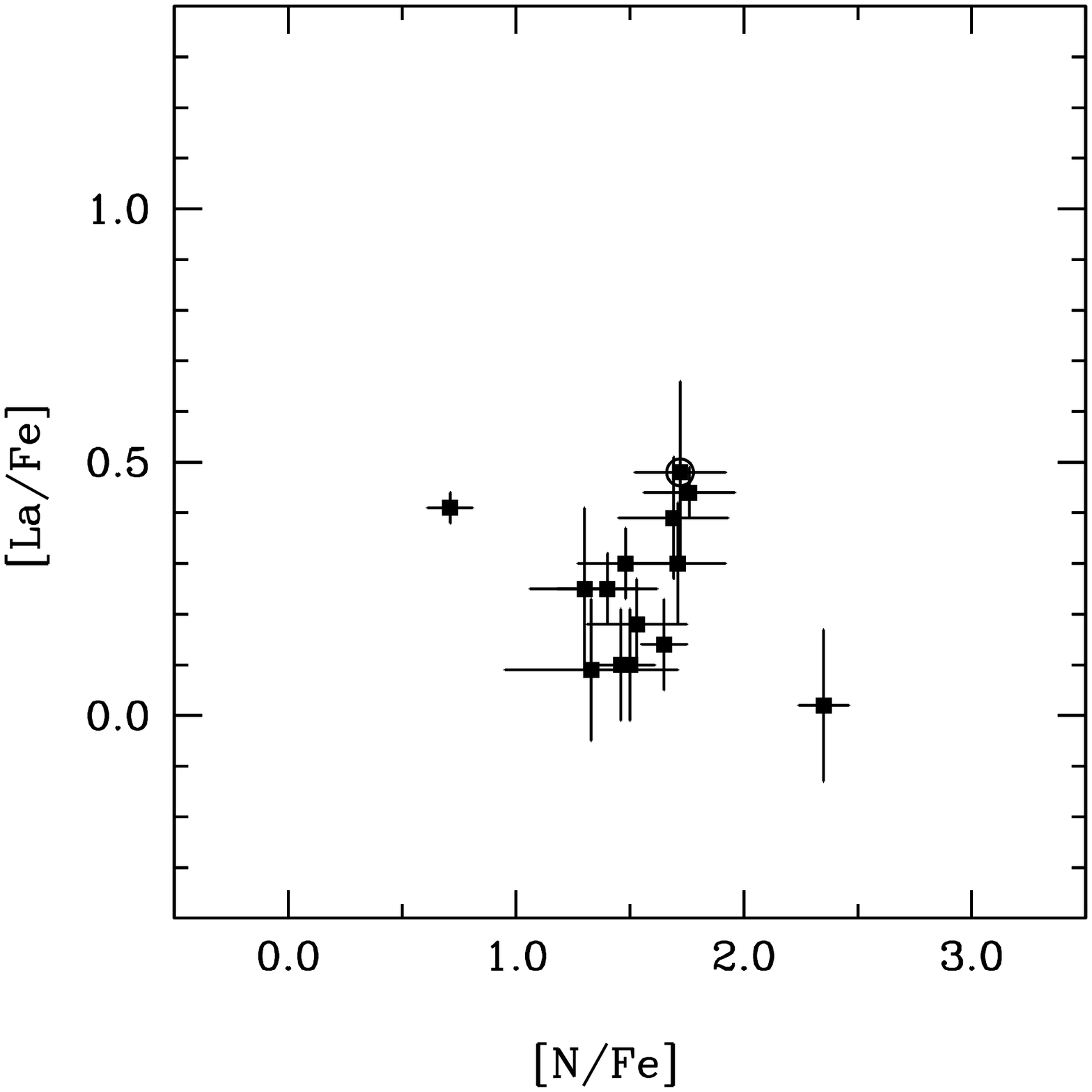} \hspace*{0.1in}
\includegraphics[angle=0,width=2.0in]{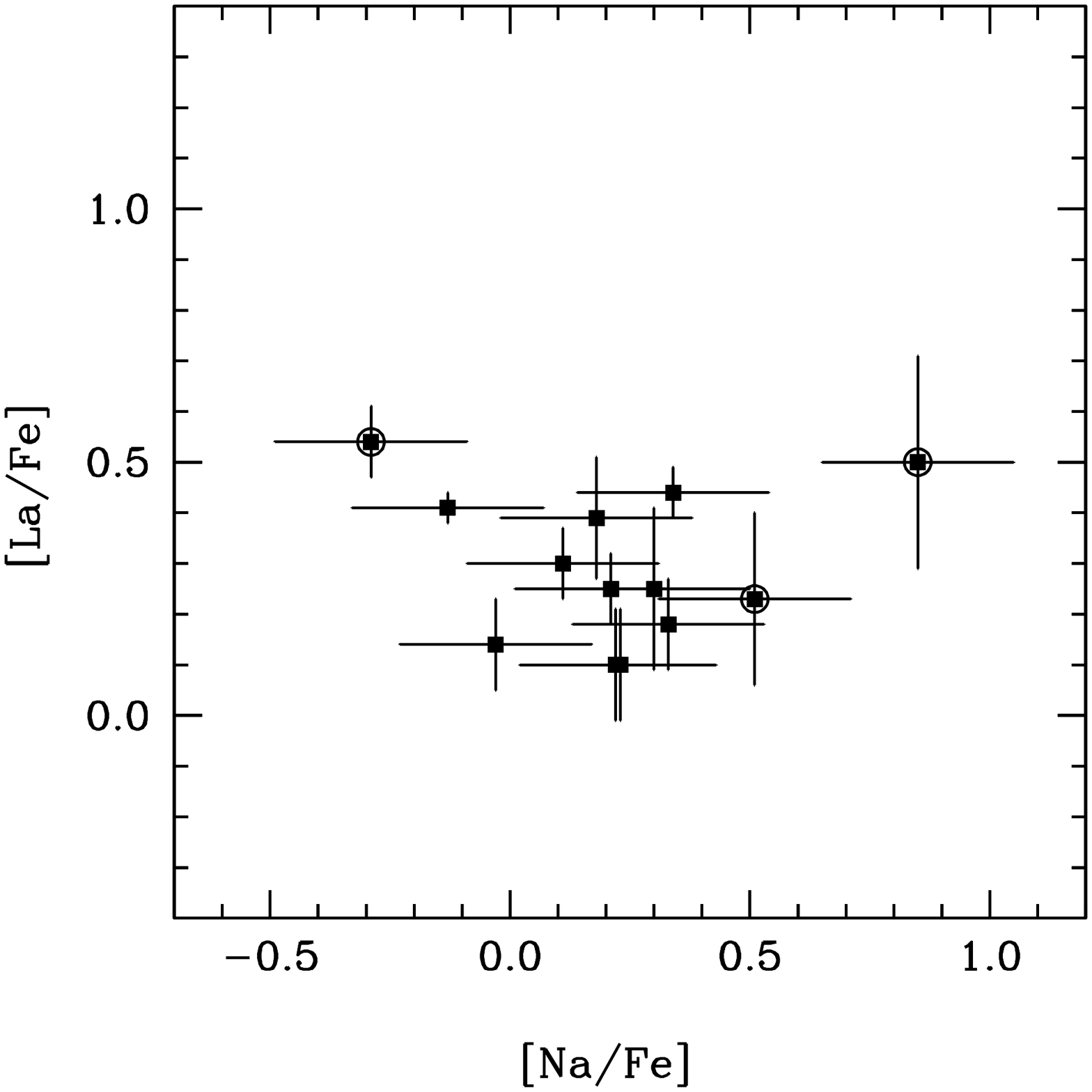} \\
\vspace*{0.2in}
\includegraphics[angle=0,width=2.0in]{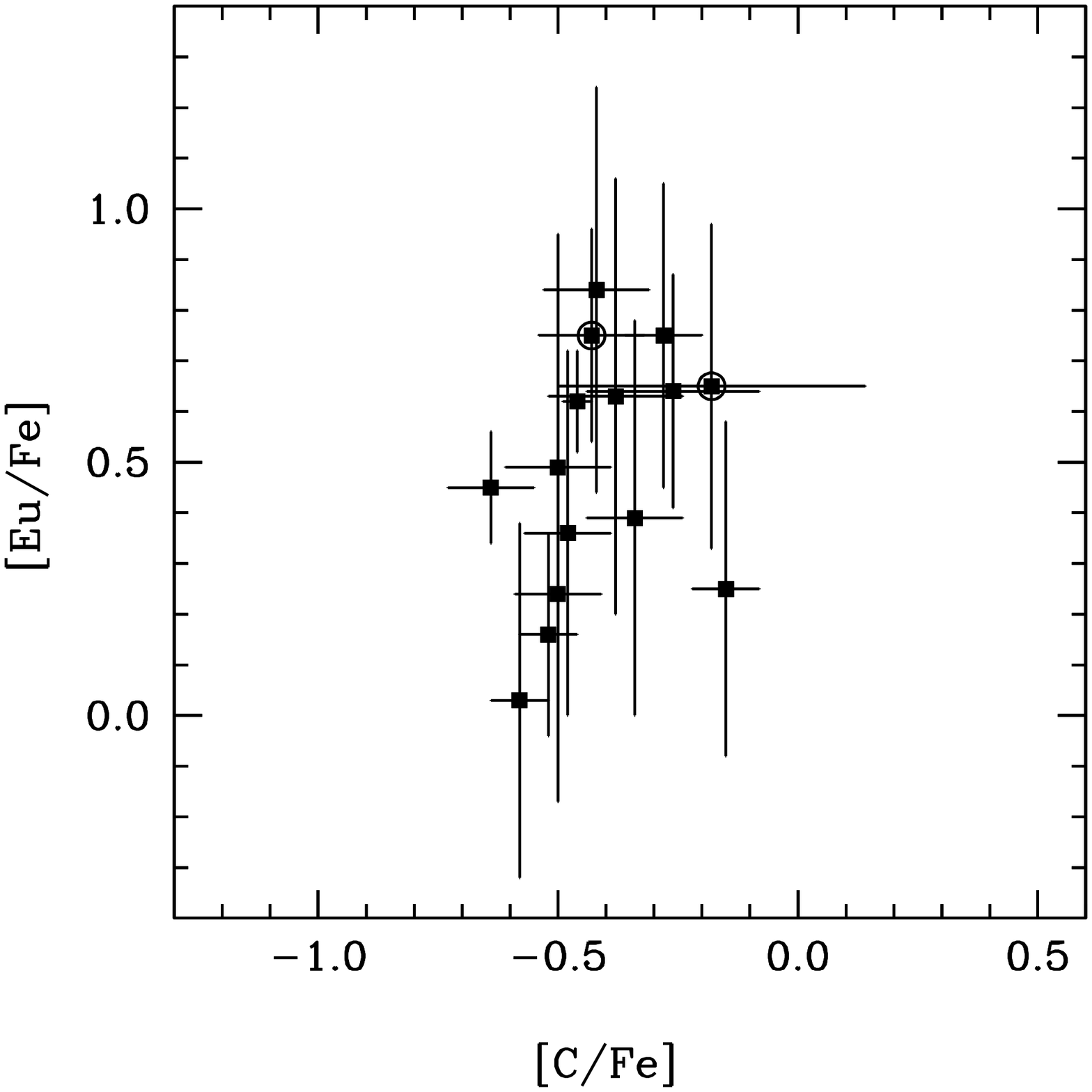} \hspace*{0.1in}
\includegraphics[angle=0,width=2.0in]{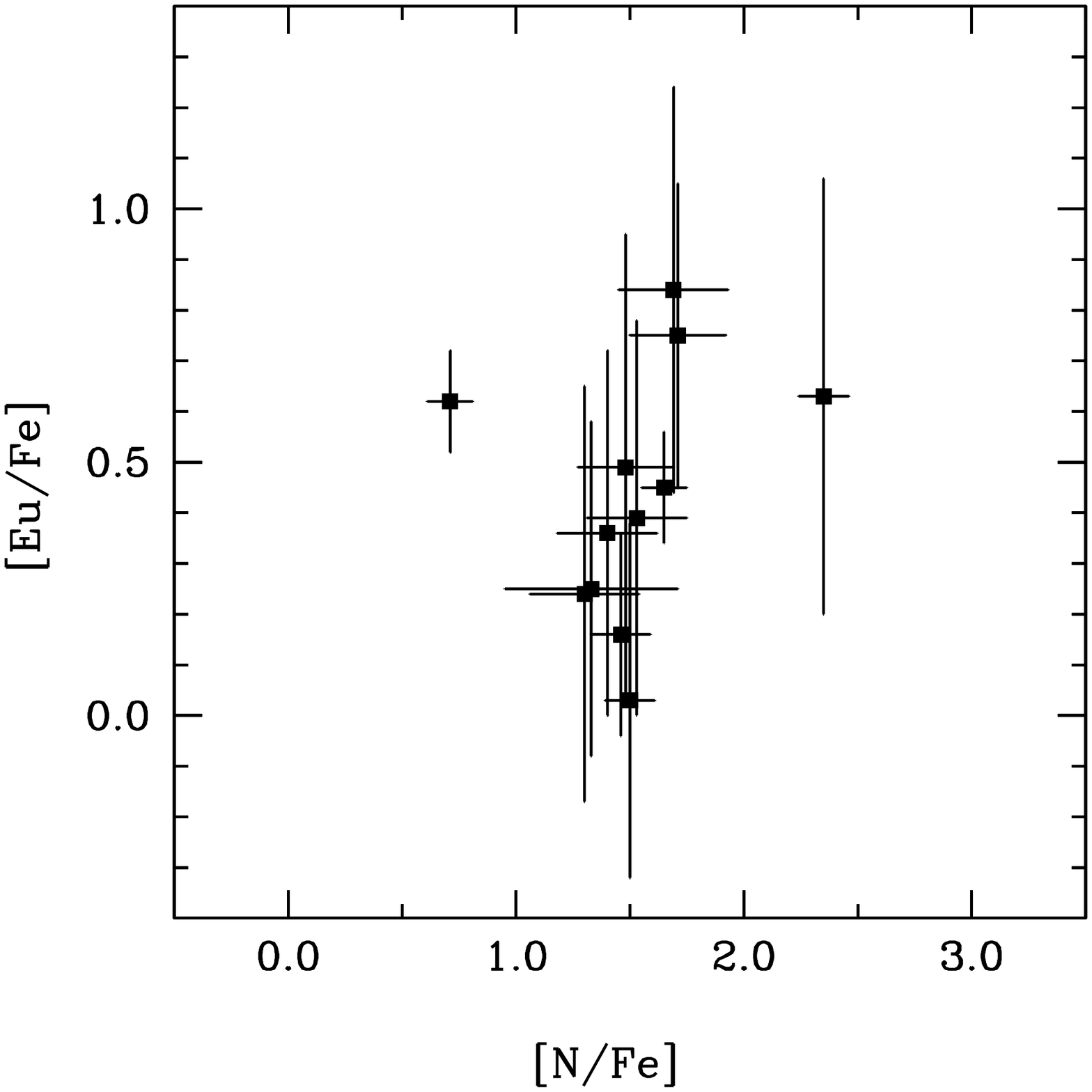} \hspace*{0.1in}
\includegraphics[angle=0,width=2.0in]{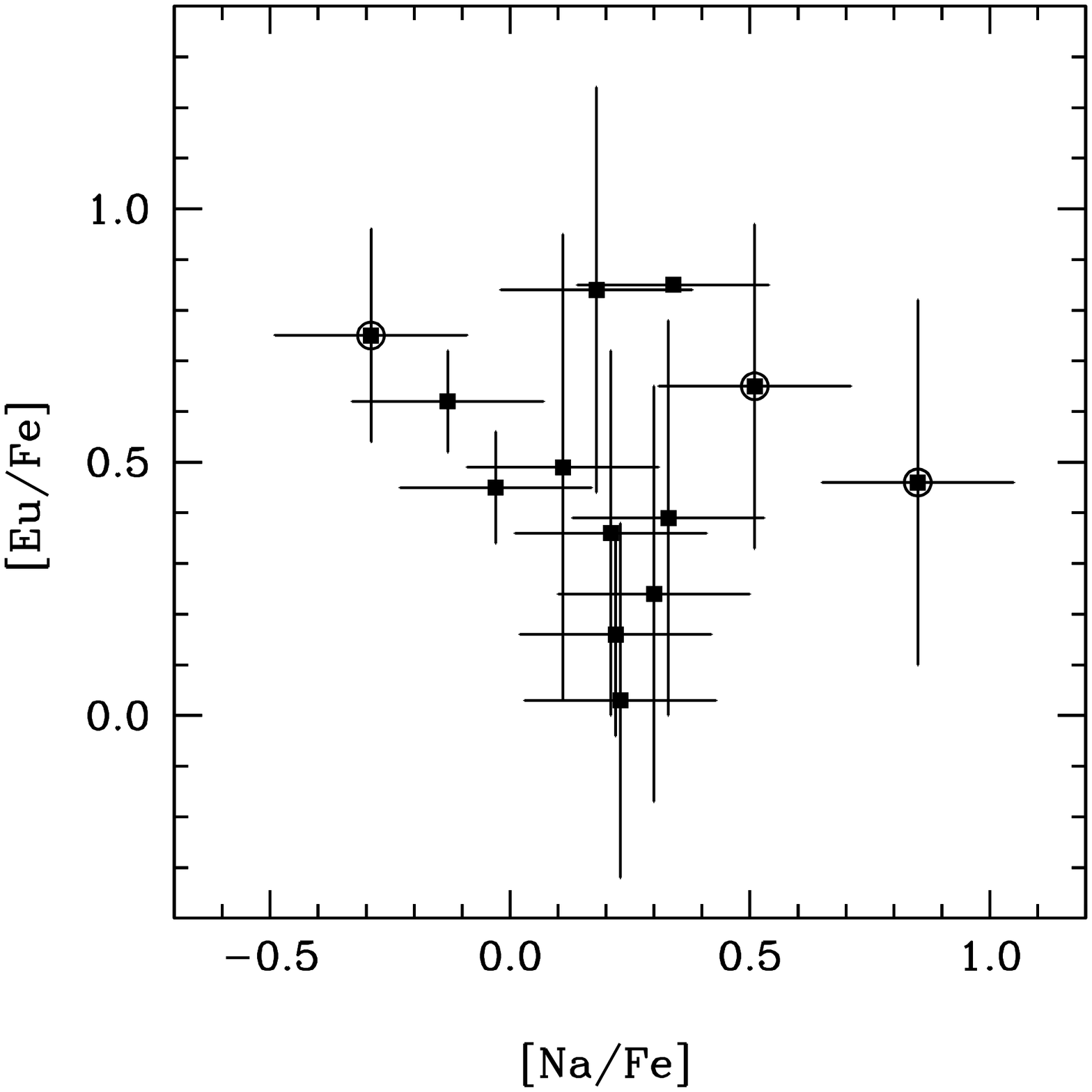} \\
\end{center}
\clearpage
\caption{
\label{comparec}
%\scriptsize
Comparison of the [La/Fe] and [Eu/Fe] ratios against [C/Fe],
[N/Fe], and [Na/Fe].
The [Na/Fe] ratios are taken from \citet{sneden00}.
Open circles indicate probable AGB stars.
None of these ratios exhibits a significant correlation.
Only the internal (i.e., star-to-star) uncertainties are shown.
}
\end{figure*}

If this is the only source of dispersion, 
we would also expect these ratios to be 
uncorrelated with one another.
Figure~\ref{comparev} demonstrates that this holds true
for Ti~\textsc{ii} and V~\textsc{ii}.
The linear correlation coefficient $r$ (e.g., \citealt{bevington03})
for these variables is 0.31, indicating a probability $P_{c}$ of 20\%
that the [Ti~\textsc{ii}/Fe~\textsc{i}] and [V~\textsc{ii}/Fe~\textsc{i}]
ratios could have come from an uncorrelated parent population.
This correlation is not significant.
Figure~\ref{comparev} indicates otherwise for 
La~\textsc{ii} and Eu~\textsc{ii}.
[Eu~\textsc{ii}/Fe~\textsc{i}] and [La~\textsc{ii}/Fe~\textsc{i}]
correlate with one another, yielding $r =$~0.49 and a
probability of only 5\% that they were drawn from an 
uncorrelated parent population.
This is perhaps moderately significant.

Figure~\ref{compareho} illustrates that La~\textsc{ii} and
Eu~\textsc{ii} also correlate strongly with Ho~\textsc{ii}.
Furthermore, La~\textsc{ii} and Eu~\textsc{ii} each 
correlate strongly with the Ba~\textsc{ii} abundances 
derived by \citet{sneden00}, also shown in Figure~\ref{compareho}.
This is an important point since the Sneden et al.\ [Ba/Fe] 
ratios were derived from an analysis that 
obtained the stellar model atmosphere parameters
from a different set of photometry (and color-\teff\ relations)
than we have used.
Such a correlation is unlikely to emerge unless it is
a genuine feature of the stars being studied.
These correlation coefficients 
are also listed in Table~\ref{correlatetab}.

Figure~\ref{overplotspec} demonstrates that dispersion
in the heavy \ncap\ elements is discernible without a
rigorous abundance analysis.  
Two spectra are shown covering the 
wavelength regions surrounding several of the 
lines used in our analysis.
These stars, 
\mbox{XII-8} and \mbox{XI-80}, have nearly identical
($V-K$) colors. % (hence \teff\ and \logg) as one another.
Their Fe-group elements have similar strengths, indicating
that the photometry and \teff\ are not in serious error.
The lines of Eu~\textsc{ii} and La~\textsc{ii} in 
\mbox{XII-8} are stronger than those in \mbox{XI-80},
suggesting an intrinsic difference.
Note that the Ti~\textsc{ii}, Y~\textsc{ii}, and Zr~\textsc{ii}
lines are not significantly different, and the Sc~\textsc{ii} 
line shows the opposite effect as La~\textsc{ii} and Eu~\textsc{ii}.
If one of these stars lies on the RGB and the other on the AGB,
we could expect to see consistently different line strengths
in all ionized species, which is not the case.

Random uncertainties in our estimation of \teff\
(or $V-K$), \logg, $v_{t}$, or [M/H] 
for many stars in the sample
could in principle lead to correlated ratios of [La/Fe] and [Eu/Fe].
We can exclude this explanation 
according to the results of our tests presented in Table~\ref{deltaatmtab}.
To account for a dispersion of 0.3~dex in [La/Fe] (0.4~dex in [Eu/Fe]),
corresponding to about half the full range of [La/Fe] ([Eu/Fe])
observed in M92, would require 
$\Delta \teff \approx$~500~K (570~K)\footnote{
This corresponds to $\Delta (V-K) \approx$~0.5~mag.},
$\Delta \logg \approx$~1.1 (3.0),
$\Delta v_{t} \approx$~4.5~\kmsec---which is clearly non-physical, and
$\Delta$[M/H]~$\approx$~1.2~dex.
Uncertainties are not expected to be linear and 
are certainly correlated, but these estimates 
are illuminating.
Furthermore, this scatter would be minimized when comparing
ratios of La~\textsc{ii} or Eu~\textsc{ii} to Ti~\textsc{ii} or V~\textsc{ii}
since these species all respond similarly to changes in the 
atmosphere; even so, the correlations between [La/Ti,V] and [Eu/Ti,V]
are still highly significant
(see Table~\ref{correlatetab}).
We conclude that it is extremely unlikely 
that random scatter in the photometry or relative
model atmosphere parameters can account for the 
observed dispersion and correlation in [La/Fe] and [Eu/Fe].

In conclusion, several lines of evidence each point to 
intrinsic star-to-star dispersion in the heavy 
\ncap\ element abundances in the red giants we have studied in M92.

\subsection{Examining Correlations with Light Element Dispersion}
\label{lightelements}

\begin{deluxetable}{ccccc}
\tablecaption{Impact of Model Atmosphere Uncertainties on Abundances
\label{deltaatmtab}}
\tablewidth{0pt}
\scriptsize
\tablehead{
\colhead{Species} &
\colhead{$\Delta T =$} &
\colhead{$\Delta \logg =$} &
\colhead{$\Delta v_{t} =$} &
\colhead{$\Delta {\rm [M/H]} =$} \\
\colhead{} &
\colhead{$\pm$~100~K} &
\colhead{$\pm$~0.3} &
\colhead{$\pm$~0.3~\kmsec} &
\colhead{$\pm$~0.4~dex} }
\startdata
\mbox{log$\epsilon$(Ti~\textsc{ii})} & $\pm$0.02 & $\pm$0.09 & $\mp$0.15 & $\pm$0.01 \\
\mbox{log$\epsilon$(V~\textsc{ii})}  & $\pm$0.02 & $\pm$0.13 & $\mp$0.05 & $\pm$0.05 \\
\mbox{log$\epsilon$(Fe~\textsc{i})}  & $\pm$0.10 & $\pm$0.01 & $\mp$0.03 & $\pm$0.05 \\
\mbox{log$\epsilon$(La~\textsc{ii})} & $\pm$0.04 & $\pm$0.09 & $\mp$0.01 & $\pm$0.05 \\
\mbox{log$\epsilon$(Eu~\textsc{ii})} & $\pm$0.03 & $\pm$0.05 & $\mp$0.01 & $\pm$0.05 \\
\enddata
\end{deluxetable}

\begin{deluxetable}{ccccc}
\tablecaption{M92 Mean Abundance Ratios
\label{meantab}}
\tablewidth{0pt}
%\scriptsize
\tablehead{
\colhead{Ratio} &
\colhead{Mean} &
\colhead{$\sigma_{\mu}$} &
\colhead{$\sigma$} &
\colhead{$N$} }
\startdata
\mbox{[C/Fe]}              & $-$0.40 & 0.04 & 0.16 & 18 \\
\mbox{[N/Fe]}              & $+$1.55 & 0.10 & 0.36 & 14 \\
\mbox{[Si~\textsc{i}/Fe]}  & $+$0.63 & 0.04 & 0.16 & 19 \\
\mbox{[Sc~\textsc{i}/Fe]}  & $+$0.16 & 0.04 & 0.07 &  4 \\
\mbox{[Sc~\textsc{ii}/Fe]} & $+$0.62 & 0.03 & 0.09 &  8 \\
\mbox{[Ti~\textsc{i}/Fe]}  & $+$0.08 & 0.03 & 0.11 & 19 \\
\mbox{[Ti~\textsc{ii}/Fe]} & $+$0.71 & 0.03 & 0.12 & 19 \\
\mbox{[V~\textsc{ii}/Fe]}  & $+$0.53 & 0.02 & 0.10 & 19 \\
\mbox{[Cr~\textsc{i}/Fe]}  & $-$0.38 & 0.08 & 0.16 &  4 \\
\mbox{[Mn~\textsc{i}/Fe]}  & $-$0.28 & 0.04 & 0.16 & 19 \\
\mbox{[Fe~\textsc{i}/H]}\tablenotemark{a} & $-$2.70 & 0.03 & 0.14 & 19 \\
\mbox{[Co~\textsc{i}/Fe]}  & $+$0.15 & 0.03 & 0.14 & 19 \\
\mbox{[Ni~\textsc{i}/Fe]}  & $-$0.10 & 0.03 & 0.08 &  9 \\
\mbox{[Y~\textsc{ii}/Fe]}  & $-$0.07 & 0.03 & 0.12 & 19 \\
\mbox{[Zr~\textsc{ii}/Fe]} & $+$0.47 & 0.03 & 0.14 & 19 \\
\mbox{[La~\textsc{ii}/Fe]} & $+$0.36 & 0.04 & 0.17 & 19 \\
\mbox{[Ce~\textsc{ii}/Fe]} & $+$0.46 & 0.04 & 0.07 &  4 \\ 
\mbox{[Nd~\textsc{ii}/Fe]} & $+$0.43 & 0.08 & 0.20 &  6 \\
\mbox{[Eu~\textsc{ii}/Fe]} & $+$0.54 & 0.06 & 0.23 & 16 \\
\mbox{[Ho~\textsc{ii}/Fe]} & $+$0.56 & 0.09 & 0.28 &  9 \\
\mbox{[Er~\textsc{ii}/Fe]} & $+$0.76 & 0.07 & 0.17 &  7 \\
\enddata
\tablenotetext{a}{The absolute value here is tied to the
[Fe~\textsc{i}/H] derived for \mbox{XII-8}.  
The undertainties on that quantity are 
$\sigma_{\mu} =$~0.06 and $\sigma =$~0.18.
See Section~\ref{abund} and Appendix~\ref{appendix}
for details.}
\end{deluxetable}

\begin{deluxetable*}{cccccc}
\tablecaption{Correlations among Abundance Ratios
\label{correlatetab}}
\tablewidth{0pt}
%\rotate
\tabletypesize{\scriptsize}
\tablehead{
\colhead{} &
\multicolumn{2}{c}{All Stars} &
\colhead{} &
\multicolumn{2}{c}{Excluding Probable AGB Stars} \\
\cline{2-3} \cline{5-6}
\colhead{} &
\colhead{[La~\textsc{ii}/Fe~\textsc{i}]} &
\colhead{[Eu~\textsc{ii}/Fe~\textsc{i}]} &
\colhead{} &
\colhead{[La~\textsc{ii}/Fe~\textsc{i}]} &
\colhead{[Eu~\textsc{ii}/Fe~\textsc{i}]} }
\startdata
\mbox{[C/Fe~\textsc{i}]}                               & ($-$0.11, 18, 0.66) & (0.35, 15, 0.21)    & & ($-$0.24, 15, 0.39) & (0.32, 13, 0.28) \\
\mbox{[N/Fe~\textsc{i}]}                               & ($-$0.22, 14, 0.45) & (0.27, 12, 0.40)    & & ($-$0.36, 13, 0.23) & (0.25, 12, 0.43) \\
\mbox{[Na~\textsc{i}/Fe~\textsc{i}]}\tablenotemark{a}  & ($-$0.02, 13, 0.95) & ($-$0.24, 12, 0.45) & & ($-$0.13, 10, 0.71) & ($-$0.46, 9, 0.21) \\
\mbox{[Ba~\textsc{ii}/Fe~\textsc{i}]}\tablenotemark{a} & (0.78, 13, 0.0015)  & (0.66, 12, 0.020)   & & (0.57, 10, 0.08)    & (0.78, 9, 0.014) \\
\mbox{[La~\textsc{ii}/Fe~\textsc{i}]}                  & \nodata             & (0.49, 16, 0.052)   & & \nodata             & (0.51, 13, 0.075) \\
\mbox{[Eu~\textsc{ii}/Fe~\textsc{i}]}                  & (0.49, 16, 0.052)   & \nodata             & & (0.51, 13, 0.075)   & \nodata          \\
\mbox{[Ho~\textsc{ii}/Fe~\textsc{i}]}                  & (0.81, 9, 0.0080)   & (0.72, 8, 0.042)    & & (0.36, 6, 0.48)     & (0.68, 6, 0.13)  \\
\hline
                  & \mbox{[La~\textsc{ii}/Ti~\textsc{ii}]} & \mbox{[La~\textsc{ii}/V~\textsc{ii}]} & & \mbox{[La~\textsc{ii}/Ti~\textsc{ii}]} & \mbox{[La~\textsc{ii}/V~\textsc{ii}]} \\
\hline
\mbox{[Eu~\textsc{ii}/Ti~\textsc{ii}]}                 & (0.69, 16, 0.0033)  & \nodata             & & (0.66, 13, 0.013)   & \nodata          \\
\mbox{[Eu~\textsc{ii}/V~\textsc{ii}]}                  & \nodata             & (0.73, 16, 0.0014)  & & \nodata             & (0.80, 13, 0.0011) \\
\enddata
\tablenotetext{a}{\citet{sneden00}}
\tablecomments{Each set of data indicates $r$, $N$, and $P_{c}(r; N)$.
If two element ratios of a parent distribution are uncorrelated, 
the probability that a random sample of $N$ stars will yield a
correlation coefficient $\geq |r|$ is 
given by $P_{c}(r; N)$.
}
\end{deluxetable*}

\begin{figure}
\begin{center}
\includegraphics[angle=0,width=3.41in]{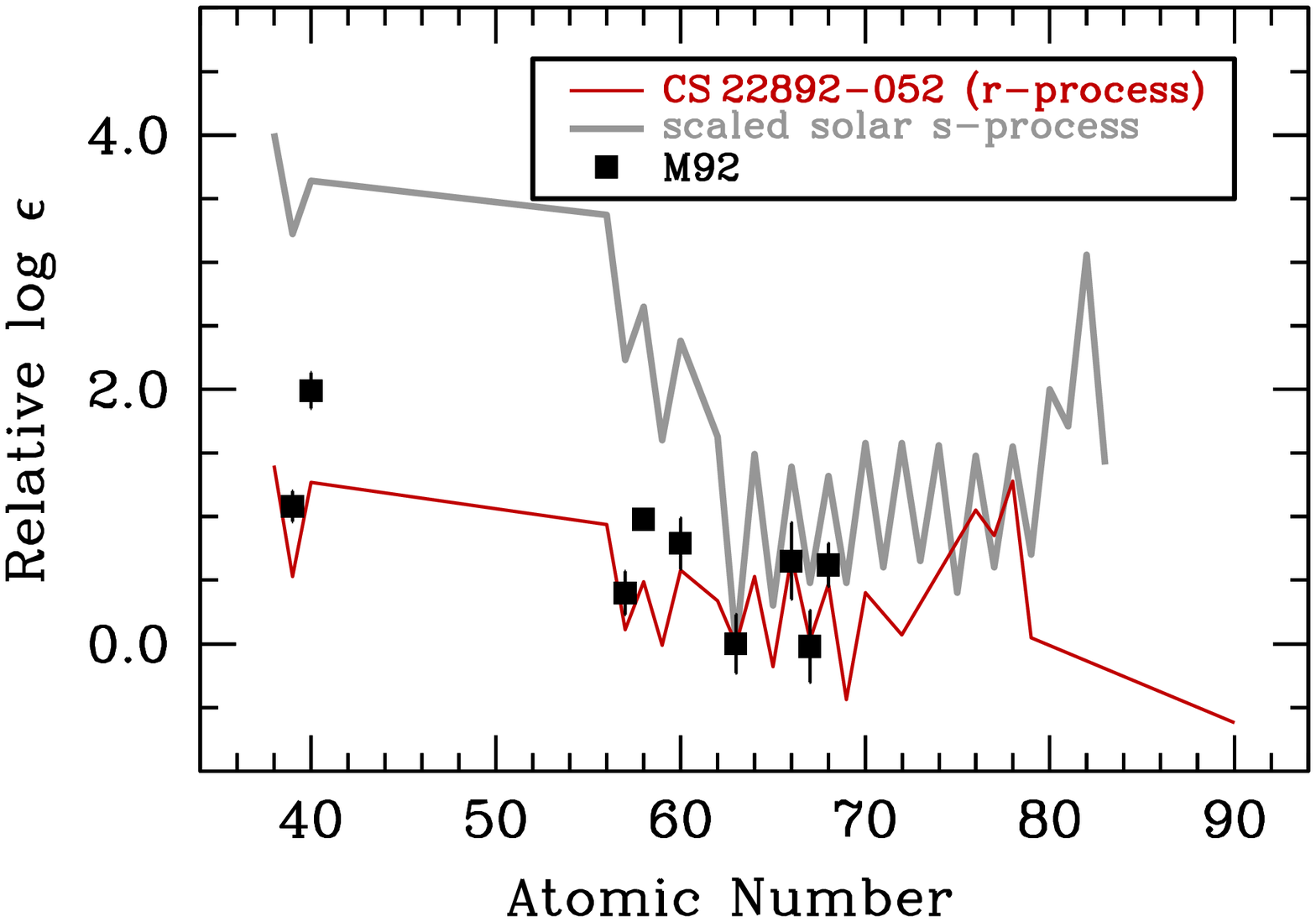} \\
\vspace*{0.2in}
\includegraphics[angle=0,width=3.41in]{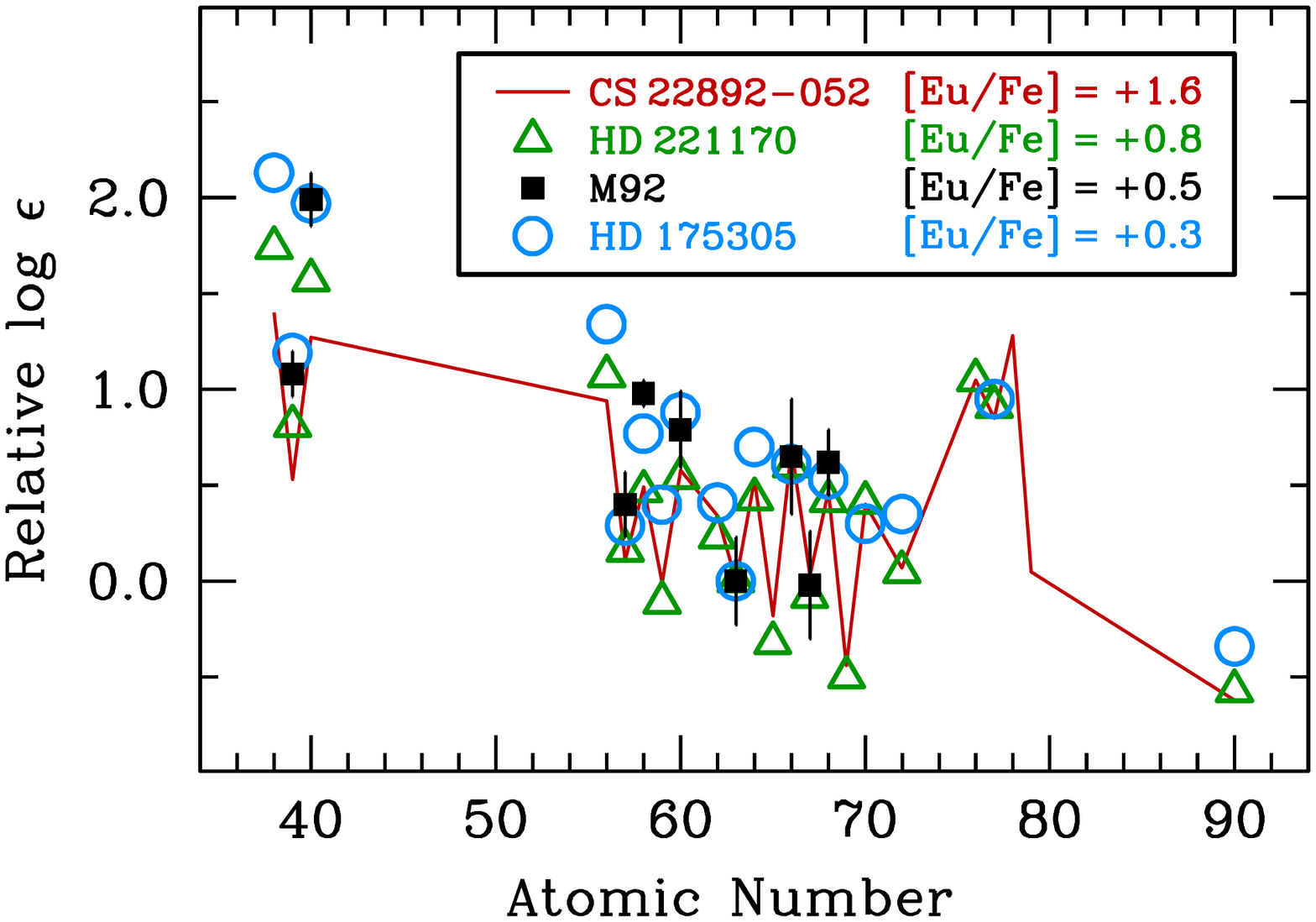}
\end{center}
\caption{
\label{rproplot}
\scriptsize
\textit{Top}: 
Comparison of the mean heavy element abundances in M92 with
the metal-poor \rpro\ standard star \mbox{CS~22892--052}
and the scaled S.S.\
\spro\ pattern.
The abundances are normalized to Eu.
\textit{Bottom}:
Comparison of the mean heavy element abundances in M92 with
three metal-poor field stars with differing levels of
\rpro\ enrichment.
The abundances are normalized to Eu.
The M92 abundance pattern is very similar to that in 
\mbox{HD~175305}, which has a similar level of [Eu/Fe]
and similar ratios of [Y/Eu] and [Zr/Eu].
Since all abundances are normalized to Eu, 
only the internal (i.e., star-to-star) uncertainties are shown.
}
\end{figure}

Does the \ncap\ element dispersion correlate with the 
light element dispersion?
Figure~\ref{comparec} shows [La/Fe] and [Eu/Fe]
as a function of [C/Fe].
Neither exhibits a significant correlation
(see Table~\ref{correlatetab}).
Figure~\ref{comparec} also shows [La/Fe] and [Eu/Fe]
as a function of [N/Fe], 
and again there is no significant correlation.
\citet{sneden00} derived [Na/Fe] ratios for 13 and 12 stars 
whose [La/Fe] and [Eu/Fe] ratios we have derived.
Figure~\ref{comparec} illustrates that no correlation exists
between either [La/Fe] or [Eu/Fe] and [Na/Fe].

\citet{smith08} points out that in a limited number of 
GCs, including M92 and M15, 
the AGB stars with the highest [Na/Fe] ratios are often 
those with the highest [Ba/Fe] ratios.
It is evident from Figures~\ref{comparev}--\ref{compareho}
that the 4 probable RHB/AGB stars we have analyzed in M92
often are among those with the highest [Ba/Fe], [La/Fe], 
[Eu/Fe], and [Ho/Fe] ratios.
Several authors whose GC data was 
reexamined by Smith
noted that their [Ba/Fe] ratios may be unreliable in
the AGB stars. 
This could arise if the strong, saturated
Ba~\textsc{ii} lines are formed (at least in part) in the 
chromospheric layers not accounted for
in the models \citep{shetrone00} or 
if the microturbulent velocity parameter derived from 
Fe~\textsc{i} lines is not appropriate for the
Ba~\textsc{ii} line-forming layers \citep{ivans01}.\footnote{
\citet{ivans01} also note that their M5 [La/Eu] ratios are lower
for the AGB stars than the RGB stars,
$\langle$[La/Eu]$\rangle_{\rm AGB} = -$0.46~$\pm$~0.05 ($\sigma =$~0.12)
and
$\langle$[La/Eu]$\rangle_{\rm RGB} = -$0.37~$\pm$~0.05 ($\sigma =$~0.18).
This is the opposite sense of what would be expected if 
the AGB stars contained a larger fraction of \spro\ material
than the RGB stars.
From this Ivans et al.\ conclude that the enhanced [Ba/Fe]
ratios in their AGB stars are not likely due to \spro\ enrichment.}
Both explanations are likely true to some extent and
may apply to other analyses,
and we regard the evidence for \spro\ self-enrichment
in low mass AGB stars as inconclusive.

Nevertheless, we conservatively reexamine the \ncap\
dispersion in M92 with the AGB stars excluded.
Table~\ref{correlatetab} lists the correlation coefficients and probabilities
for the remaining 15 stars.
There is still no evidence that [La/Fe] or [Eu/Fe] correlate
with [C/Fe], [N/Fe], or [Na/Fe].
The correlations among the heavy \ncap\ elements are still
significant, though generally less so because fewer stars
are included.
The 
[La~\textsc{ii}/Ti~\textsc{ii}] versus
[Eu~\textsc{ii}/Ti~\textsc{ii}] and
[La~\textsc{ii}/V~\textsc{ii}] versus
[Eu~\textsc{ii}/V~\textsc{ii}]
correlations are still highly significant.

In summary, 
these data indicate that the \ncap\ dispersion in M92 is robust and
independent of the light element dispersion.

\section{Discussion}
\label{discussion}

We have demonstrated that the heavy \ncap\ abundances vary 
together relative to Fe in M92.
What is the nucleosynthetic origin of the \ncap\ material, 
and how does this phenomenon relate to the \rpro\ dispersion
observed in M15?
Other matters concerning the astrophysical mechanism(s)
that lead to star-to-star dispersion are not so straightforward.
In this section we discuss each of these matters.

\subsection{The $r$-process Abundance Pattern in M92}
\label{rpropattern}

Figure~\ref{rproplot} shows the abundance distribution for the 
$Z \geq$~39 elements in M92.\footnote{
In an effort to detect weak lines of additional \ncap\ species,
we have co-added the spectra of 21 individual stars on the RGB
with 4730~$\leq \teff \leq$~5080~K and
1.7~$\leq \logg \leq$~2.1.
This combined spectrum has S/N~$\sim$~270 at 4000\AA.
Unfortunately, we are only able to detect one new line
from this spectrum, Dy~\textsc{ii} 3944.68\AA.
From this spectrum we determine a mean 
$\log \epsilon$~(Dy/La)~$= +$0.25~$\pm$~0.3.
This value is shown in Figure~\ref{rproplot}.}
In the top panel, we compare the M92 abundances to that of the 
S.S.\ \spro\ pattern \citep{sneden08} and the 
\rpro\ standard star \mbox{CS~22892--052}
\citep{sneden03,sneden09}.
The M92 abundances clearly resemble \rpro\ nucleosynthesis more than
\spro\ nucleosynthesis.
In the bottom panel, the M92 abundances are compared 
with three metal-poor $r$-enriched field stars, 
\mbox{CS~22892--052},
\mbox{HD~221170} \citep{ivans06,sneden09}, and
\mbox{HD~175305} \citep{roederer10a}.
These three comparison stars have a range of heavy element abundances
that effectively bracket the mean [Eu/Fe] ratio of M92, and 
\citet{roederer09,roederer10b} demonstrated that the low [Pb/Eu] ratios
(or upper limits) in these three stars suggest that they contain
no detectable trace of \spro\ material. 
Ba, Ce, and Nd in \mbox{HD~221170} 
and \mbox{HD~175305} are slightly higher than their abundances
in \mbox{CS~22892--052} when normalized to Eu,
which \citet{roederer10b} argued to be a result of
intrinsic variations in \rpro\ nucleosynthesis, perhaps a result of
different physical conditions at the nucleosynthesis site.
M92 has an abundance pattern nearly identical to that of
\mbox{HD~175305}, which has a similar [Eu/Fe] ratio, $+$0.35~$\pm$~0.15,
as what we have derived for M92. 
Other heavy element abundances are similar between these two stars.
Figure~\ref{rproplot} implies that the heavy elements ($Z >$~56)
in M92 originated in an \rpro.

At low metallicity, the \spro\ produces large [Pb/Eu] ratios
due to the high ratio of neutrons 
to Fe-group seed nuclei (e.g., \citealt{clayton88,gallino98}), so
[Pb/Eu] is a good diagnostic of \spro\ material
in metal-poor stars.
Our spectra just miss the Pb~\textsc{i} line at 4057\AA.
\citet{shetrone01} obtained high resolution 
blue spectra with Keck HIRES of two 
stars in M92, \mbox{III-13} and \mbox{III-65}. 
From these spectra (M.\ Shetrone, 2011, private communication)
we use the Eu~\textsc{ii} 4129\AA\ line (detected) and the Pb~\textsc{i}
4057\AA\ line (not detected) to derive an approximate upper limit on 
Pb, [Pb/Eu]~$\lesssim +$0.3.
This is low enough \citep{roederer10b} to rule out contributions from
low-metallicity intermediate-mass 
AGB stars to the gas from which the M92 stars formed,
reinforcing our assertion that the
$Z >$~56 material in M92 originated only in an \rpro.

Nucleosynthesis of the Sr-Y-Zr group of elements is more complex.
While only true \ncap\ processes can produce elements
heavier than the $A \simeq$~130 peak in significant quantities, 
several other charged-particle reaction mechanisms
like the $\nu p$ process or $\alpha$-rich freezeout
may also contribute to---if not dominate 
production of---the lighter Sr-Y-Zr group
(e.g., \citealt{woosley94}, \citealt{freiburghaus99}, 
\citealt{frohlich06}, \citealt{arcones10}, \citealt{farouqi10}).
In M92, these elements display a dispersion similar to that
of the Fe-group elements. 
The predictable nature of the [Y/Eu] or [Zr/Eu] ratios based on
the [Eu/Fe] ratio (inferred from Figure~\ref{rproplot})
is therefore a consequence of rather similar [Y/Fe] and [Zr/Fe] ratios.
The fact that the heavy (La--Er) elements in M92 have a larger
dispersion than Y or Zr implies that these groups
are produced mainly by different nucleosynthetic mechanisms.

\subsection{Comparison with Globular Cluster M15}
\label{m15}

\begin{figure}
\begin{center}
\includegraphics[angle=0,width=3.4in]{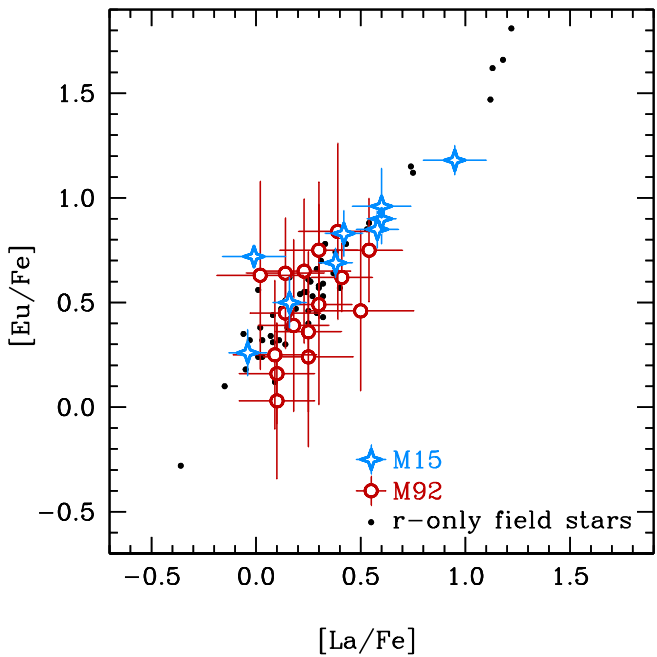}
\end{center}
\caption{
\label{m92m15plot}
Comparison of the [Eu/Fe] and [La/Fe] ratios (and total
uncertainties) in M92 with
those in M15 \citep{sobeck11} and a sample of field stars
whose only enrichment in La and Eu has come from \rpro\ nucleosynthesis
\citep{roederer10b}.
}
\end{figure}

M15 is more massive than M92 
(7.6~$\times$~10$^{5}$~\msun\ and 3.1~$\times$~10$^{5}$~\msun, respectively,
assuming $M/L_{V} =$~2~\msun/\lsun),
but both have nearly identical metallicities and ages.
\citet{sneden97} identified a dispersion
in [Ba/Fe] in M15.
\citet{sneden00}, \citet{otsuki06}, and \citet{sobeck11}
confirmed this result and extended it to all 
$Z \geq$~56 elements that have been studied in M15.
These studies have shown that the heavy elements in M15 
are produced by \rpro\ nucleosynthesis, and there are 
no hints of \spro\ contamination.
M15 is the only other GC where %an unambiguous
dispersion of \rpro\ material has been reported.

Figure~\ref{m92m15plot} compares [La/Fe] and [Eu/Fe]
in M15 and M92.
Both ratios vary over a wide range in each GC
($-$0.1~$<$~[La/Fe]~$< +$0.9 and 
 $+$0.2~$<$~[Eu/Fe]~$< +$1.2 in M15, 
    0.0~$<$~[La/Fe]~$< +$0.6 and
    0.0~$<$~[Eu/Fe]~$< +$0.9 in M92,
but note that the ranges of M15 ratios 
include a systematic offset of 0.35~dex
between the RGB and RHB stars examined by \citealt{sobeck11}).
[La/Fe] and [Eu/Fe] correlate with each other
in both M15 and M92, similar to the correlation found 
in metal-poor halo field stars enriched by \rpro\ nucleosynthesis.
\citet{armosky94} and \citet{sneden00} showed that the mean [Ba/Fe] ratios
are higher in M15 than in M92, but the [Ba/Eu] ratios in both
GCs reflect only \rpro\ nucleosynthesis.
Based on comparisons between our results and previous studies
(Section~\ref{compare}), we confirm that the mean [Eu/Fe] ratio in M92
is lower than that in M15.
These data do not suggest why M15
contains a larger mean \rpro\ overabundance relative to M92.
Nevertheless, it is clear that the stars in both M15 and M92
exhibit a range of \rpro\ abundances.

\citet{sneden97} looked for and found no correlation between
the light (among C, N, O, Na, Mg, and Al) and heavy element dispersion 
(among Ba and Eu) in M15.
We confirm that M92 behaves similarly.
The data imply that the \rpro\ dispersion in M15 and M92 was
imprinted in the gas from which all present-day GC stars, 
including those of the second generation, were formed.

\subsection{Explanations for the Origin of the $r$-process Dispersion}

Some mechanism(s) must account for the ability of M15 and M92
to arrive at a homogeneous set of Ca and Fe-group abundances
and an inhomogeneous set of heavy \ncap\ abundances
before the source of the light element dispersion becomes
an active participant in the chemical evolution of these GCs.
Since the \rpro\ dispersion is also present in later 
generations of stars, some mechanism(s)
must also preserve this inhomogeneity over long ($\gtrsim$~10~Myr)
timescales (see also \citealt{dorazi10}).
%%%%%%
%% add this footnote to the resubmitted version.  clean the sentence up some.
%%
%\footnote{Furthermore, \citet{dorazi10} have demonstrated that
%there is no correlation between [Na/Fe] and [Ba/Fe] or between
%the Ba-rich/Ba-poor stars and [Na/O] in their sample of
%more than 1200~stars in 15~GCs.
%Thus, the disconnect between the mechanism(s) that
%produce the heavy \ncap\ elements and 
%those that produce the light element variations
%appears to be a common characteristic.}
%%%%%%

Despite the time that has passed since \citet{sneden97} first
reported an \rpro\ dispersion in M15, we are aware of no
published attempts to explain this phenomenon.
Variations in the [La/Eu] ratios---observed in 
$r$-only field stars and M92---suggests that
dilution of the yields from rare but identical \rpro\ events
cannot alone account for the inhomogeneous distribution of
\rpro\ material.
Both M15 and M92 are on moderately eccentric (but unrelated)
Galactic orbits,
and each is currently located near its apogalactic radius
(approximately 10~kpc, \citealt{dinescu99}).
Other massive GCs (e.g., $\omega$~Cen)
exhibit a complex variety of abundance patterns;
these GCs likely formed in much larger 
parent systems, since disrupted by the Milky Way,
that were capable of driving chemical evolution within themselves.
Neither M15 nor M92 has been associated with tidal debris from
a dwarf galaxy or stellar streams in the Galactic halo
(e.g., \citealt{smith09}),
and neither CMD exhibits multiple main sequences or subgiant branches.
We conclude that there is no convincing explanation at present 
for the observed \rpro\ dispersion in M15 and M92.

\section{Conclusions}
\label{conclusions}

We have obtained new high S/N spectra covering 3850--4050\AA\ 
for 19~stars in the metal-poor GC M92
using the Hydra spectrograph on the WIYN Telescope.
We perform a detailed differential abundance analysis 
and quantify the chemical homogeneity in M92
for 21~species of 19~elements from
carbon to erbium.
%The S/N in 19 of these stars is sufficient to perform a detailed
%differential abundance analysis and quantify the chemical
%homogeneity in M92 for 21~species of 19~elements from
%carbon to erbium.
Our main results are summarized as follows.

(1) These stars are chemically homogeneous at the level of 0.07--0.16~dex
for Sc, Ti, V, Cr, Fe, Co, Ni, Y, and Zr.
The absolute metallicity and [X/H] ratios should be
treated with caution, but
the ratios among metals are quite robust.

(2) The heavy \ncap\ elements La, Eu, and Ho are \textit{not}
chemically homogeneous throughout these 19~stars in M92.
The [La/Fe], [Eu/Fe], and [Ho/Fe] ratios have dispersions
of 0.17--0.28~dex 
and span ranges of 0.5--0.8~dex
(a factor of 3--6). 
This dispersion is not due to observational uncertainty since
these ratios correlate with each other and with the 
[Ba/Fe] ratios derived by \citet{sneden00}.

(3) The elements Y and Zr show dispersion similar to 
that of the Fe-group and less than that of Ba, La, Eu, and Ho.
This suggests that that the Y and Zr were not
formed primarily by \rpro\ nucleosynthesis and were more uniformly
mixed at the time of star formation.

(4) The heaviest elements originate in \rpro\ nucleosynthesis
without contributions from the \spro.
The \rpro\ dispersion does not correlate 
with the light element dispersion (C, N, and Na),
indicating that the \rpro\ dispersion was 
present in the gas throughout star formation.

(5) The \rpro\ dispersion in M92 is similar---but not identical to---that 
observed previously in the massive, metal-poor GC M15
(e.g., \citealt{sneden97}). 
Both GCs show unmistakable star-to-star dispersion of
\rpro\ material relative to Fe. 
The dispersion in M15 is larger and the mean \rpro\ level
is higher in M15 than in M92.
Sneden et al.\ demonstrated that 
the \rpro\ dispersion in M15, like M92, also does not correlate
with the light element dispersion.

There are at least two (perhaps several; \citealt{roederer11}) 
massive, metal-poor Milky Way GCs that formed
from material with inhomogeneous distributions of \rpro\ material.
At present there exists no explanation for the 
astrophysical mechanism(s) responsible for this phenomenon.
Attempts to understand and incorporate this
into the rapidly-evolving theory of GC formation and evolution
will surely prove rewarding.

\acknowledgments

We thank 
C.I.\ Johnson and C.\ Pilachowski for their helpful advice and
generous assistance in preparing for our observing run,
M.\ Shetrone for sharing his M92 spectra,
J.A.\ Johnson for useful discussions, 
J.\ Sobeck for conveying abundance results in advance of publication,
and the referee for a careful review of our work.
%A.\ McWilliam, G.\ Preston, and D.\ Yong for encouraging discussions
%throughout the course of this work.
We are extremely grateful to M.\ Spite and the ``First Stars'' team,
J.A.\ Johnson, as well as
G.\ Preston, S.\ Shectman, and I.\ Thompson
for permitting us to compare portions of their spectra with our own data.
This research has made use of the 
NASA Astrophysics Data System (ADS), the
NIST Atomic Spectra Database, and the 
Two Micron All Sky Survey, which is a joint project of the 
University of Massachusetts and the Infrared Processing and Analysis 
Center/California Institute of Technology, 
funded by the National Aeronautics and Space Administration 
and the National Science Foundation.
I.U.R.\ is supported by the Carnegie Institution of Washington 
through the Carnegie Observatories Fellowship.
C.S.\ is supported by the U.S.\ National Science Foundation 
(grant AST~09-08978).

{\it Facilities:} 
\facility{WIYN (Hydra)}

\appendix

\section{The Metallicity of M92}
\label{appendix}

Our absolute metallicity for M92 is anchored to the
Fe~\textsc{i} abundance in \mbox{XII-8}, 
the reference star used in our analysis.
The metallicities of all other M92 stars in our analysis have been 
computed differentially with respect to \mbox{XII-8}.
The mean metallicity derived from 19 RGB stars, 
[Fe~\textsc{i}/H]~$= -$2.70~$\pm$~0.03,
is lower by more than a factor of 2 than that derived from 33 RGB stars 
by \citet{sneden00}, 
[Fe~\textsc{i}/H]~$= -$2.34~$\pm$~0.01.
Based on equivalent width (EW) measurement of 
Fe~\textsc{i} or Fe~\textsc{ii} lines from high resolution spectra,
numerous studies over the last 20~years
have derived metallicities ranging from 
$-$2.4~$<$~[Fe/H]~$< -$2.1 for M92, though
\citet{peterson90} and \citet{king98} have presented evidence for 
[Fe/H]~$\lesssim -$2.5 in M92.

There are three significant differences between our study and \citet{sneden00}
that in principle may account for portions of this offset, including
(1) different laboratory sources for the Fe~\textsc{i} log($gf$) values
and different sets of Fe~\textsc{i} lines available for analysis;
(2) different grids of model atmospheres
(we use the most recent set of MARCS $\alpha$-enhanced models,
\citealt{gustafsson08}, while Sneden et al.\ used the 
\citealt{gustafsson75} set available at the time); and
(3) different versions of the MOOG code, with the most notable
difference being the explicit calculation of the Rayleigh scattering
contribution to the blue continuous opacity as described in \citet{sobeck11}.
We check each of these effects below.

\citet{obrian91a}, our preferred source for Fe~\textsc{i} log($gf$) values, 
did not report log($gf$) values 
for 5 of the 7~lines covered by Sneden et al.;
for the remaining 2~lines, the O'Brian et al.\ values are higher
by 0.12 and 0.19 dex.
Naively extrapolating these offsets 
suggests that the Sneden et al.\ Fe~\textsc{i} abundance could be
lower by 0.1--0.2~dex on the O'Brian et al.\ log($gf$) scale.
Rederiving the Fe~\textsc{i} abundance of \mbox{XII-8} 
using the Sneden et al.\ EWs and the MARCS model used in 
the present study (which accounts for differences in
model parameters and grids) decreases the abundance by 0.12~dex.
Rederiving the \mbox{XII-8} abundance from the two versions
of MOOG decreases the Sneden et al.\ abundance by 0.01~dex.
Together, these effects can
produce a decrease of $\sim$~0.2--0.3~dex in the abundance derived
by Sneden et al.
We have derived [Fe/H] lower by 0.55~dex for \mbox{XII-8},
so these effects can account for about half of the discrepancy.
The standard deviation of the 9 Fe~\textsc{i} lines we have
examined in \mbox{XII-8} is 0.18~dex
and Sneden et al.\ produced a standard deviation of
0.25~dex from 4 Fe~\textsc{i} lines.
This could, in principle, 
account for another significant portion of the discrepancy.

In Section~\ref{comments}
we found [Ti~\textsc{ii}/Fe~\textsc{i}] to be higher than
[Ti~\textsc{i}/Fe~\textsc{i}] by $\sim$~0.5~dex,
and in Section~\ref{compare} we found that our 
[X~\textsc{ii}/Fe~\textsc{i}] ratios were higher
by 0.2--0.6~dex than had been found in previous studies
of \mbox{VII-18}.
We have not forced Fe (or Ti) ionization equilibrium when deriving
our atmospheric parameters. 
The singly-ionized species are the dominant ones for
Fe-group elements in these stellar atmospheres, 
and neglecting to account for
departures from LTE in our analysis would tend to underestimate
the abundance of the neutral species.
By this reasoning it is
plausible that our [Fe~\textsc{i}/H] abundances have 
been underestimated by several tenths of a dex.

Adopting a different photometric temperature scale would 
not have altered our results significantly.
For these M92 stars , the \citet{alonso99} $V-K$ scale
predicts no significant difference for stars with 
\teff~$> 4650$~K, but for the cooler giants it predicts
temperatures systematically lower by 60--130~K.
For \mbox{XII-8}, Sneden et al.\ (who used
$B-V$ versus \teff\ relations derived by \citealt{carbon82})
derived \teff~$=$~4490~K.
Using the Alonso et al.\ scale we would derive 4380~K, 
and using the \citet{ramirez05b} scale we have derived 4450~K.
For \mbox{XII-8}, our tests indicate that 
$\Delta\teff = \pm$~100~K translates to
$\Delta$[Fe~\textsc{i}/H]~$= \pm$~0.10~dex.
This corresponds to [Fe/H] differences of $+$0.04~dex and
$-$0.08~dex with respect to Sneden et al.\ and
the Alonso et al.\ scale, respectively.
Thus, adopting any of these three temperature scales would
have produced similar metallicity results.

\begin{figure*}
\begin{center}
\includegraphics[angle=270,width=5.3in]{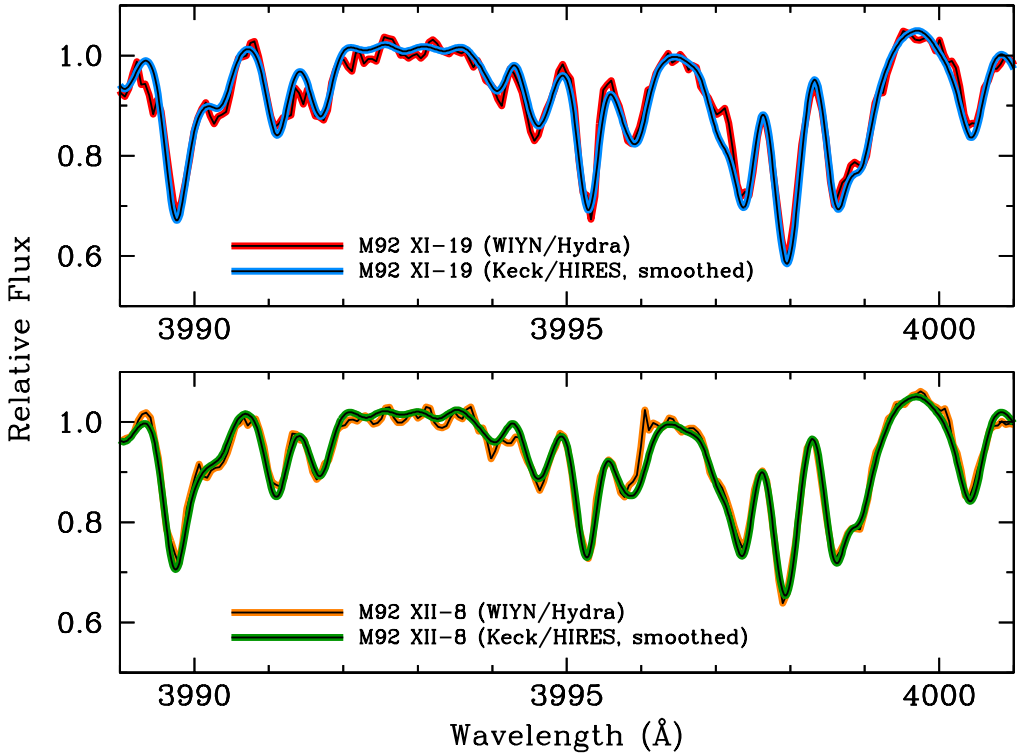}
\caption{
\label{jajplot}
Comparison of our WIYN/Hydra spectra for \mbox{XI-19} and \mbox{XII-8}
with Keck/HIRES spectra (J.A.\ Johnson, private communication, 2011)
that have been Gaussian smoothed to the same resolution.
Except for the lower S/N in our spectra, the two sets of spectra
are nearly identical.
}
\end{center}
\end{figure*}

We have compared our WIYN/Hydra spectra for two stars,
\mbox{XI-19} and \mbox{XII-8} (our reference star)
with spectra of these two stars obtained by J.A.\ Johnson
(2011, private communication) using Keck/HIRES.
Figure~\ref{jajplot} illustrates this comparison for 
a representative wavelength range.
We have smoothed the HIRES spectra down to our Hydra resolution.
Our spectra have lower S/N, but otherwise the spectra for
these two stars are essentially identical.
This gives us confidence that we have not made serious errors
during the extraction procedure (e.g., poor 
subtraction of sky or scattered light from the image frames).

Recently, others have found a similarly low metallicity for stars in M15:
[Fe~\textsc{i}/H]~$= -$2.66 and [Fe~\textsc{ii}/H]~$= -$2.60 
from six RHB stars \citep{preston06},
[Fe~\textsc{i}/H]~$= -$2.69 and [Fe~\textsc{ii}/H]~$= -$2.64
from the same six RHB stars or
[Fe~\textsc{i}/H]~$= -$2.56 and [Fe~\textsc{ii}/H]~$= -$2.53
from three RGB stars \citep{sobeck11}.
Tests conducted by Preston et al.\ and Sobeck et al.\ 
indicate that the persistent
metallicity offset between the RGB and RHB stars and the 
metallicity offset between the RGB stars studied by them and 
\citet{sneden97,sneden00}
are not a result of the choice of atmospheric parameters, line lists,
model atmosphere grids, or recent upgrades to MOOG.
The offset between the cool RGB stars and the 
warm RGB/RHB stars in our M92 study runs the opposite direction.
We note that
5 of the 6~RHB stars in M15 studied by Sobeck et al.\ 
are significantly warmer than our warmest star in M92,
so considering the \teff-[Fe/H] slopes
in this simple manner may not afford a fair comparison

\begin{figure*}
\begin{center}
\includegraphics[angle=270,width=5.3in]{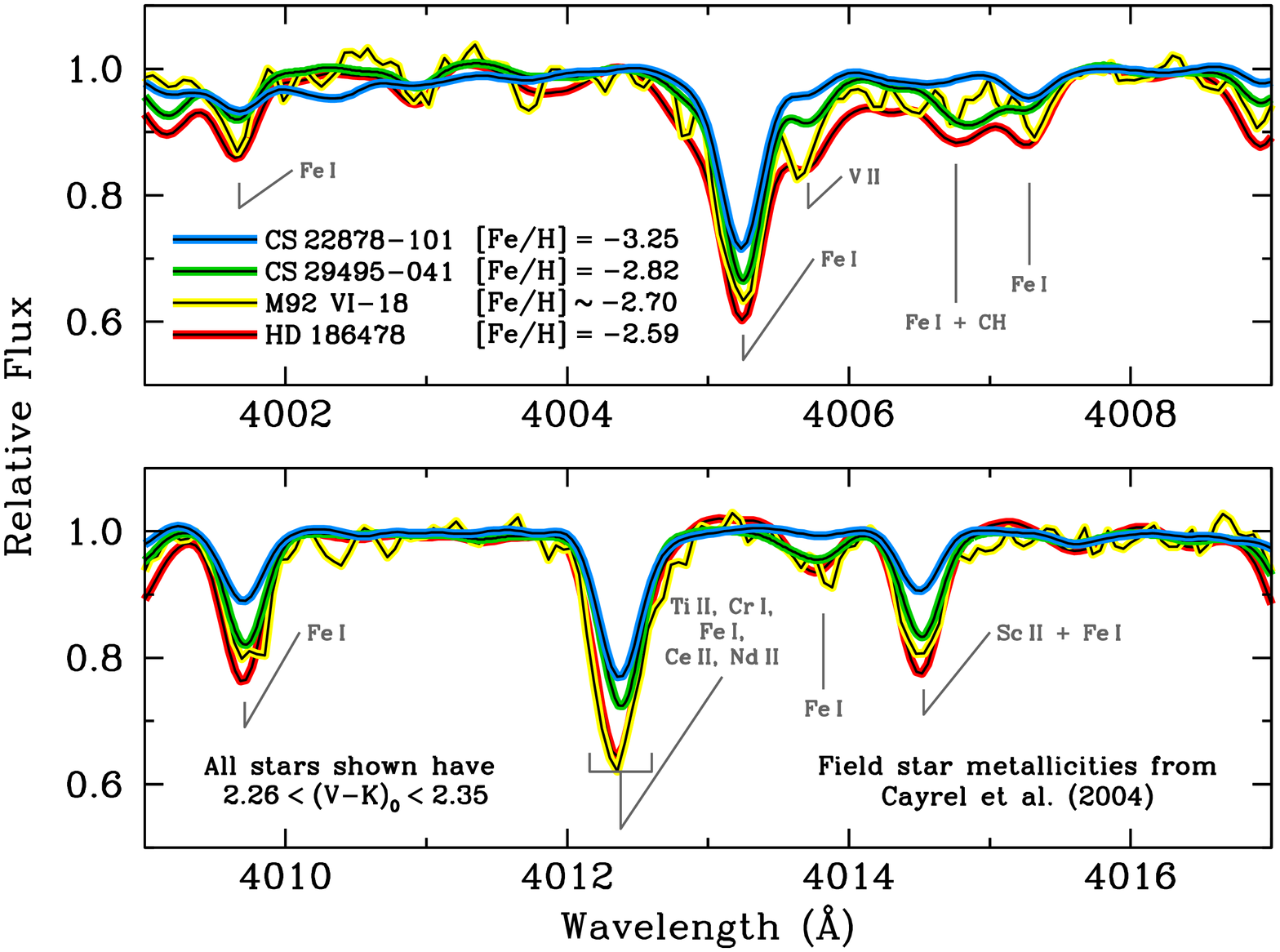}
\caption{
\label{metallicityplot}
Comparison of spectra for 4 stars with very similar colors but
differing metallicities.
The spectrum of \mbox{VI-18} was obtained in our study,
the spectrum of \mbox{CS~29495--041} was obtained as part of the
``First Stars'' project using UVES on the VLT
(M.\ Spite, private communication, 2010), 
and the spectra of \mbox{CS~22878--101} and \mbox{HD~186478} 
were obtained in followup to the ``HK Survey'' using MIKE on 
the Magellan-Clay Telescope.
All spectra have been Gaussian smoothed to the resolution of
\mbox{VI-18}.
Note that most absorption features in this wavelength
range are due to Fe-group elements.
}
\end{center}
\end{figure*}

Finally, in Figure~\ref{metallicityplot}
we present an empirical verification that our
metallicities are consistent with those derived for other
metal-poor field RGB stars.
All four stars shown in Figure~\ref{metallicityplot} have
very similar colors, and
in all four cases the reddening is small, $E(B-V) <$~0.1.
Most of the absorption lines in the wavelength region shown
are due to Fe-group species.
The metallicities listed for \mbox{CS~22878--101}, 
\mbox{CS~29495--041}, and \mbox{HD~186478} are reported directly
from \citet{cayrel04}, whose abundances are frequently used
as abundance standards for low-metallicity field RGB stars
in the Solar neighborhood.
(The metallicities derived by \citealt{mcwilliam95} for these stars
are very similar.)
The metallicity of \mbox{VI-18} inferred from the spectra of
these other three stars, [Fe/H]~$\approx -$2.7, 
agrees well with the metallicity we have derived
in our own analysis, [Fe/H]~$= -$2.78~$\pm$~0.06 ($\sigma =$~0.18).
(\mbox{VI-18} is chosen for comparison because
it has a color very similar to several stars in the
Cayrel et al.\ sample with higher and lower metallicities.)
Performing the same tests on these stars as described above
(comparing log($gf$) values, rederiving [Fe~\textsc{i}/H]
using the Cayrel et al.\ EWs and atmospheric parameters 
but our analysis tools) only leads to a lowering 
of their [Fe/H] by 0.07~dex.
Our derived metallicities thus appear reasonable
when compared with metal-poor field RGB stars.

\clearpage

\clearpage


\begin{thebibliography}{}


\bibitem[Alonso et al.(1999)]{alonso99} Alonso, A., Arribas, S., \& 
Mart{\'{\i}}nez-Roger, C.\ 1999, \aaps, 139, 335 

\bibitem[Arcones \& Montes(2011)]{arcones10} Arcones, A., \& 
Montes, F.\ 2011, \apj, 731, 5 

\bibitem[Armosky et al.(1994)]{armosky94} Armosky, B.~J., Sneden, 
C., Langer, G.~E., \& Kraft, R.~P.\ 1994, \aj, 108, 1364 

\bibitem[Asplund et al.(2009)]{asplund09} Asplund, M., Grevesse, N., 
Sauval, A.~J., \& Scott, P.\ 2009, \araa, 47, 481 

\bibitem[Barden \& Armandroff(1995)]{barden95} Barden, S.~C., \& 
Armandroff, T.\ 1995, \procspie, 2476, 56 

\bibitem[Bergemann(2011)]{bergemann11} Bergemann, M.\ 2011, \mnras, in press
 (arXiv:1101.0828)

\bibitem[Bershady et al.(2008)]{bershady08} Bershady, M., et al.\ 
2008, \procspie, 7014, 70140H 

\bibitem[Bevington \& Robinson(2003)]{bevington03} 
Bevington, P.~R., \& Robinson, D.~K.\ 2003, 
Data Reduction and Error Analysis for the Physical Sciences, 3rd ed., 
Boston, MA: McGraw-Hill, p.\ 197--201, 252--255

\bibitem[Bi{\'e}mont et al.(1989)]{biemont89} Bi{\'e}mont, E., Grevesse, N., 
Faires, L.~M., Marsden, G., \& Lawler, J.~E.\ 1989, \aap, 209, 391 

\bibitem[Blackwell et al.(1982)]{blackwell82} Blackwell, D.~E., 
Petford, A.~D., Shallis, M.~J., \& Leggett, S.\ 1982, \mnras, 199, 21 

\bibitem[Blackwell-Whitehead \& Bergemann(2007)]{blackwellwhitehead07} 
Blackwell-Whitehead, R., \& Bergemann, M.\ 2007, \aap, 472, L43 

\bibitem[Bonifacio et al.(2009)]{bonifacio09} Bonifacio, P., et al.\ 2009, 
\aap, 501, 519 

\bibitem[Booth et al.(1984)]{booth84} Booth, A.~J., Blackwell, 
D.~E., Petford, A.~D., \& Shallis, M.~J.\ 1984, \mnras, 208, 147 

\bibitem[Buonanno et al.(1983)]{buonanno83} Buonanno, R., Buscema, G., 
Corsi, C.~E., Iannicola, G., Smriglio, F., \& Pecci, F.~F.\ 1983, \aaps, 53, 1 

\bibitem[Carbon et al.(1982)]{carbon82} Carbon, D.~F., 
Romanishin, W., Langer, G.~E., Butler, D., Kemper, E., Trefzger, C.~F., 
Kraft, R.~P., \& Suntzeff, N.~B.\ 1982, \apjs, 49, 207 

\bibitem[Carretta et al.(2010a)]{carretta10a} Carretta, E., et al.\ 
2010a, \aap, 520, A95 

\bibitem[Carretta et al.(2010b)]{carretta10b} Carretta, E., et al.\
2010b, \apjl, 722, L1

\bibitem[Cayrel et al.(2004)]{cayrel04} Cayrel, R., et al.\ 2004, \aap, 
416, 1117 

\bibitem[Clayton(1988)]{clayton88} Clayton, D.~D.\ 1988, \mnras, 234, 1

\bibitem[Cohen(1979)]{cohen79} Cohen, J.~G.\ 1979, \apj, 231, 
751

\bibitem[Cohen \& McCarthy(1997)]{cohen97} Cohen, J.~G., \& McCarthy, J.~K.\ 
1997, \aj, 113, 1353 

\bibitem[Cohen et al.(2004)]{cohen04} Cohen, J.~G., et al.\ 
2004, \apj, 612, 1107 

\bibitem[Cohen et al.(2010)]{cohen10} Cohen, J.~G., Kirby, 
E.~N., Simon, J.~D., \& Geha, M.\ 2010, \apj, 725, 288 

\bibitem[Cudworth(1976)]{cudworth76} Cudworth, K.~M.\ 1976, \aj, 81, 975 

\bibitem[Den Hartog et al.(2003)]{denhartog03} Den Hartog, E.~A., 
Lawler, J.~E., Sneden, C., \& Cowan, J.~J.\ 2003, \apjs, 148, 543 

\bibitem[Dinescu et al.(1999)]{dinescu99} Dinescu, D.~I., Girard, 
T.~M., \& van Altena, W.~F.\ 1999, \aj, 117, 1792 

\bibitem[D'Orazi et al.(2010)]{dorazi10} D'Orazi, V., Gratton, 
R., Lucatello, S., Carretta, E., Bragaglia, A., 
\& Marino, A.~F.\ 2010, \apjl, 719, L213 

\bibitem[Drukier et al.(2007)]{drukier07} Drukier, G.~A., Cohn, 
H.~N., Lugger, P.~M., Slavin, S.~D., Berrington, R.~C., 
\& Murphy, B.~W.\ 2007, \aj, 133, 1041 

\bibitem[Farouqi et al.(2010)]{farouqi10} Farouqi, K., Kratz, 
K.-L., Pfeiffer, B., Rauscher, T., Thielemann, F.-K., 
\& Truran, J.~W.\ 2010, \apj, 712, 1359 

\bibitem[Freiburghaus et al.(1999)]{freiburghaus99} Freiburghaus, C., 
Rembges, J.-F., Rauscher, T., Kolbe, E., Thielemann, F.-K., Kratz, K.-L., 
Pfeiffer, B., \& Cowan, J.~J.\ 1999, \apj, 516, 381 

\bibitem[Fr{\"o}hlich et al.(2006)]{frohlich06} Fr{\"o}hlich, C., 
Mart{\'{\i}}nez-Pinedo, G., Liebend{\"o}rfer, M., Thielemann, F.-K., Bravo, 
E., Hix, W.~R., Langanke, K., 
\& Zinner, N.~T.\ 2006, Physical Review Letters, 96, 142502 

\bibitem[Gallino et al.(1998)]{gallino98} Gallino, R., Arlandini, 
C., Busso, M., Lugaro, M., Travaglio, C., Straniero, O., Chieffi, A., 
\& Limongi, M.\ 1998, \apj, 497, 388 

\bibitem[Gratton et al.(1996)]{gratton96} Gratton, R.~G., Carretta, E., \& 
Castelli, F.\ 1996, \aap, 314, 191 

\bibitem[Grevesse et al.(1989)]{grevesse89} Grevesse, N., Blackwell, D.~E., 
\& Petford, A.~D.\ 1989, \aap, 208, 157 

\bibitem[Gustafsson et al.(1975)]{gustafsson75} Gustafsson, B., Bell, R.~A., 
Eriksson, K., \& Nordlund, A.\ 1975, \aap, 42, 407 

\bibitem[Gustafsson et al.(2008)]{gustafsson08} Gustafsson, B., 
Edvardsson, B., Eriksson, K., J{\o}rgensen, U.~G., Nordlund, {\AA}., 
\& Plez, B.\ 2008, \aap, 486, 951 

\bibitem[Hannaford et al.(1982)]{hannaford82} Hannaford, P., Lowe, 
R.~M., Grevesse, N., Bi{\'e}mont, E., \& Whaling, W.\ 1982, \apj, 261, 736 

\bibitem[Harris(1996)]{harris96} Harris, W.~E.\ 1996, \aj, 112, 1487 

\bibitem[Ivans et al.(1999)]{ivans99} Ivans, I.~I., Sneden, C., 
Kraft, R.~P., Suntzeff, N.~B., Smith, V.~V., Langer, G.~E., 
\& Fulbright, J.~P.\ 1999, \aj, 118, 1273 

\bibitem[Ivans et al.(2001)]{ivans01} Ivans, I.~I., Kraft, 
R.~P., Sneden, C., Smith, G.~H., Rich, R.~M., 
\& Shetrone, M.\ 2001, \aj, 122, 1438 

\bibitem[Ivans et al.(2006)]{ivans06} Ivans, I.~I., Simmerer, 
J., Sneden, C., Lawler, J.~E., Cowan, J.~J., Gallino, R., 
\& Bisterzo, S.\ 2006, \apj, 645, 613 

\bibitem[Johnson(2002)]{johnson02} Johnson, J.~A.\ 2002, \apjs, 139, 219 

\bibitem[Johnson \& Pilachowski(2010)]{johnson10} Johnson, C.~I., \& 
Pilachowski, C.~A.\ 2010, \apj, 722, 1373 

\bibitem[King et al.(1998)]{king98} King, J.~R., Stephens, A., 
Boesgaard, A.~M., \& Deliyannis, C.\ 1998, \aj, 115, 666 

\bibitem[Kurucz \& Bell(1995)]{kurucz95} Kurucz, R.~L., \& Bell, B.\ 
1995, Kurucz CD-ROM, Cambridge, MA: Smithsonian Astrophysical Observatory

\bibitem[Lai et al.(2008)]{lai08} Lai, D.~K., Bolte, M., 
Johnson, J.~A., Lucatello, S., Heger, A., 
\& Woosley, S.~E.\ 2008, \apj, 681, 1524 

\bibitem[Langer et al.(1998)]{langer98} Langer, G.~E., Fischer, 
D., Sneden, C., \& Bolte, M.\ 1998, \aj, 115, 685 

\bibitem[Lawler \& Dakin(1989)]{lawler89} Lawler, J.~E., \& Dakin, J.~T.\ 
1989, Journal of the Optical Society of America B Optical Physics, 6, 1457 

\bibitem[Lawler et al.(2001a)]{lawler01a} Lawler, J.~E., 
Bonvallet, G., \& Sneden, C.\ 2001a, \apj, 556, 452 

\bibitem[Lawler et al.(2001b)]{lawler01b} Lawler, J.~E., 
Wickliffe, M.~E., den Hartog, E.~A., \& Sneden, C.\ 2001b, \apj, 563, 1075 

\bibitem[Lawler et al.(2004)]{lawler04} Lawler, J.~E., Sneden, 
C., \& Cowan, J.~J.\ 2004, \apj, 604, 850 

\bibitem[Lawler et al.(2008)]{lawler08} Lawler, J.~E., Sneden, 
C., Cowan, J.~J., Wyart, J.-F., Ivans, I.~I., Sobeck, J.~S., Stockett, 
M.~H., \& Den Hartog, E.~A.\ 2008, \apjs, 178, 71 

\bibitem[Lawler et al.(2009)]{lawler09} Lawler, J.~E., Sneden, 
C., Cowan, J.~J., Ivans, I.~I., \& Den Hartog, E.~A.\ 2009, \apjs, 182, 51 

\bibitem[Lupton et al.(1985)]{lupton85} Lupton, R., Gunn, J.~E., 
\& Griffin, R.~F.\ 1985, Proceedings of the 113th Symposium of the IAU,
``Dynamics of Star Clusters,''
Goodman, J.\ \& Hut, P., eds., 113, 19

\bibitem[Malcheva et al.(2006)]{malcheva06} Malcheva, G., Blagoev, 
K., Mayo, R., Ortiz, M., Xu, H.~L., Svanberg, S., Quinet, P., 
\& Bi{\'e}mont, E.\ 2006, \mnras, 367, 754 

\bibitem[Marino et al.(2009)]{marino09} Marino, A.~F., Milone, A.~P., 
Piotto, G., Villanova, S., Bedin, L.~R., Bellini, A., \& Renzini, A.\ 
2009, \aap, 505, 1099 

\bibitem[Marino et al.(2011)]{marino11} Marino, A.~F., Sneden, C., 
Kraft, R.~P., Wallerstein, G., Norris, J.~E., Da Costa, G., 
Milone, A.~P., Ivans, I.~I., et al.\ 2011, \aap, in press (arXiv:1105.1523)

\bibitem[McCall(2004)]{mccall04} McCall, M.~L.\ 2004, \aj, 128, 2144 

\bibitem[McWilliam et al.(1995)]{mcwilliam95} McWilliam, A., 
Preston, G.~W., Sneden, C., \& Searle, L.\ 1995, \aj, 109, 2757 

\bibitem[M{\'e}sz{\'a}ros et al.(2009)]{meszaros09} 
M{\'e}sz{\'a}ros, S., Dupree, A.~K., \& Szalai, T.\ 2009, \aj, 137, 4282 

\bibitem[Nitz et al.(1999)]{nitz99} Nitz, D.~E., Kunau, A.~E., 
Wilson, K.~L., \& Lentz, L.~R.\ 1999, \apjs, 122, 557 

\bibitem[Norris \& Zinn(1977)]{norris77} Norris, J., \& Zinn, R.\ 
1977, \apj, 215, 74 

\bibitem[O'Brian et al.(1991)]{obrian91a} O'Brian, T.~R., Wickliffe, M.~E., 
Lawler, J.~E., Whaling, W., \& Brault, J.~W.\ 1991, J.\ Opt.\ Soc.\ Am.\ B 
Optical Physics, 8, 1185 

\bibitem[O'Brian \& Lawler(1991)]{obrian91b} O'Brian, T.~R., \& Lawler, J.~E.\ 
1991, \pra, 44, 7134 

\bibitem[Otsuki et al.(2006)]{otsuki06} Otsuki, K., Honda, S., 
Aoki, W., Kajino, T., \& Mathews, G.~J.\ 2006, \apjl, 641, L117 

\bibitem[Peterson \& Reed(1987)]{peterson87} Peterson, C.~J., \& 
Reed, B.~C.\ 1987, \pasp, 99, 20 

\bibitem[Peterson et al.(1990)]{peterson90} Peterson, R.~C., 
Kurucz, R.~L., \& Carney, B.~W.\ 1990, \apj, 350, 173 

\bibitem[Pickering et al.(2002)]{pickering02} Pickering, J.~C., 
Thorne, A.~P., \& Perez, R.\ 2002, \apjs, 138, 247 

\bibitem[Pilachowski et al.(2000)]{pilachowski00} Pilachowski, C.~A., 
Sneden, C., Kraft, R.~P., Harmer, D., 
\& Willmarth, D.\ 2000, \aj, 119, 2895 

\bibitem[Pont et al.(1998)]{pont98} Pont, F., Mayor, M., Turon, C., \& 
Vandenberg, D.~A.\ 1998, \aap, 329, 87 

\bibitem[Preston et al.(2006)]{preston06} Preston, G.~W., Sneden, 
C., Thompson, I.~B., Shectman, S.~A., \& Burley, G.~S.\ 2006, \aj, 132, 85 

\bibitem[Pryor \& Meylan(1993)]{pryor93} Pryor, C., \& Meylan, G.\ 
1993, ASP Conference Series,
``Structure and Dynamics of Globular Clusters,''
Djorgovski, S.G., \& Meylan, G., eds. 50, 357 

\bibitem[Ram{\'{\i}}rez \& Mel{\'e}ndez(2005a)]{ramirez05a} Ram{\'{\i}}rez, I., 
\& Mel{\'e}ndez, J.\ 2005a, \apj, 626, 446 

\bibitem[Ram{\'{\i}}rez \& Mel{\'e}ndez(2005b)]{ramirez05b} Ram{\'{\i}}rez, I., 
\& Mel{\'e}ndez, J.\ 2005b, \apj, 626, 465 

\bibitem[Reed et al.(1988)]{reed88} Reed, B.~C., Hesser, 
J.~E., \& Shawl, S.~J.\ 1988, \pasp, 100, 545 

\bibitem[Rees(1992)]{rees92} Rees, R.~F., Jr.\ 1992, \aj, 103, 1573 

\bibitem[Roederer et al.(2009)]{roederer09} Roederer, I.~U., 
Kratz, K.-L., Frebel, A., Christlieb, N., Pfeiffer, B., Cowan, J.~J., 
\& Sneden, C.\ 2009, \apj, 698, 1963 

\bibitem[Roederer et al.(2010a)]{roederer10a} Roederer, I.~U., 
Sneden, C., Thompson, I.~B., Preston, G.~W., 
\& Shectman, S.~A.\ 2010a, \apj, 711, 573 

\bibitem[Roederer et al.(2010b)]{roederer10b} Roederer, I.~U., 
Cowan, J.~J., Karakas, A.~I., Kratz, K.-L., Lugaro, M., Simmerer, J., 
Farouqi, K., \& Sneden, C.\ 2010b, \apj, 724, 975 

\bibitem[Roederer(2011)]{roederer11} Roederer, I.~U.\ 2011, \apj, 732, L17

\bibitem[Sadakane et al.(2004)]{sadakane04} Sadakane, K., Arimoto, 
N., Ikuta, C., Aoki, W., Jablonka, P., 
\& Tajitsu, A.\ 2004, \pasj, 56, 1041 

\bibitem[Sandage \& Walker(1966)]{sandage66} Sandage, A., \& Walker, M.~F.\ 
1966, \apj, 143, 313 

\bibitem[Sandage(1969)]{sandage69} Sandage, A.\ 1969, \apj, 157, 515 

\bibitem[Schlegel et al.(1998)]{schlegel98} Schlegel, D.~J., 
Finkbeiner, D.~P., \& Davis, M.\ 1998, \apj, 500, 525 

\bibitem[Shetrone(1996)]{shetrone96} Shetrone, M.~D.\ 1996, \aj, 112, 1517 

\bibitem[Shetrone et al.(1998)]{shetrone98} Shetrone, M.~D., 
Bolte, M., \& Stetson, P.~B.\ 1998, \aj, 115, 1888 

\bibitem[Shetrone \& Keane(2000)]{shetrone00} Shetrone, M.~D., \& 
Keane, M.~J.\ 2000, \aj, 119, 840 

\bibitem[Shetrone et al.(2001)]{shetrone01} Shetrone, M.~D., 
C{\^o}t{\'e}, P., \& Sargent, W.~L.~W.\ 2001, \apj, 548, 592 

\bibitem[Skrutskie et al.(2006)]{skrutskie06} Skrutskie, M.~F., et 
al.\ 2006, \aj, 131, 1163 

\bibitem[Smith et al.(2000)]{smith00} Smith, V.~V., Suntzeff,
N.~B., Cunha, K., Gallino, R., Busso, M., Lambert, D.~L.,
\& Straniero, O.\ 2000, \aj, 119, 1239

\bibitem[Smith(2008)]{smith08} Smith, G.~H.\ 2008, \pasp, 120, 952 

\bibitem[Smith et al.(2009)]{smith09} Smith, M.~C., et al.\ 
2009, \mnras, 399, 1223 

\bibitem[Sneden(1973)]{sneden73} Sneden, C.~A.\ 1973, 
Ph.D.~Thesis, Univ.\ of Texas at Austin

\bibitem[Sneden et al.(1991)]{sneden91} Sneden, C., Kraft, 
R.~P., Prosser, C.~F., \& Langer, G.~E.\ 1991, \aj, 102, 2001 

\bibitem[Sneden et al.(1997)]{sneden97} Sneden, C., Kraft, 
R.~P., Shetrone, M.~D., Smith, G.~H., Langer, G.~E., 
\& Prosser, C.~F.\ 1997, \aj, 114, 1964 

\bibitem[Sneden et al.(2000)]{sneden00} Sneden, C., Pilachowski, 
C.~A., \& Kraft, R.~P.\ 2000, \aj, 120, 1351 

\bibitem[Sneden et al.(2003)]{sneden03} Sneden, C., et al.\ 
2003, \apj, 591, 936 

\bibitem[Sneden et al.(2008)]{sneden08} Sneden, C., Cowan, J.~J., \& 
Gallino, R.\ 2008, \araa, 46, 241 

\bibitem[Sneden et al.(2009)]{sneden09} Sneden, C., Lawler, 
J.~E., Cowan, J.~J., Ivans, I.~I., 
\& Den Hartog, E.~A.\ 2009, \apjs, 182, 80 

\bibitem[Sobeck et al.(2007)]{sobeck07} Sobeck, J.~S., Lawler, 
J.~E., \& Sneden, C.\ 2007, \apj, 667, 1267 

\bibitem[Sobeck et al.(2011)]{sobeck11} Sobeck, J.~S., et al.\ 
2011, \aj, 141, 175 

\bibitem[Soderberg et al.(1999)]{soderberg99} Soderberg, A.~M., 
Pilachowski, C.~A., Barden, S.~C., Willmarth, D., 
\& Sneden, C.\ 1999, \pasp, 111, 1233 

\bibitem[Strom \& Strom(1971)]{strom71} Strom, S.~E., \& Strom, K.~M.\ 
1971, \aap, 14, 111 

\bibitem[Tucholke et al.(1996)]{tucholke96} Tucholke, H.-J., 
Scholz, R.-D., \& Brosche, P.\ 1996, \aap, 312, 74 

\bibitem[van den Bergh et al.(1991)]{vandenbergh91} van den Bergh, 
S., Morbey, C., \& Pazder, J.\ 1991, \apj, 375, 594 

\bibitem[Villanova et al.(2010)]{villanova10} Villanova, S.,
Geisler, D., \& Piotto, G.\ 2010, \apjl, 722, L18

\bibitem[Webbink(1985)]{webbink85} Webbink, R.~F.\ 1985, 
Proceedings of the 113th Symposium of the IAU,
``Dynamics of Star Clusters,''
Goodman, J.\ \& Hut, P., eds., 113, 541 

\bibitem[Woosley et al.(1994)]{woosley94} Woosley, S.~E., Wilson, 
J.~R., Mathews, G.~J., Hoffman, R.~D., 
\& Meyer, B.~S.\ 1994, \apj, 433, 229 

\bibitem[Yong et al.(2008a)]{yong08a} Yong, D., Lambert, D.~L., 
Paulson, D.~B., \& Carney, B.~W.\ 2008a, \apj, 673, 854 

\bibitem[Yong et al.(2008b)]{yong08b} Yong, D., Karakas, A.~I., 
Lambert, D.~L., Chieffi, A., \& Limongi, M.\ 2008b, 689, 1031 

\bibitem[Yong \& Grundahl(2008)]{yong08c} Yong, D., \& Grundahl, F.\ 2008,
\apjl, 672, L29

\bibitem[Yong et al.(2009)]{yong09} Yong, D., Grundahl, F.,
D'Antona, F., Karakas, A.~I., Lattanzio, J.~C.,
\& Norris, J.~E.\ 2009, \apjl, 695, L62

\bibitem[Zinn(1973)]{zinn73} Zinn, R.\ 1973, \apj, 182, 183 

\bibitem[Zinn(1980)]{zinn80} Zinn, R.\ 1980, \apjs, 42, 19 




\end{thebibliography}
\end{document}